\date{\today}

\documentclass[aps,pra,twocolumn,superscriptaddress,groupedaddress]{revtex4}  

\usepackage{dcolumn}
\usepackage{graphicx}
\usepackage{amsmath}
\usepackage{amssymb}
\usepackage{algorithm}
\usepackage{algpseudocode}
\usepackage{xcolor}
\usepackage{tikz}
\usepackage{qcircuit}
\usepackage{mathtools} 
\usepackage{braket} 
\usepackage{verbatim}
\usepackage{pdflscape}
\usepackage{paralist}
\usepackage[colorlinks=true,hyperindex,breaklinks=true,allcolors=blue]{hyperref}

\begin{document}

\definecolor{brickred}{rgb}{.72,0,0} 
\definecolor{darkblue}{rgb}{0,0,0.5} 
\definecolor{darkgreen}{rgb}{0,0.5,0} 

\title{
Local, Expressive, Quantum-Number-Preserving VQE Ans\"atze for Fermionic Systems
}

\author{Gian-Luca R. Anselmetti}
\email{gian-luca.anselmetti@covestro.com}
\affiliation{
Covestro Deutschland AG, Leverkusen 51373, Germany
}

\author{David Wierichs}
\email{wierichs@thp.uni-koeln.de}
\affiliation{
Institute for Theoretical Physics, University of Cologne, Germany
}

\author{Christian Gogolin}
\email{christian.gogolin@covestro.com}
\affiliation{
Covestro Deutschland AG, Leverkusen 51373, Germany
}

\author{Robert M. Parrish}
\email{rob.parrish@qcware.com}
\affiliation{
QC Ware Corporation, Palo Alto, CA 94301, USA
}

\begin{abstract} 
We propose VQE circuit fabrics with advantageous properties for the simulation
of strongly correlated ground and excited states of molecules and materials
under the Jordan-Wigner mapping that can be implemented linearly locally and
preserve all relevant quantum numbers: the number of spin up ($\alpha$) and down
($\beta$) electrons and the total spin squared. We demonstrate that our
entangler circuits are expressive already at low depth and parameter count,
appear to become universal, and may be trainable without having to cross regions
of vanishing gradient, when the number of parameters becomes sufficiently large
and when these parameters are suitably initialized.  One particularly appealing
construction achieves this with just orbital rotations and pair exchange gates.
We derive optimal four-term parameter shift rules for and provide explicit
decompositions of our quantum number preserving gates and perform numerical
demonstrations on highly correlated molecules on up to 20 qubits.
\end{abstract}

\maketitle

\section{Introduction}
Hybrid quantum classical variational algorithms, including those of the
variational quantum eigensolver (VQE) type \cite{
peruzzo2014variational,
mcclean2016theory},
are among the leading candidates for quantum algorithms that may yield quantum
advantage in areas such as computational chemistry or machine learning already
in the era of noisy intermediate scale quantum (NISQ) computing
\cite{preskill2018quantum}. A foundational issue in VQE \cite{
peruzzo2014variational,
mcclean2016theory},
and in many of its extensions and alternatives
\cite{
mcclean2017hybrid,
parrish2019quantum,
nakanishi2019subspace,
urbanek2020chemistry,
ollitrault2020quantum,
huggins2020non,
parrish2019qfd,
stair2020multireference},
is finding a ``good'' definition of the entangler circuit. Here the qualifier
``good'' has many facets, possibly including:
(1) Providing an efficient approximate representation of the target quantum
states in the limit of an intermediate (ideally polynomial-scaling) depth
(2) Consisting of a low number of distinct physically realizable gate elements
(3) Exhibiting a simple pattern of how these gate elements are applied
(4) Exhibiting sparse spatial locality that is further compatible with device connectivity
(5) Exhibiting simple analytical gradient recipes and robust numerical convergence behavior during optimization of the
VQE energy, e.g., by mitigating the effects of barren plateaus \cite{mcclean2018barren} 
(6) Respecting exactly the natural particle and spin quantum number symmetries
of the target quantum states, e.g., as notably explored in
\cite{gard2020efficient}
(7) Providing an exact representation of the target quantum states in the limit
of sufficient (usually exponential-scaling) depth.

Especially within the context of VQE for spin-$1/2$ fermions governed by real,
spin-free Hamiltonian operators (e.g., electrons in molecules and materials, the
prime application of VQE), a variety of compelling VQE entangler circuit recipes
have been discussed in the literature.  Prominent examples include 
UCCSD \cite{peruzzo2014variational,PhysRevX.6.031007,Ryabinkin2018},   
$k$-UpCCGSD \cite{lee2018generalized,o2019generalized},
Jastrow-Factor VQE \cite{matsuzawa2020jastrow},
the symmetry-preserving ans\"atze \cite{gard2020efficient},
the hardware efficient ans\"atze \cite{Kandala2017,Bian2019}, 
ADAPT-VQE \cite{Grimsley2019},
pUCCD \cite{elfving2020simulating},
and additional methods discussed below \cite{
evangelista2019exact,
ganzhorn2019gate,
xia2020qubit,
yordanov2020efficient,
salis2019short,
khamoshi2020correlating}.
Each of these generally satisfies a subset of the ``good'' facets listed above,
though no extant ansatz that we are aware of obtains all of them, with the
notable exception of the generalized swap network form of $k$-UpCCGSD
of \cite{o2019generalized}.

In this work we develop a VQE entangler circuit recipe for fermions in the
Jordan-Wigner representation and show, or at least provide evidence, that it
obtains all facets, with facet (5) partially left to future numerical studies.
Perhaps the most notable property of our fabrics is the exact preservation of
all relevant quantum numbers the individual gate elements of the fabric,
which is why we refer to them as quantum number
preserving (QNP). This property may be critical for employment of VQE in larger
systems, where contaminations from or even variational collapse onto states
with different particle or spin quantum numbers can severely degrade the
quality of the VQE wavefunction. 

Note that after we posted the first version of our manuscript, we became aware
of the generalized swap network reformulation of $k$-UpCCGSD of
\cite{o2019generalized}. This paper refactors $k$-UpCCGSD to use
nearest-neighbor connectivity, yielding a circuit fabric that could be written
in terms of four-qubit gates containing diagonal pair exchange and orbital
rotation elements in a very similar manner as our $\hat Q$-type QNP gate fabric
discussed below.  There are some tactical differences in the qubit ordering and
the generalized swap network paper does not emphasize the role of quantum number
symmetry as much as the present manuscript.  Moreover, the origin of the $\hat
Q$-type QNP gate fabric as a simplification of our more-complete $\hat F$-type
QNP gate fabric of Appendix \ref{appendix:other_qnp_gate_fabrics} provides a
markedly different approach to developing this gate fabric. In any case, we
encourage any readers interested in the present manuscript to also explore
\cite{o2019generalized}.

\section{Gate fabrics}
\begin{figure}[t]
\centering
\begin{equation*}
\label{eq:SU2N}
\begin{array}{l}
\Qcircuit @R=0.3em @C=0.3em @!R {
 & \multigate{1}{\mathit{SU}(4)}
 & \qw
 & \multigate{1}{\mathit{SU}(4)}
 & \qw
 & \qw \\
 & \ghost{\mathit{SU}(4)}
 & \multigate{1}{\mathit{SU}(4)}
 & \ghost{\mathit{SU}(4)}
 & \multigate{1}{\mathit{SU}(4)}
 & \qw \\
 & \multigate{1}{\mathit{SU}(4)}
 & \ghost{\mathit{SU}(4)}
 & \multigate{1}{\mathit{SU}(4)}
 & \ghost{\mathit{SU}(4)}
 & \qw \\
 & \ghost{\mathit{SU}(4)}
 & \multigate{1}{\mathit{SU}(4)}
 & \ghost{\mathit{SU}(4)}
 & \multigate{1}{\mathit{SU}(4)}
 & \qw \\
 & \multigate{1}{\mathit{SU}(4)}
 & \ghost{\mathit{SU}(4)}
 & \multigate{1}{\mathit{SU}(4)}
 & \ghost{\mathit{SU}(4)}
 & \qw \\
 & \ghost{\mathit{SU}(4)}
 & \qw
 & \ghost{\mathit{SU}(4)}
 & \qw
 & \qw \\
}
\end{array}
\ldots
\phantom{}
\cong
\phantom{}
\begin{array}{l}
\Qcircuit @R=0.3em @C=0.3em @!R {
 & \multigate{5}{\mathit{SU}(2^6)}
 & \qw \\
 & \ghost{\mathit{SU}(2^6)}
 & \qw \\
 & \ghost{\mathit{SU}(2^6)}
 & \qw \\
 & \ghost{\mathit{SU}(2^6)}
 & \qw \\
 & \ghost{\mathit{SU}(2^6)}
 & \qw \\
 & \ghost{\mathit{SU}(2^6)}
 & \qw \\
}
\end{array}
\end{equation*}
\begin{equation*}
\mathit{SU}(4)
\coloneqq
\exp(\hat X) : \hat X = -\hat X^\dagger, \mathrm{Tr} (\hat X) = 0, \hat X \in \mathbb{C}^{4} \times
\mathbb{C}^{4}
\end{equation*}
\begin{equation*}
\cong
\begin{array}{l}
\Qcircuit @R=0.3em @C=0.3em @!R {
 & \gate{SU(2)}
 & \ctrl{1}
 & \gate{SU(2)}
 & \ctrl{1}
 & \gate{SU(2)}
 & \ctrl{1}
 & \gate{SU(2)}
 & \qw \\
 & \gate{SU(2)}
 & \ctrl{-1}
 & \gate{SU(2)}
 & \ctrl{-1}
 & \gate{SU(2)}
 & \ctrl{-1}
 & \gate{SU(2)}
 & \qw \\
}
\end{array}
\end{equation*}
\caption{Sketch for $N=6$ of a gate fabric universal for $\mathcal{SU}(2^N)$ providing
inspiration for the fermionic quantum number preserving gate fabrics developed here.
The gate fabric is a 2-local-nearest-neighbor tessellation of alternating
15-parameter, 2-qubit $SU(4)$ gates. The $SU(2)$ gate
in the $SU(4)$ gate decomposition on the bottom line is the 3-parameter universal
gate for the 1-qubit Bloch sphere. The indicated 24-parameter decomposition of
$SU(4)$ is overcomplete for the 15-parameter $\mathcal{SU}(4)$ group.}
\label{fig:SU4}
\end{figure}
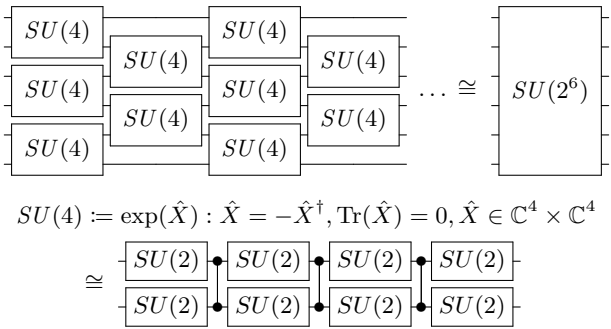

Our VQE entangler circuit recipe draws inspiration from 
the well known fact that the qubit Hilbert space (without any fermionic symmetries)
$\mathcal{SU}(2^{N})$ can be spanned by a tessellation of 2-qubit gates
universal for $\mathcal{SU}(4)$ in alternating layers (see Figure~\ref{fig:SU4}).
This tessellation can formally be
repeated to infinite depth. However, one finds that after some finite,
$N$-dependent critical depth of order $\mathcal{O}(2^{2N})$, additional gate
layers do not increase the expressiveness of the circuit, as formal completeness
(denoted ``universality'') in $\mathcal{SU}(2^{N})$ is achieved.
In practice usually shorter circuit depths are of interest.
For instance, one may consider the case where the tessellation is restricted to
be polynomial scaling in $N$, in which case universality cannot be exactly
achieved. However, a good approximation of specialized (e.g., physically relevant) parts of some
subgroup may still be achievable in a way that is tractable to compute even on a
NISQ computer but intractable to compute with a classical device.

Particularly striking in Figure \ref{fig:SU4} is the locality
(alternating nearest neighbor connectivity) and simplicity (single gate element) of the
circuit, properties of what we call a ``gate fabric.'' More precisely,
throughout this manuscript, we define a gate fabric for a subgroup of
$\mathcal{SU}(2^{N})$ to be a tessellation of gates over $N$-qubits with the
following properties:
\begin{compactenum}
\item Simplicity: Composed of a single type of
$k$-qubit, $l$-parameter gate element (with a known decomposition into 
elementary gates), where $k$ and $l$ are independent of $N$.
\item Linear Locality: When the qubits are thought of as arranged on a vertical line
the gate elements are arranged in layers and connect up to $k$ contiguous qubits.
\item Universality: Achieving universality within the target subgroup
of $\mathcal{SU}(2^{N})$ within a finite number of layers depending on $N$.
\item Symmetry: commuting with all symmetry operators used to define the
subgroup of $\mathcal{SU}(2^{N})$, i.e., $[\hat U, \hat N] = 0$, where $\hat U$
is the circuit unitary for any set of parameters and $\hat N$ is the
symmetry operator.
\end{compactenum}
Depending on the subgroup of $\mathcal{SU}(2^{N})$ of interest it can be more or
less difficult to find fabrics akin to the one shown in Figure~\ref{fig:SU4}.
In Appendix~\ref{sec:subgroups} we discuss the trivial restriction to
$\mathcal{SO}(2^{N})$ and the less-trivial restriction to subspaces of definite
Hamming weight.

\section{Gate fabrics for fermions under the Jordan-Wigner mapping}
The focus of this work is the construction of gate fabrics for the subgroup
$\mathcal{F}(2^{2M}) \in \mathcal{SU}(2^{N})$ constrained to spin-restricted fermionic symmetry under
the Jordan-Wigner representation.
To make this more precise, we define $M$ real orthogonal spatial
orbitals $\{| \phi_{p} \rangle\}_{p=0}^M$. For each spatial orbital, we define
corresponding $\alpha$($\beta$) spin orbitals $| \psi_{p} \rangle \coloneqq |\phi_{p}
\rangle | \alpha \rangle$ ($| \psi_{\bar p} \rangle \coloneqq |\phi_{p} \rangle | \beta
\rangle$) for a total of $N \coloneqq 2M$ spin orbitals in a spin-restricted
formalism. We associate the occupation numbers of these spin orbitals with the
occupation numbers of $N$ qubits. We number the qubits in
``interleaved'' ordering
$\ldots|1_{\beta}\rangle|1_{\alpha}\rangle|0_{\beta}\rangle|0_{\alpha}\rangle$.
The fermionic creation/annihilation operators are defined in terms of the qubit
creation/annihilation operators via the Jordan-Wigner mapping in
``$\alpha$-then-$\beta$'' ordering, 
$p^{\pm} \coloneqq (\hat X_{p} \mp i
\hat Y_{p}) / 2 \bigotimes_{q = 0}^{p-1} \hat Z_{q}$ 
and
$\bar p^{\pm} \coloneqq (\hat X_{\bar p} \mp i
\hat Y_{\bar p}) / 2 \bigotimes_{q = 0}^{p-1} \hat Z_{\bar q}
\bigotimes_{q = 0}^{M-1} \hat Z_{q}$. 
We note that for the majority of applications in the space of
spin-$1/2$ fermions, the governing Hamiltonians are real (e.g., for
non-relativistic electronic structure theory), and so we restrict
from complex to real unitary operators, i.e., $\mathcal{SU}(2^{N}) \rightarrow
\mathcal{SO}(2^{N})$. The spin-restricted fermionic subgroup is then defined as
the subgroup of $\hat U \in \mathcal{SO}(2^{N})$ that respect the commutation
relations 
$[ \hat U, \hat N_{\alpha} ]  = 0$,
$[ \hat U, \hat N_{\beta} ]  = 0$, and
$[ \hat U, \hat S^2] = 0$.
Here the $\alpha$($\beta$) number operator is 
$\hat N_{\alpha} \coloneqq \sum_{p} p^\dagger p = \sum_{p} (\hat I - \hat Z_p) /
2$
[
$\hat N_{\beta} \coloneqq \sum_{p} \bar p^\dagger \bar p = \sum_{p} (\hat I - \hat
Z_{\bar p}) /
2$
].
\cite{ntnote}
The spin-squared operator is $\hat S^2 \coloneqq \sum_{pq} p q^\dagger \bar
p^\dagger \bar q + (\hat N_{\alpha} - \hat N_{\beta}) / 2 + (\hat N_{\alpha} -
\hat N_{\beta})^2 / 4$, and does not admit a local description in terms of Pauli
operators in the Jordan-Wigner basis (we provide further details in Appendices~\ref{appendix:jw_details} and \ref{appendix:two_mode_fermions}). 
We denote this real subgroup, preserving $\hat
N_{\alpha}$, $\hat N_{\beta}$, and $\hat S^2$, as $\mathcal{F}(2^{2M})$.

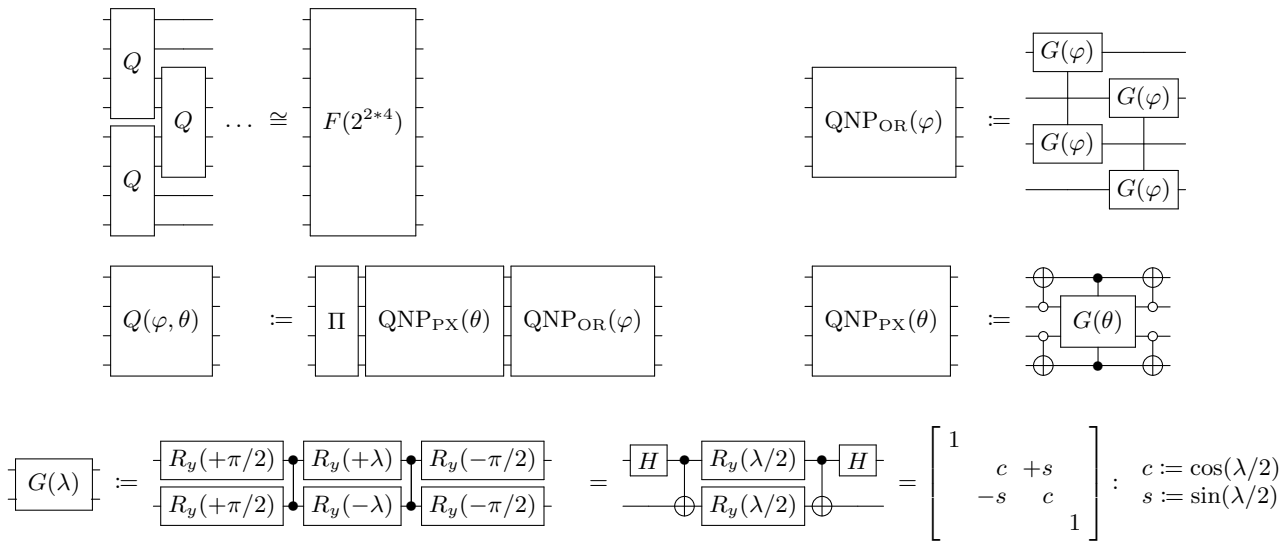
\begin{figure*}[tb]
\centering

\begin{equation*}
\begin{array}{llll}
\begin{array}{l}
\Qcircuit @R=0.3em @C=0.3em @!R {
 & \multigate{3}{Q}
 & \qw
 & \qw \\
 & \ghost{Q}
 & \qw
 & \qw \\
 & \ghost{Q}
 & \multigate{3}{Q}
 & \qw \\
 & \ghost{Q}
 & \ghost{Q}
 & \qw \\
 & \multigate{3}{Q}
 & \ghost{Q}
 & \qw \\
 & \ghost{Q}
 & \ghost{Q}
 & \qw \\
 & \ghost{Q}
 & \qw
 & \qw \\
 & \ghost{Q}
 & \qw
 & \qw \\
}
\end{array}
\ldots
&
\cong
\begin{array}{l}
\Qcircuit @R=0.3em @C=0.3em @!R {
 & \multigate{7}{F(2^{2 * 4})}
 & \qw \\
 & \ghost{F(2^{2 * 4})}
 & \qw \\
 & \ghost{F(2^{2 * 4})}
 & \qw \\
 & \ghost{F(2^{2 * 4})}
 & \qw \\
 & \ghost{F(2^{2 * 4})}
 & \qw \\
 & \ghost{F(2^{2 * 4})}
 & \qw \\
 & \ghost{F(2^{2 * 4})}
 & \qw \\
 & \ghost{F(2^{2 * 4})}
 & \qw \\
}
\end{array}
&
\begin{array}{l}
\Qcircuit @R=0.3em @C=0.3em @!R {
 & \multigate{3}{\mathrm{QNP_{OR}} (\varphi)}
 & \qw \\
 & \ghost{\mathrm{QNP_{OR}} (\varphi)}
 & \qw \\
 & \ghost{\mathrm{QNP_{OR}} (\varphi)}
 & \qw \\
 & \ghost{\mathrm{QNP_{OR}} (\varphi)}
 & \qw \\
}
\end{array}
&
\coloneqq
\begin{array}{l}
\Qcircuit @R=0.3em @C=0.3em @!R {
 & \gate{G (\varphi)} 
 & \qw
 & \qw \\
 & \qw \qwx[1] \qwx[-1]
 & \gate{G (\varphi)} 
 & \qw \\
 & \gate{G (\varphi)} 
 & \qw \qwx[1] \qwx[-1]
 & \qw \\
 & \qw
 & \gate{G (\varphi)} 
 & \qw \\
}
\end{array}
\\
\\
\begin{array}{l}
\Qcircuit @R=0.3em @C=0.3em @!R {
 & \multigate{3}{Q (\varphi, \theta)}
 & \qw \\
 & \ghost{Q (\varphi, \theta)}
 & \qw \\
 & \ghost{Q (\varphi, \theta)}
 & \qw \\
 & \ghost{Q (\varphi, \theta)}
 & \qw \\
}
\end{array}

&
\coloneqq
\begin{array}{l}
\Qcircuit @R=0.3em @C=0.3em @!R {
 & \multigate{3}{\Pi}
 & \multigate{3}{\mathrm{QNP_{PX}} (\theta)}
 & \multigate{3}{\mathrm{QNP_{OR}} (\varphi)}
 & \qw \\
 & \ghost{\Pi}
 & \ghost{\mathrm{QNP_{PX}} (\theta)}
 & \ghost{\mathrm{QNP_{OR}} (\varphi)}
 & \qw \\
 & \ghost{\Pi}
 & \ghost{\mathrm{QNP_{PX}} (\theta)}
 & \ghost{\mathrm{QNP_{OR}} (\varphi)}
 & \qw \\
 & \ghost{\Pi}
 & \ghost{\mathrm{QNP_{PX}} (\theta)}
 & \ghost{\mathrm{QNP_{OR}} (\varphi)}
 & \qw \\
}
\end{array}
\phantom{QISFTW}
&
\begin{array}{l}
\Qcircuit @R=0.3em @C=0.3em @!R {
 & \multigate{3}{\mathrm{QNP_{PX}} (\theta)}
 & \qw \\
 & \ghost{\mathrm{QNP_{PX}} (\theta)}
 & \qw \\
 & \ghost{\mathrm{QNP_{PX}} (\theta)}
 & \qw \\
 & \ghost{\mathrm{QNP_{PX}} (\theta)}
 & \qw \\
}
\end{array}
&
\coloneqq
\begin{array}{l}
\Qcircuit @R=0.3em @C=0.3em @!R {
 & \targ
 & \ctrl{1}
 & \targ
 & \qw \\
 & \ctrlo{-1}
 & \multigate{1}{G (\theta)}
 & \ctrlo{-1}
 & \qw \\
 & \ctrlo{+1}
 & \ghost{G (\theta)}
 & \ctrlo{+1}
 & \qw \\
 & \targ
 & \ctrl{-1}
 & \targ
 & \qw \\
}
\end{array}
\\
\end{array}
\end{equation*}

\begin{align*}
\begin{array}{l}
\Qcircuit @R=0.3em @C=0.3em @!R {
 & \multigate{1}{G (\lambda)}
 & \qw \\
 & \ghost{G (\theta)}
 & \qw \\
}
\end{array}
&\coloneqq
\begin{array}{l}
\Qcircuit @R=0.3em @C=0.3em @!R {
 & \gate{R_{y} (+\pi / 2)}
 & \ctrl{1}
 & \gate{R_{y} (+\lambda)}
 & \ctrl{1}
 & \gate{R_{y} (-\pi / 2)}
 & \qw \\
 & \gate{R_{y} (+\pi / 2)}
 & \ctrl{-1}
 & \gate{R_{y} (-\lambda)}
 & \ctrl{-1}
 & \gate{R_{y} (-\pi / 2)}
 & \qw \\
}
\end{array}
&=
\begin{array}{l}
\Qcircuit @R=0.3em @C=0.3em @!R {
 & \gate{H}
 & \ctrl{1}
 & \gate{R_{y} (\lambda / 2)}
 & \ctrl{1}
 & \gate{H}
 & \qw \\
 & \qw
 & \targ
 & \gate{R_{y} (\lambda / 2)}
 & \targ
 & \qw
 & \qw \\
}
\end{array}
=
\left [
\begin{array}{rrrr}
1 & & & \\
 & c & +s & \\
 & -s & c & \\
 & & & 1 \\
\end{array}
\right ]
:
\
\begin{array}{l}
c \coloneqq \cos(\lambda/2)
\\
s \coloneqq \sin(\lambda/2)
\\
\end{array}
\end{align*}
\caption{Proposed gate fabric for $\mathcal{F}(2^{2M})$
(sketched for $M=4$). The spin orbitals in Jordan-Wigner representation are
in ``interleaved'' ordering with even (odd) qubit indices
denoting $\alpha$ ($\beta$) spin orbitals. The Jordan-Wigner strings are taken to be 
in ``$\alpha$-then-$\beta$'' order. 
The gate fabric is a 4-local-nearest-neighbor- tessellation of alternating even
and odd spatial-orbital-pair 2-parameter, 4-qubit $\hat Q$ gates.
Each $\hat Q$ gate has two independent parameters and contains a 1-parameter, 4-qubit spatial orbital rotation
gate $\mathrm{QNP_{OR}}(\varphi)$ and a 1-parameter, 4-qubit diagonal pair
exchange gate $\mathrm{QNP_{PX}}(\theta)$. The order of $\mathrm{QNP_{OR}}$
and $\mathrm{QNP_{PX}}$ (note $[\mathrm{QNP_{OR}}, \mathrm{QNP_{PX}}]
\neq 0$) does not seem to substantially change expressiveness at intermediate depths.
$\mathrm{QNP_{OR}} (\varphi)$
implements the spatial orbital Givens rotation 
$|\phi_{0} \rangle = c |\phi_{0}\rangle + s |\phi_{1}\rangle$
and
$|\phi_{1} \rangle = -s |\phi_{0}\rangle + c |\phi_{1}\rangle$, with the same
orbital rotation applied in the $\alpha$ and $\beta$ spin orbitals. 
$\mathrm{QNP_{PX}}(\theta)$ implements the diagonal pair Givens rotation,
$|0011\rangle = c |0011\rangle + s |1100\rangle$
and
$|1100\rangle = -s |0011\rangle + c |1100\rangle$.
The real 1-parameter, 1-qubit rotation gate is $\hat R_{y} (\lambda) \coloneqq e^{-i
\lambda \hat Y/2}$.
In $\hat Q$, we include the optional constant $\hat \Pi$ gate, for which natural choices include
the 4-qubit identity gate, i.e., $\hat \Pi = \hat I$, or the fixed spin-adapted
orbital rotation gate $\hat \Pi = \mathrm{QNP_{OR}} (\pi)$. In the latter
case, the gate fabric with all parameters $\{ \theta = 0\}$ and $\{\varphi = 0\}$ promotes exchange of orbitals.
We find that the choice of $\hat \Pi \in \{\hat I,
\mathrm{QNP_{OR}} (\pi)\}$ does not appear to affect the expressiveness of the quantum
circuit, but the latter choice has turned out to be advantageous during gradient-based parameter
optimization. Regardless of the choice of $\hat \Pi$, 
this gate fabric exactly preserves the real nature of the subgroup,
exactly commutes with the $\hat N_{\alpha}$, $\hat N_{\beta}$, and $\hat S^2$
symmetry operators, and numerically appears to provide universality at
sufficient parameter count.
}
\label{fig:F}
\end{figure*}

Naively one might expect that there should not be any local gate fabric exactly
preserving all three fermionic quantum numbers, since the $\hat{S}^2$ operator
is non-local.  The crux of this work is thus the simple
quantum-number-preserving gate fabric of Figure~\ref{fig:F}. This gate fabric is
composed of 2-parameter 4-qubit gate elements $\hat Q$, each composed of a
1-parameter 4-qubit spin-adapted spatial orbital rotation gate
$\mathrm{QNP_{OR}} (\varphi)$ and a 1-parameter 4-qubit diagonal pair exchange
gate $\mathrm{QNP_{PX}} (\theta)$.  We describe further related
quantum-number-preserving gate fabrics for $\mathcal{F}(2^{2M})$ in
Appendix~\ref{appendix:other_qnp_gate_fabrics} - these were the progenitors of
the simpler gate fabrics shown in the main text, and may have advantageous
properties in specific realizations of VQE entangler circuits.

Facets (2-4) of gate fabrics are manifestly fulfilled for all these proposals
and facet (6) holds by construction as all gates individually preserve all
quantum numbers. For facets (1) and (7) we provide numerical evidence below and
in Appendix~\ref{sec:additional numerics}. It is worth noting that these tests
numerically indicate that our gate fabrics are universal in the vast bulk of
quantum number irreps (with the exception of a few high-spin edge cases for the
$\hat Q$-type gate fabrics of the main text, see Appendix~\ref{sec:additional
numerics}), i.e., that they may be used for cases where $S \neq 0$ and/or where
$N_{\alpha} \neq N_{\beta}$ (including both even and odd spin cases).
We believe that the methods from
\cite{brandao2016local,oszmaniec2020epsilon} or \cite{evangelista2019exact} can
be used to rigorously show universality in the (bulk of the) quantum number
sectors of $\mathcal{F}(2^{2M})$ in all cases (as well as that our circuits are
polynomial depth $\epsilon$-approximate unitary $t$-designs and form
$\epsilon$-nets).  Working out the details of a rigorous proof, which we believe
has to be done spin sector by spin sector in some cases, is however beyond the
scope of this work.

For $\mathrm{QNP_{OR}}$, $\mathrm{QNP_{PX}}$ (and all other parametrized quantum
number preserving gates introduced in
Appendix~\ref{appendix:other_qnp_gate_fabrics}) we provide explicit
decompositions into elementary gates in Appendix~\ref{sec:qnp_decompositions}.
We further provide generalized parameter shift rules
\cite{Mitarai_Fujii_18,farhi2018classification, Schuld2019, PennyLane} for
theses gates in Appendix~\ref{sec:four-term-rule}, enabling a computation of the
gradients with respect to their circuit parameters with a maximum of four
distinct circuits and without increasing circuit depth, gate count, or qubit
number.  We compare this gradient recipe to the generalization presented in
\cite{Kottmann_Guzik_20}, extend the variance minimization technique from
\cite{Mari_Killoran_20} to the $\mathrm{QNP}$ gate gradients and note that our
new rule can be applied to a large variety of other gates.

\section{Numerical demonstrations}
To numerically investigate the properties of the gate fabric from
Figure~\ref{fig:F} and to collect evidence that it satisfies all facets of a
``good'' entangler circuit, we consider two prototypical examples of highly
correlated molecular ground states: The first is p-benzyne, which exhibits a
biradical open-shell singlet ground state, with two unpaired electrons indicated
by significant deviations from Hartree-Fock natural orbital occupation numbers,
and four other moderate deviations from Hartree-Fock natural orbital occupation
numbers. We use the geometry from \cite{keller2015selection}, build the orbitals
at RHF/cc-pVDZ, and construct a (6e, 6o) active space Hamiltonian with the
orbitals ranging from HOMO-2 to LUMO+2. This corresponds to a case of $M=6$
spatial orbitals (i.e., $N=12$ qubits), and we focus on the ground state
irreducible representation $(N_{\alpha} = 3, N_{\beta} = 3, S = 0)$. For a
larger test case, we consider naphthalene, which while not intrinsically
biradical, has multiple natural orbitals with significant deviations from
Hartree-Fock natural orbital occupation numbers. We build the orbitals at
RHF/STO-3G, and then construct a (10e, 10o) active space Hamiltonian consisting
of the $\pi$ and $\pi^*$ orbitals. This corresponds to a case of $M=10$ spatial
orbitals (i.e., $N=20$ qubits), and we focus on the ground state irreducible
representation $(N_{\alpha} = 5, N_{\beta} = 5, S = 0)$. In both cases, we
consider VQE gate fabrics of the form of Figure \ref{fig:F}. 

\begin{figure*}[ht!]
\centering
\includegraphics[width=\linewidth]{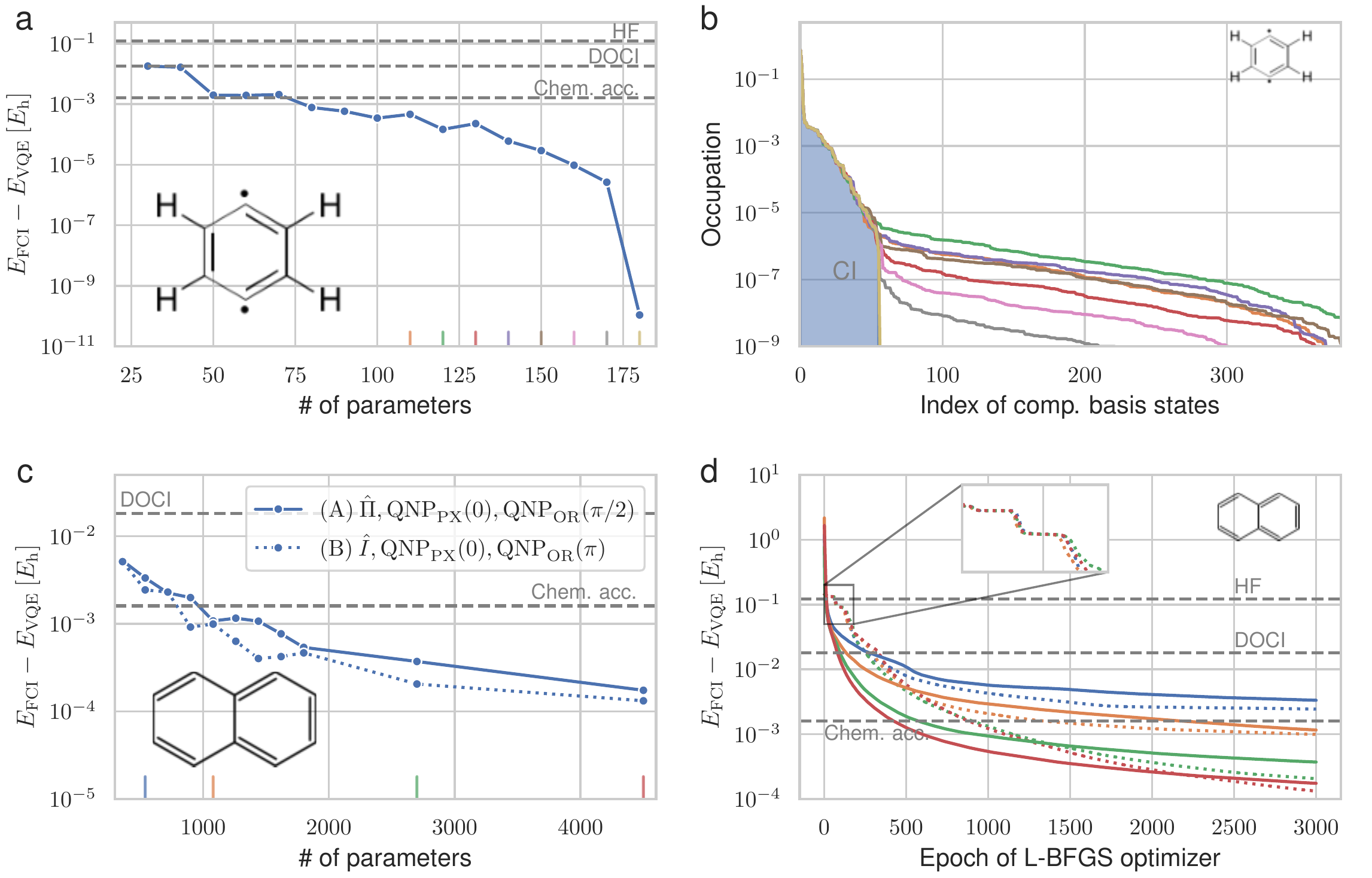}
\caption{Results of the discussed VQE fabric for representative molecular test
cases:
(a) Convergence of the VQE energy relative to the exact ground state energy
$E_\mathrm{FCI}$ of the $12$-spin-orbital active space of p-benzyne as a
function of the number of parameters in the fabric, 
(b) Occupation $\left|\braket{I|\Psi}\right|^2$ of each computational basis
state $\langle I|$ in the optimized VQE state $|\Psi\rangle$ at the
color-indicated parameter counts in (a). Blue area indicates the computational
basis states of the FCI ground state in the active space.
Each set of computational basis states is sorted in descending order, we show 
a figure with consistent ordering between sets in see Appendix~\ref{sec:additional numerics}, 
(c) Convergence of the VQE energy to the exact ground state energy
$E_\mathrm{FCI}$ in the 20-spin-orbital active space of naphthalene as a
function of the number of parameters an for the two different initialization schemes,
(d) Convergence under the L-BFGS optimizer for the color coded parameter
counts indicated in (c), dotted (solid) convergence lines correspond to a data
point from the dotted (solid) curve in (c), Inset highlights the plateaus encountered
with initialization method B during the first 180 epochs.}
\label{fig:experiments}
\end{figure*}

Our final VQE circuit starts with the preparation of an uncorrelated product state by
applying local Pauli $\hat X$ gates to appropriate qubits of an all-zero state depending on the
number of alpha and beta electrons.
The qubits are chosen such that for all parameters equal to zero in the following fabric
the state is transformed to the state with the energetically lowest orbitals occupied.
We then consider two parameter initialization strategies:
(A) The fabric is initialized with all $\theta = 0$ and all $\varphi = \pi/2$ and
$\hat \Pi = \mathrm{QNP_{OR}} (\pi)$ (solid lines in \ref{fig:experiments})
(B) The fabric is initialized with all $\theta = 0$ and all $\varphi = \pi$ and
$\hat \Pi = \hat I$ (dotted lines in \ref{fig:experiments}).

In both cases we optimize the VQE ground state energy with respect to the
VQE gate fabric parameters via L-BFGS. As the purpose of this study is
to explore the expressive power of this gate fabric, we consider neither shot
noise nor decoherence. This restriction permits the use of
analytical expressions for the Hamiltonian expectation values and VQE parameter
gradients thereof, greatly accelerating the classical statevector simulation of
the VQE.

Figure~\ref{fig:experiments} shows the salient results of this study. For the
case of p-benzyne, Figure~\ref{fig:experiments}a shows the VQE ground state
energy vs. full configuration interaction (FCI) with respect to the depth of the
gate fabric. The first notable point is that the fabric is able to
provide higher accuracy than either Hartree-Fock (HF) (a fabric of
$\mathrm{QNP_{OR}}$ gates - redundant here due to the use of Hartree-Fock
orbitals in the active space) or doubly-occupied configuration interaction
(DOCI) (a fabric of $\mathrm{QNP_{PX}}$ gates - equivalent to the pUCCD ansatz
from \cite{elfving2020simulating}). Focusing on the early convergence
behavior on the left side of the plot, even with only a few layers of the VQE
gate fabric, e.g. $\sim 50-80$ parameters, absolute accuracy of $1$ kcal
mol$^{-1}$
is achieved, which is commonly referred to as chemical accuracy. As the gate fabric depth is increased, roughly
geometric (exponential) convergence of the absolute energy is achieved, modulo some minor
aberrations due to difficulties in tightly converging the L-BFGS-based numerical
optimizations of the VQE gate fabric parameters. Focusing on the later
convergence behavior on the right side of the plot, as the number of parameters
in the VQE gate fabric approaches the number of parameters in the FCI problem
(note that in this irrep there are $175$ configuration state functions (CSFs), see Appendix \ref{appendix:two_mode_fermions}.3), the error convergence turns
sharply downward. At 180 parameters we are able to achieve very tight convergence
to errors of $\sim 10^{-10}$ $E_{\mathrm{h}}$ relative to FCI, numerically
indicating the onset of universality. Figure~\ref{fig:experiments}b shows the
sorted power spectra of the computational basis state (determinant) amplitudes
of the various VQE gate fabrics and the FCI state. The exact zeros in
the FCI state amplitudes are an artifact of the $D_{\mathrm{2h}}$ spatial point
group symmetry of this molecule, which our VQE gate fabric was not optimized to
capture. Even for low VQE gate fabric circuit depths, we see that all
determinants are populated by nonzero amplitudes, with a compromise apparently
being made to allow for some nonzero error in all amplitudes to provide for the
best variational energy. As more layers are added to the gate fabrics, the
precision of the amplitude spectra increases, as indicated by, e.g., significant
attenuation of the symmetry-driven zero block of the amplitude spectrum.
The tail of amplitudes exactly zero in FCI is exactly extinguished in the VQE state only when numerical
universality is achieved at a 180-parameter VQE gate fabric. This behavior is
reminiscent of the nonzero but structured tensor factorized representation of
the determinant amplitudes in coupled cluster theories, where here the tensor
structure is provided by the local quantum gate fabric.

Moving to the larger test case of naphthalene, Figure~\ref{fig:experiments}c
tells a similar story as the corresponding plot for
p-benzyne. Here we see similar and roughly geometric convergence of energy error
vs. VQE gate fabric depth and parameter count, albeit with a smaller prefactor. As with p-benzyne,
the VQE gate fabric rather quickly outstrips both the HF and DOCI ans\"atze, which
its primitive gates are constituted from, and achieves chemical accuracy of
$\sim 1$ kcal mol$^{-1}$ in absolute energy at just $\sim 800-1000$ parameters (there
are $19404$ CSFs in this irrep, so universality is not reached for any of the depth explored here).
Figure~\ref{fig:experiments}d considers
convergence of the energy with respect to the L-BFGS epoch for a number of different
VQE gate fabric depths.
A first key finding is that making the gate fabric deeper 
decreases the epoch count needed to converge to chemical accuracy.
A second key insight is that, while initialization strategy (B) 
has shallower circuits and ultimately achieves lower energy error at very high
epoch count, plateaus are visible during the optimization with the L-BFGS
optimizer (Figure~\ref{fig:experiments}).  Strategy (A), which exchanges
orbitals by means of the non trivial choice $\hat \Pi = \mathrm{QNP_{OR}}
(\pi)$, appears to circumvent the plateaus entirely and for deeper circuits
speeds up (the power-law like) convergence to below chemical accuracy.

The fabric presented here has favorable properties for implementation on NISQ hardware:
The $12$ qubit ansatz at $110$ parameters is without (with) $\hat \Pi$ gates
decomposable into elementary gates (2-qubit controlled Pauli and 1-qubit gates)
with resulting depth of $507$ ($617$).  The $20$ qubit ansatz at $1080$
parameters without (with) $\hat \Pi$ gates has depth $2761$ ($3361$) in such
decomposition.  To put this into perspective, a single trotter step of a naive
UCCSD circuit has gate depth $\approx 6600$ ($12$ qubits), respectively $\approx
57600$ ($20$ qubits).  Another considerable advantage is that only $N-2$ unique
$4$-qubit gates have to be calibrated on hardware as the structure is repetitive
after the first two layers.

\section{Comparison with other entangler circuits}
Having numerically demonstrated the features of the VQE gate fabric, it is worth
considering the relationship of this gate fabric to other proposed VQE entangler
circuits. There has been substantial prior work along these lines in the past
few years. 

For one instance, the hardware efficient ansatz \cite{Kandala2017,Bian2019} is
manifestly a local gate fabric, using essentially $\mathcal{SU}(4)$ entangler
elements or subsets thereof from the native gate set of the underlying
quantum circuit architecture. However, this gate fabric does not respect the
particle or spin quantum number symmetries, and therefore is likely to encounter
substantial difficulties in locating low-lying states within a target quantum
number irrep, particularly in larger active spaces. 

In another direction, there are myriad proposed entangler circuit
constructions which are either already explicitly or in principle could be
adapted to real amplitudes and strict commutation with the number and/or
spin-squared symmetry operators, but which are either nonlocal circuits or
composed of heterogeneous gate layers. For instance, UCCSD
\cite{peruzzo2014variational,PhysRevX.6.031007,Ryabinkin2018}, (here referring
to the Trotterized version thereof) and its sparse and/or low-rank cousins
$k$-UpCCGSD \cite{lee2018generalized,huggins2020non}, ADAPT-VQE \cite{Grimsley2019}, and Jastrow-Factor VQE
\cite{matsuzawa2020jastrow} all may have the power to achieve
universality at sufficient depth, e.g., as proved in a recent analysis of
distangled UCC \cite{evangelista2019exact} and have been either partially or
completely symmetrized already. However, as written, all of these ans\"atze
require nonlocal gate elements that, e.g., mediate excitations among non-adjacent 
spin orbitals in UCCSD, and thus are not gate fabrics. Moreover, many of these
constructions involve heterogeneous gate layers. For a canonical instance,
Jastrow-factor VQE \cite{matsuzawa2020jastrow} involves alternating circuit
layers of orbital rotations and substitutions (with the last of these being
nonlocal and complex-valued in the usual formulation). Of
all methods discussed in the prior literature, $k$-UpCCGSD is likely closest
to our proposed method, with products of single and diagonal double substitution
operators comprising the method. $k$-UpCCGSD as described in \cite{lee2018generalized} involves
nonlocal pair substitutions and therefore does not yield a local gate fabric.
Note however that the generalized swap network reformulation of $k$-UpCCGSD
described in and around Equation 12 and Figure 7 of \cite{o2019generalized}
(noticed after the first version of this manuscript was posted) appears to
realize $k$-UpCCGSD by means of a local circuit composed of 4-qubit gates that is
essentially a gate fabric.

Yet another interesting direction to consider is previously proposed true gate
fabrics that preserve quantum number symmetry, but do not achieve
universality. Orbital rotation fabrics \cite{
reck1994experimental,
wecker2015solving,
google2020hartree}, 
(i.e., Hartree-Fock) are clearly one example here, but so too is doubly
occupied configuration interaction (DOCI), for which a gate fabric was developed
with the pUCCD ansatz \cite{elfving2020simulating}. Both of these ans\"atze have the
interesting property that they can be mapped into gate fabrics requiring only
$M$ qubits, but both fail to reach FCI universality as the parameter depth is
increased.

Another interesting gate fabric construction is the ``gate-efficient ansatz''
presented in \cite{ganzhorn2019gate}, which presents as a gate fabric that
preserves total particle number $\hat N_{\alpha} + N_{\beta}$, but does not
appear to respect high-spin particle number $\hat N_{\alpha} - \hat N_{\beta}$
or $\hat S^2$ symmetry. 
Yet another interesting entangler is the ``qubit coupled cluster'' approach
presented in \cite{xia2020qubit}, which essentially implements a partial spin
adaption of UCCSD to preserve $\hat N_{\alpha}$ and $\hat N_{\beta}$ symmetry
within the single and double excitation operations, but neither preserves $\hat
S^2$ symmetry nor attains the structure of a local gate fabric.
Another related approach presented in \cite{yordanov2020efficient} constructs
fermion-adapted excitation operators which preserve particle number symmetry
but not spin symmetry, and additionally are aimed at optimizing the number of
CNOT gates in non-gate-fabric UCCSD methods.
An entangler that has the potential to preserve all quantum number symmetries
with additional spin-adaption work is the QAOA-inspired Pauli-term approach,
presented in \cite{salis2019short}, but this approach yields highly nonlocal
circuits which do not resemble gate fabrics.
Another approach which has some intersection with the present work is the
correlating antisymmetric geminal power (AGP) approach explored in
\cite{khamoshi2020correlating}, which first implements classically-tractable
APG to provide state preparation in quantum circuits and then augments AGP with
an anti-Hermitian pair hopping entangler which resembles our pair exchange gate.
However, the correlating AGP is not written in the form of a local gate fabric.

Another interesting direction to explore that we propose here is alternative local gate fabrics that
fully preserve quantum number symmetry, but which exhibit different gate
constructions than the $\mathrm{QNP_{PX}}$ and $\mathrm{QNP_{OR}}$ gates used in
Figure~\ref{fig:F}. Examples of such gate fabrics using generic 4-qubit 5-parameter
FCI gates and decompositions of these gates into $\mathrm{QNP_{PX}}$,
$\mathrm{QNP_{OR}}$, 1-hole/particle substitution, and pair-break up/down gates
are described in Appendix~\ref{appendix:other_qnp_gate_fabrics}. 

One additional interesting direction is the ``symmetry preserving state
preparation circuits'' of \cite{gard2020efficient}. This work primarily focuses
on total number symmetry, but does introduce a four-qubit gate that preserves
particle and spin quantum numbers via a hyperspherical parametrization. 

Note that an alternative approach to the exact
symmetry preservation explored in this manuscript is symmetry projection
\cite{tsuchimochi2020spin,lacroix2020symmetry}, which often requires ancilla
qubits and extra measurements due to the necessarily non-unitary nature of the
projection operation.

\section{Summary and Outlook} 

In this work, we set out to construct doppelg\"angers of the well known gate
fabric (i.e., a potentially infinitely repeatable, simple and geometrically
local pattern of gate elements that span the parent group at sufficient depth)
for the unrestricted qubit Hilbert space 
$\mathcal{SU}(2^{N})$
consisting of simple 2-qubit gate elements $SU(4)$.
Our major result is the construction of a gate fabric for the
important special case of spin-1/2 fermionic systems in the Jordan-Wigner representation
$\mathcal{F} (2^{2M})$
consisting of simple 4-qubit gate elements $\hat Q$.
Each 2-parameter $\hat Q$ gate comprises a
1-parameter spatial orbital rotation gate $\mathrm{QNP_{OR}} (\varphi)$
and a 1-parameter diagonal pair exchange gate $\mathrm{QNP_{PX}}
(\theta)$.
A fabric made of either of these gate elements alone does not achieve FCI
universality with sufficient parameter depth, but our VQE gate fabric, being an amalgamation
of the two appears to be able to do so.
Moreover, at intermediate depths, the VQE gate fabric appears to be
pragmatically expressive as evidenced by tests of the ground state energy
convergence in strongly correlated molecular systems. 
It is worth emphasizing that these properties seem to hold in the vast bulk of
quantum number irreps, i.e., that these fabric circuits can be applied for cases
where $S \neq 0$ (including even or odd spin cases) and/or where $N_{\alpha}
\neq N_{\beta}$ (see Appendix~\ref{sec:additional numerics} for details on
specific high-spin edge cases that are not universal with the $\hat Q$-type QNP
gate fabrics of the main text, but that can be addressed with elements of the
$\hat F$-type QNP gate fabrics of
Appendix~\ref{appendix:other_qnp_gate_fabrics}).
Many important questions remain
regarding our quantum number preserving gate fabrics. These include:
(1) How does the numerical optimization of parameters for such gate fabrics behave in the
presence of shot and/or decoherence noise?
(2) How can numerical optimization algorithms be adapted to exploit the
knowledge that the VQE entangler circuit is a gate fabric?
(3) Is the fixed $\hat \Pi$ gate construction or an extension thereof an
effective way to mitigate barren plateaus during numerical optimization?
(4) How does the VQE gate fabric perform for relative properties, for properties
at different nuclear geometries, and for properties in different quantum number
irreps?
(5) Is the construction of the VQE gate fabric in terms of $\hat Q$ gates
optimal, or do more elaborate constructions, e.g., using the $\hat F$ gates of
Appendix \ref{appendix:other_qnp_gate_fabrics} provide additional benefits?
(6) What is the scaling behavior of the error in absolute and/or relative
properties as a function of parameter depth for representative interesting
molecular systems?
(7) Can the gate fabric be adapted to additionally exploit external symmetries
such as spatial point group symmetries, e.g., as explored in
\cite{setia2020reducing}?  Taken together, the results of this work might
provide an interesting guide for the required symmetries and limiting
simplicities when constructing more elaborate VQE entanglers for fermionic
systems.

\textbf{Acknowledgements:} RMP is grateful to Dr.~Edward Hohenstein for many
discussions on the structure of the $\hat S^2$ operator. The authors further
thank Fotios Gkritsis for discussions.  QC Ware Corp.  acknowledges generous
research funding from Covestro Deutschland AG for this project. Covestro
acknowledges funding from the German Ministry for Education and Research (BMBF)
under the funding program quantum technologies as part of project HFAK
(13N15630).  DW acknowledges funding by the Deutsche Forschungsgemeinschaft
(DFG, German Research Foundation) under Germany’s Excellence Strategy Cluster of
Excellence Matter and Light for Quantum Computing (ML4Q) EXC2004/1 390534769.
 
\textbf{Conflict of Interest:} The VQE gate fabrics described in this work are
elements of two US provisional applications for patents both filed jointly
by QC Ware Corp. and Covestro Deutschland AG. RMP owns stock/options in QC Ware
Corp.

\bibliographystyle{unsrturl}
\bibliography{jrncodes,par_shift_note,cg,rmp}

\begin{thebibliography}{10}

\bibitem{peruzzo2014variational}
Alberto Peruzzo, Jarrod McClean, Peter Shadbolt, Man-Hong Yung, Xiao-Qi Zhou,
  Peter~J Love, Al{\'a}n Aspuru-Guzik, and Jeremy~L O’brien.
\newblock A variational eigenvalue solver on a photonic quantum processor.
\newblock {\em Nature communications}, 5(1):1--7, 2014.
\newblock \href {http://arxiv.org/abs/1304.3061} {\path{arXiv:1304.3061}}.

\bibitem{mcclean2016theory}
Jarrod~R McClean, Jonathan Romero, Ryan Babbush, and Al{\'a}n Aspuru-Guzik.
\newblock The theory of variational hybrid quantum-classical algorithms.
\newblock {\em New Journal of Physics}, 18(2):023023, 2016.
\newblock URL: \url{https://doi.org/10.1088/1367-2630/18/2/023023}, \href
  {http://dx.doi.org/10.1088/1367-2630/18/2/023023}
  {\path{doi:10.1088/1367-2630/18/2/023023}}.

\bibitem{preskill2018quantum}
John Preskill.
\newblock Quantum computing in the nisq era and beyond.
\newblock {\em Quantum}, 2:79, 2018.
\newblock URL: \url{https://doi.org/10.22331/q-2018-08-06-79}, \href
  {http://dx.doi.org/10.22331/q-2018-08-06-79}
  {\path{doi:10.22331/q-2018-08-06-79}}.

\bibitem{mcclean2017hybrid}
Jarrod~R McClean, Mollie~E Kimchi-Schwartz, Jonathan Carter, and Wibe~A
  De~Jong.
\newblock Hybrid quantum-classical hierarchy for mitigation of decoherence and
  determination of excited states.
\newblock {\em Physical Review A}, 95(4):042308, 2017.
\newblock URL: \url{https://doi.org/10.1103/PhysRevA.95.042308}, \href
  {http://dx.doi.org/10.1103/PhysRevA.95.042308}
  {\path{doi:10.1103/PhysRevA.95.042308}}.

\bibitem{parrish2019quantum}
Robert~M Parrish, Edward~G Hohenstein, Peter~L McMahon, and Todd~J
  Mart{\'\i}nez.
\newblock Quantum computation of electronic transitions using a variational
  quantum eigensolver.
\newblock {\em Physical review letters}, 122(23):230401, 2019.
\newblock URL: \url{https://doi.org/10.1103/PhysRevLett.122.230401}, \href
  {http://dx.doi.org/10.1103/PhysRevLett.122.230401}
  {\path{doi:10.1103/PhysRevLett.122.230401}}.

\bibitem{nakanishi2019subspace}
Ken~M Nakanishi, Kosuke Mitarai, and Keisuke Fujii.
\newblock Subspace-search variational quantum eigensolver for excited states.
\newblock {\em Physical Review Research}, 1(3):033062, 2019.
\newblock URL: \url{https://doi.org/10.1103/PhysRevResearch.1.033062}, \href
  {http://dx.doi.org/10.1103/PhysRevResearch.1.033062}
  {\path{doi:10.1103/PhysRevResearch.1.033062}}.

\bibitem{urbanek2020chemistry}
Miroslav Urbanek, Daan Camps, Roel Van~Beeumen, and Wibe~A de~Jong.
\newblock Chemistry on quantum computers with virtual quantum subspace
  expansion.
\newblock {\em Journal of Chemical Theory and Computation}, 16(9):5425--5431,
  2020.
\newblock URL: \url{https://doi.org/10.1021/acs.jctc.0c00447}, \href
  {http://dx.doi.org/10.1021/acs.jctc.0c00447}
  {\path{doi:10.1021/acs.jctc.0c00447}}.

\bibitem{ollitrault2020quantum}
Pauline~J Ollitrault, Abhinav Kandala, Chun-Fu Chen, Panagiotis~Kl Barkoutsos,
  Antonio Mezzacapo, Marco Pistoia, Sarah Sheldon, Stefan Woerner, Jay~M
  Gambetta, and Ivano Tavernelli.
\newblock Quantum equation of motion for computing molecular excitation
  energies on a noisy quantum processor.
\newblock {\em Physical Review Research}, 2(4):043140, 2020.
\newblock URL: \url{https://doi.org/10.1103/PhysRevResearch.2.043140}, \href
  {http://dx.doi.org/10.1103/PhysRevResearch.2.043140}
  {\path{doi:10.1103/PhysRevResearch.2.043140}}.

\bibitem{huggins2020non}
William~J Huggins, Joonho Lee, Unpil Baek, Bryan O’Gorman, and K~Birgitta
  Whaley.
\newblock A non-orthogonal variational quantum eigensolver.
\newblock {\em New Journal of Physics}, 22(7):073009, 2020.
\newblock \href {http://arxiv.org/abs/1909.09114} {\path{arXiv:1909.09114}}.

\bibitem{parrish2019qfd}
Robert~M Parrish and Peter~L McMahon.
\newblock Quantum filter diagonalization: Quantum eigendecomposition without
  full quantum phase estimation.
\newblock {\em arXiv preprint arXiv:1909.08925}, 2019.
\newblock URL: \url{https://arxiv.org/abs/1909.08925}.

\bibitem{stair2020multireference}
Nicholas~H Stair, Renke Huang, and Francesco~A Evangelista.
\newblock A multireference quantum krylov algorithm for strongly correlated
  electrons.
\newblock {\em Journal of chemical theory and computation}, 16(4):2236--2245,
  2020.
\newblock URL: \url{https://doi.org/10.1021/acs.jctc.9b01125}, \href
  {http://dx.doi.org/10.1021/acs.jctc.9b01125}
  {\path{doi:10.1021/acs.jctc.9b01125}}.

\bibitem{mcclean2018barren}
Jarrod~R McClean, Sergio Boixo, Vadim~N Smelyanskiy, Ryan Babbush, and Hartmut
  Neven.
\newblock Barren plateaus in quantum neural network training landscapes.
\newblock {\em Nature communications}, 9(1):1--6, 2018.
\newblock URL: \url{https://doi.org/10.1038/s41467-018-07090-4}, \href
  {http://dx.doi.org/10.1038/s41467-018-07090-4}
  {\path{doi:10.1038/s41467-018-07090-4}}.

\bibitem{gard2020efficient}
Bryan~T Gard, Linghua Zhu, George~S Barron, Nicholas~J Mayhall, Sophia~E
  Economou, and Edwin Barnes.
\newblock Efficient symmetry-preserving state preparation circuits for the
  variational quantum eigensolver algorithm.
\newblock {\em npj Quantum Information}, 6(1):1--9, 2020.
\newblock \href {http://arxiv.org/abs/1904.10910} {\path{arXiv:1904.10910}}.

\bibitem{PhysRevX.6.031007}
P.~J.~J. O'Malley, R.~Babbush, I.~D. Kivlichan, J.~Romero, J.~R. McClean,
  R.~Barends, J.~Kelly, P.~Roushan, A.~Tranter, N.~Ding, B.~Campbell, Y.~Chen,
  Z.~Chen, B.~Chiaro, A.~Dunsworth, A.~G. Fowler, E.~Jeffrey, E.~Lucero,
  A.~Megrant, J.~Y. Mutus, M.~Neeley, C.~Neill, C.~Quintana, D.~Sank,
  A.~Vainsencher, J.~Wenner, T.~C. White, P.~V. Coveney, P.~J. Love, H.~Neven,
  A.~Aspuru-Guzik, and J.~M. Martinis.
\newblock Scalable quantum simulation of molecular energies.
\newblock {\em Phys. Rev. X}, 6:031007, Jul 2016.
\newblock URL: \url{https://link.aps.org/doi/10.1103/PhysRevX.6.031007}, \href
  {http://dx.doi.org/10.1103/PhysRevX.6.031007}
  {\path{doi:10.1103/PhysRevX.6.031007}}.

\bibitem{Ryabinkin2018}
Ilya~G. Ryabinkin, Tzu-Ching Yen, Scott~N. Genin, and Artur~F. Izmaylov.
\newblock Qubit coupled cluster method: A systematic approach to quantum
  chemistry on a quantum computer.
\newblock {\em Journal of Chemical Theory and Computation}, 14(12):6317--6326,
  November 2018.
\newblock URL: \url{https://doi.org/10.1021/acs.jctc.8b00932}, \href
  {http://dx.doi.org/10.1021/acs.jctc.8b00932}
  {\path{doi:10.1021/acs.jctc.8b00932}}.

\bibitem{lee2018generalized}
Joonho Lee, William~J Huggins, Martin Head-Gordon, and K~Birgitta Whaley.
\newblock Generalized unitary coupled cluster wave functions for quantum
  computation.
\newblock {\em Journal of chemical theory and computation}, 15(1):311--324,
  2018.
\newblock URL: \url{https://pubs.acs.org/doi/abs/10.1021/acs.jctc.8b01004}.

\bibitem{o2019generalized}
Bryan O'Gorman, William~J Huggins, Eleanor~G Rieffel, and K~Birgitta Whaley.
\newblock Generalized swap networks for near-term quantum computing.
\newblock {\em arXiv preprint arXiv:1905.05118}, 2019.
\newblock URL: \url{https://arxiv.org/abs/1905.05118}.

\bibitem{matsuzawa2020jastrow}
Yuta Matsuzawa and Yuki Kurashige.
\newblock Jastrow-type decomposition in quantum chemistry for low-depth quantum
  circuits.
\newblock {\em Journal of chemical theory and computation}, 16(2):944--952,
  2020.
\newblock URL: \url{https://doi.org/10.1021/acs.jctc.9b00963}, \href
  {http://dx.doi.org/10.1021/acs.jctc.9b00963}
  {\path{doi:10.1021/acs.jctc.9b00963}}.

\bibitem{Kandala2017}
Abhinav Kandala, Antonio Mezzacapo, Kristan Temme, Maika Takita, Markus Brink,
  Jerry~M. Chow, and Jay~M. Gambetta.
\newblock Hardware-efficient variational quantum eigensolver for small
  molecules and quantum magnets.
\newblock {\em Nature}, 549(7671):242--246, September 2017.
\newblock \href {http://dx.doi.org/10.1038/nature23879}
  {\path{doi:10.1038/nature23879}}.

\bibitem{Bian2019}
Teng Bian, Daniel Murphy, Rongxin Xia, Ammar Daskin, and Sabre Kais.
\newblock Quantum computing methods for electronic states of the water
  molecule.
\newblock {\em Molecular Physics}, 117(15-16):2069--2082, February 2019.
\newblock URL: \url{https://doi.org/10.1080/00268976.2019.1580392}, \href
  {http://dx.doi.org/10.1080/00268976.2019.1580392}
  {\path{doi:10.1080/00268976.2019.1580392}}.

\bibitem{Grimsley2019}
Harper~R. Grimsley, Sophia~E. Economou, Edwin Barnes, and Nicholas~J. Mayhall.
\newblock An adaptive variational algorithm for exact molecular simulations on
  a quantum computer.
\newblock {\em Nature Communications}, 10(1), July 2019.
\newblock URL: \url{https://doi.org/10.1038/s41467-019-10988-2}, \href
  {http://dx.doi.org/10.1038/s41467-019-10988-2}
  {\path{doi:10.1038/s41467-019-10988-2}}.

\bibitem{elfving2020simulating}
Vincent~E Elfving, Marta Millaruelo, Jos{\'e}~A G{\'a}mez, and Christian
  Gogolin.
\newblock Simulating quantum chemistry in the seniority-zero space on
  qubit-based quantum computers.
\newblock {\em arXiv preprint arXiv:2002.00035}, 2020.
\newblock \href {http://arxiv.org/abs/2002.00035} {\path{arXiv:2002.00035}}.

\bibitem{evangelista2019exact}
Francesco~A Evangelista, Garnet Kin-Lic Chan, and Gustavo~E Scuseria.
\newblock Exact parameterization of fermionic wave functions via unitary
  coupled cluster theory.
\newblock {\em The Journal of chemical physics}, 151(24):244112, 2019.
\newblock URL: \url{https://doi.org/10.1063/1.5133059}, \href
  {http://dx.doi.org/10.1063/1.5133059} {\path{doi:10.1063/1.5133059}}.

\bibitem{ganzhorn2019gate}
Marc Ganzhorn, Daniel~J Egger, Panagiotis Barkoutsos, Pauline Ollitrault, Gian
  Salis, Nikolaj Moll, M~Roth, A~Fuhrer, P~Mueller, Stefan Woerner, et~al.
\newblock Gate-efficient simulation of molecular eigenstates on a quantum
  computer.
\newblock {\em Physical Review Applied}, 11(4):044092, 2019.
\newblock URL: \url{https://doi.org/10.1103/PhysRevApplied.11.044092}, \href
  {http://dx.doi.org/10.1103/PhysRevApplied.11.044092}
  {\path{doi:10.1103/PhysRevApplied.11.044092}}.

\bibitem{xia2020qubit}
Rongxin Xia and Sabre Kais.
\newblock Qubit coupled cluster singles and doubles variational quantum
  eigensolver ansatz for electronic structure calculations.
\newblock {\em Quantum Science and Technology}, 6(1):015001, 2020.
\newblock URL: \url{https://www.doi.org/10.1088/2058-9565/abbc74}, \href
  {http://dx.doi.org/10.1088/2058-9565/abbc74}
  {\path{doi:10.1088/2058-9565/abbc74}}.

\bibitem{yordanov2020efficient}
Yordan~S Yordanov, David~RM Arvidsson-Shukur, and Crispin~HW Barnes.
\newblock Efficient quantum circuits for quantum computational chemistry.
\newblock {\em Physical Review A}, 102(6):062612, 2020.
\newblock URL: \url{https://doi.org/10.1103/PhysRevA.102.062612}, \href
  {http://dx.doi.org/10.1103/PhysRevA.102.062612}
  {\path{doi:10.1103/PhysRevA.102.062612}}.

\bibitem{salis2019short}
Gian Salis and Nikolaj Moll.
\newblock Short-depth trial-wavefunctions for the variational quantum
  eigensolver based on the problem hamiltonian.
\newblock {\em arXiv preprint arXiv:1908.09533}, 2019.
\newblock URL: \url{https://arxiv.org/abs/1908.09533}.

\bibitem{khamoshi2020correlating}
Armin Khamoshi, Francesco~A Evangelista, and Gustavo~E Scuseria.
\newblock Correlating agp on a quantum computer.
\newblock {\em Quantum Science and Technology}, 6(1):014004, 2020.
\newblock URL: \url{https://doi.org/10.1088/2058-9565/abc1bb}, \href
  {http://dx.doi.org/10.1088/2058-9565/abc1bb}
  {\path{doi:10.1088/2058-9565/abc1bb}}.

\bibitem{ntnote}
Note that in a field-free lab frame, only the total particle number $\langle
  \hat N_{\mathrm{T}} \equiv \hat N_{\alpha} + \hat N_{\beta} \rangle$ and the
  total spin $\langle \hat S^2 \rangle$ are distinguishable quantum numbers.
  The high-spin projection particle number $\langle N_{\Delta} \equiv
  N_{\alpha} - N_{\beta} \propto \hat S_{z} \rangle$ is only resolvable in
  presence of external fields. However, as $\langle N_{\alpha} \rangle$ and
  $\langle N_{\beta} \rangle$ are relatively easy to separately constrain, we
  will require that the entangler circuits respect all three quantum number
  symmetries $\langle \hat N_{\alpha} \rangle$, $\langle \hat N_{\beta}
  \rangle$, and $\langle \hat S^2 \rangle$ (and thereby also $\langle
  N_{\mathrm{T}} \rangle$ and $\langle N_{\Delta} \rangle$ ) throughout this
  manuscript.

\bibitem{brandao2016local}
Fernando~GSL Brandao, Aram~W Harrow, and Micha{\l} Horodecki.
\newblock Local random quantum circuits are approximate polynomial-designs.
\newblock {\em Communications in Mathematical Physics}, 346(2):397--434, 2016.
\newblock \href {http://arxiv.org/abs/1208.0692} {\path{arXiv:1208.0692}}.

\bibitem{oszmaniec2020epsilon}
Micha{\l} Oszmaniec, Adam Sawicki, and Micha{\l} Horodecki.
\newblock Epsilon-nets, unitary designs and random quantum circuits.
\newblock 2020.
\newblock \href {http://arxiv.org/abs/2007.10885} {\path{arXiv:2007.10885}}.

\bibitem{Mitarai_Fujii_18}
K.~Mitarai, M.~Negoro, M.~Kitagawa, and K.~Fujii.
\newblock Quantum circuit learning.
\newblock {\em Phys. Rev. A}, 98:032309, Sep 2018.
\newblock URL: \url{https://link.aps.org/doi/10.1103/PhysRevA.98.032309}, \href
  {http://dx.doi.org/10.1103/PhysRevA.98.032309}
  {\path{doi:10.1103/PhysRevA.98.032309}}.

\bibitem{farhi2018classification}
Edward Farhi and Hartmut Neven.
\newblock Classification with quantum neural networks on near term processors.
\newblock 2018.
\newblock \href {http://arxiv.org/abs/1802.06002} {\path{arXiv:1802.06002}}.

\bibitem{Schuld2019}
Maria Schuld, Ville Bergholm, Christian Gogolin, Josh Izaac, and Nathan
  Killoran.
\newblock Evaluating analytic gradients on quantum hardware.
\newblock {\em Physical Review A}, 99(3), March 2019.
\newblock URL: \url{https://doi.org/10.1103/physreva.99.032331}, \href
  {http://dx.doi.org/10.1103/physreva.99.032331}
  {\path{doi:10.1103/physreva.99.032331}}.

\bibitem{PennyLane}
Ville Bergholm, Josh Izaac, Maria Schuld, Christian Gogolin, M~Sohaib Alam,
  Shahnawaz Ahmed, Juan~Miguel Arrazola, Carsten Blank, Alain Delgado, Soran
  Jahangiri, et~al.
\newblock Pennylane: Automatic differentiation of hybrid quantum-classical
  computations.
\newblock {\em arXiv preprint arXiv:1811.04968}, 2018.
\newblock URL: \url{https://arxiv.org/abs/1811.04968}.

\bibitem{Kottmann_Guzik_20}
Jakob~S. Kottmann, Abhinav Anand, and Alán Aspuru-Guzik.
\newblock A feasible approach for automatically differentiable unitary
  coupled-cluster on quantum computers.
\newblock {\em Chem. Sci.}, pages~--, 2021.
\newblock URL: \url{http://dx.doi.org/10.1039/D0SC06627C}, \href
  {http://dx.doi.org/10.1039/D0SC06627C} {\path{doi:10.1039/D0SC06627C}}.

\bibitem{Mari_Killoran_20}
Andrea Mari, Thomas~R. Bromley, and Nathan Killoran.
\newblock Estimating the gradient and higher-order derivatives on quantum
  hardware.
\newblock {\em Phys. Rev. A}, 103:012405, Jan 2021.
\newblock URL: \url{https://link.aps.org/doi/10.1103/PhysRevA.103.012405},
  \href {http://dx.doi.org/10.1103/PhysRevA.103.012405}
  {\path{doi:10.1103/PhysRevA.103.012405}}.

\bibitem{keller2015selection}
Sebastian Keller, Katharina Boguslawski, Tomasz Janowski, Markus Reiher, and
  Peter Pulay.
\newblock Selection of active spaces for multiconfigurational wavefunctions.
\newblock {\em The Journal of chemical physics}, 142(24):244104, 2015.
\newblock URL: \url{https://doi.org/10.1063/1.4922352}, \href
  {http://dx.doi.org/10.1063/1.4922352} {\path{doi:10.1063/1.4922352}}.

\bibitem{reck1994experimental}
Michael Reck, Anton Zeilinger, Herbert~J Bernstein, and Philip Bertani.
\newblock Experimental realization of any discrete unitary operator.
\newblock {\em Physical review letters}, 73(1):58, 1994.
\newblock URL: \url{https://doi.org/10.1103/PhysRevLett.73.58}, \href
  {http://dx.doi.org/10.1103/PhysRevLett.73.58}
  {\path{doi:10.1103/PhysRevLett.73.58}}.

\bibitem{wecker2015solving}
Dave Wecker, Matthew~B Hastings, Nathan Wiebe, Bryan~K Clark, Chetan Nayak, and
  Matthias Troyer.
\newblock Solving strongly correlated electron models on a quantum computer.
\newblock {\em Physical Review A}, 92(6):062318, 2015.
\newblock URL: \url{https://doi.org/10.1103/PhysRevA.92.062318}, \href
  {http://dx.doi.org/https://doi.org/10.1103/PhysRevA.92.062318}
  {\path{doi:https://doi.org/10.1103/PhysRevA.92.062318}}.

\bibitem{google2020hartree}
{Google AI Quantum and Collaborators}.
\newblock Hartree-fock on a superconducting qubit quantum computer.
\newblock {\em Science}, 369(6507):1084--1089, 2020.
\newblock URL: \url{https://doi.org/10.1126/science.abb9811}, \href
  {http://dx.doi.org/10.1126/science.abb9811}
  {\path{doi:10.1126/science.abb9811}}.

\bibitem{tsuchimochi2020spin}
Takashi Tsuchimochi, Yuto Mori, and Seiichiro~L Ten-no.
\newblock Spin-projection for quantum computation: A low-depth approach to
  strong correlation.
\newblock {\em Physical Review Research}, 2(4):043142, 2020.
\newblock URL: \url{https://doi.org/10.1103/PhysRevResearch.2.043142}, \href
  {http://dx.doi.org/10.1103/PhysRevResearch.2.043142}
  {\path{doi:10.1103/PhysRevResearch.2.043142}}.

\bibitem{lacroix2020symmetry}
Denis Lacroix.
\newblock Symmetry-assisted preparation of entangled many-body states on a
  quantum computer.
\newblock {\em Physical Review Letters}, 125(23):230502, 2020.
\newblock URL: \url{https://doi.org/10.1103/PhysRevLett.125.230502}, \href
  {http://dx.doi.org/10.1103/PhysRevLett.125.230502}
  {\path{doi:10.1103/PhysRevLett.125.230502}}.

\bibitem{setia2020reducing}
Kanav Setia, Richard Chen, Julia~E Rice, Antonio Mezzacapo, Marco Pistoia, and
  James~D Whitfield.
\newblock Reducing qubit requirements for quantum simulations using molecular
  point group symmetries.
\newblock {\em Journal of Chemical Theory and Computation}, 16(10):6091--6097,
  2020.
\newblock URL: \url{https://doi.org/10.1021/acs.jctc.0c00113}, \href
  {http://dx.doi.org/10.1021/acs.jctc.0c00113}
  {\path{doi:10.1021/acs.jctc.0c00113}}.

\bibitem{Li_Sun_17}
Jun Li, Xiaodong Yang, Xinhua Peng, and Chang-Pu Sun.
\newblock Hybrid quantum-classical approach to quantum optimal control.
\newblock {\em Phys. Rev. Lett.}, 118:150503, Apr 2017.
\newblock URL: \url{https://link.aps.org/doi/10.1103/PhysRevLett.118.150503},
  \href {http://dx.doi.org/10.1103/PhysRevLett.118.150503}
  {\path{doi:10.1103/PhysRevLett.118.150503}}.

\bibitem{Schuld_Killoran_19}
Maria Schuld, Ville Bergholm, Christian Gogolin, Josh Izaac, and Nathan
  Killoran.
\newblock Evaluating analytic gradients on quantum hardware.
\newblock {\em Phys. Rev. A}, 99:032331, Mar 2019.
\newblock URL: \url{https://link.aps.org/doi/10.1103/PhysRevA.99.032331}, \href
  {http://dx.doi.org/10.1103/PhysRevA.99.032331}
  {\path{doi:10.1103/PhysRevA.99.032331}}.

\bibitem{Banchi_Crooks_21}
Leonardo Banchi and Gavin~E. Crooks.
\newblock Measuring {A}nalytic {G}radients of {G}eneral {Q}uantum {E}volution
  with the {S}tochastic {P}arameter {S}hift {R}ule.
\newblock {\em {Quantum}}, 5:386, January 2021.
\newblock URL: \url{https://doi.org/10.22331/q-2021-01-25-386}, \href
  {http://dx.doi.org/10.22331/q-2021-01-25-386}
  {\path{doi:10.22331/q-2021-01-25-386}}.

\bibitem{Meyer_Eisert_20}
Johannes~Jakob Meyer, Johannes Borregaard, and Jens Eisert.
\newblock A variational toolbox for quantum multi-parameter estimation.
\newblock {\em arXiv preprint arXiv:2006.06303}, 2020.
\newblock URL: \url{https://arxiv.org/abs/2006.06303}.

\end{thebibliography}

\newpage
\clearpage
\appendix

\section{Symmetry-Constrained Subgroups of $\mathcal{SU}(2^{N})$}

\label{sec:subgroups}

This section discusses some of the technical hurdles encountered in developing
universal gate fabrics for certain subgroups of $\mathcal{SU}(2^N)$, using 
well-known literature results for restriction to real operators
$\mathcal{SO}(2^N)$ and further restriction to Hamming-weight-preserving
operators $\mathcal{H}(2^{N})$.

\begin{figure}[b]
\centering
\begin{equation*}
\label{eq:SO2N}
\begin{array}{l}
\Qcircuit @R=0.3em @C=0.3em @!R {
 & \multigate{1}{\mathit{SO}(4)}
 & \qw
 & \multigate{1}{\mathit{SO}(4)}
 & \qw
 & \qw \\
 & \ghost{\mathit{SO}(4)}
 & \multigate{1}{\mathit{SO}(4)}
 & \ghost{\mathit{SO}(4)}
 & \multigate{1}{\mathit{SO}(4)}
 & \qw \\
 & \multigate{1}{\mathit{SO}(4)}
 & \ghost{\mathit{SO}(4)}
 & \multigate{1}{\mathit{SO}(4)}
 & \ghost{\mathit{SO}(4)}
 & \qw \\
 & \ghost{\mathit{SO}(4)}
 & \multigate{1}{\mathit{SO}(4)}
 & \ghost{\mathit{SO}(4)}
 & \multigate{1}{\mathit{SO}(4)}
 & \qw \\
 & \multigate{1}{\mathit{SO}(4)}
 & \ghost{\mathit{SO}(4)}
 & \multigate{1}{\mathit{SO}(4)}
 & \ghost{\mathit{SO}(4)}
 & \qw \\
 & \ghost{\mathit{SO}(4)}
 & \qw
 & \ghost{\mathit{SO}(4)}
 & \qw
 & \qw \\
}
\end{array}
\ldots
\phantom{}
\cong
\phantom{}
\begin{array}{l}
\Qcircuit @R=0.3em @C=0.3em @!R {
 & \multigate{5}{\mathit{SO}(2^6)}
 & \qw \\
 & \ghost{\mathit{SO}(2^6)}
 & \qw \\
 & \ghost{\mathit{SO}(2^6)}
 & \qw \\
 & \ghost{\mathit{SO}(2^6)}
 & \qw \\
 & \ghost{\mathit{SO}(2^6)}
 & \qw \\
 & \ghost{\mathit{SO}(2^6)}
 & \qw \\
}
\end{array}
\end{equation*}
\begin{equation*}
\mathit{SO}(4)
\coloneqq
\exp(\hat X) : \hat X = -\hat X^\dagger, \hat X \in \mathbb{R}^{4} \times
\mathbb{R}^{4}
\end{equation*}
\begin{equation*}
=
\begin{array}{l}
\Qcircuit @R=0.3em @C=0.3em @!R {
 & \gate{R_{y} (\theta_1)}
 & \ctrl{1}
 & \gate{R_{y} (\theta_3)}
 & \ctrl{1}
 & \gate{R_{y} (\theta_5)}
 & \qw \\
 & \gate{R_{y} (\theta_2)}
 & \ctrl{-1}
 & \gate{R_{y} (\theta_4)}
 & \ctrl{-1}
 & \gate{R_{y} (\theta_6)}
 & \qw \\
}
\end{array}
\end{equation*}
\caption{Gate fabric universal for $\mathcal{SO}(2^N)$ (sketched for $N=6)$.
The gate fabric
is a 2-local-nearest-neighbor  tessellation of
alternating even and odd qubit-pair 6-parameter, 4-qubit $SO(4)$
gates. 
}
\label{fig:SO4}
\end{figure}
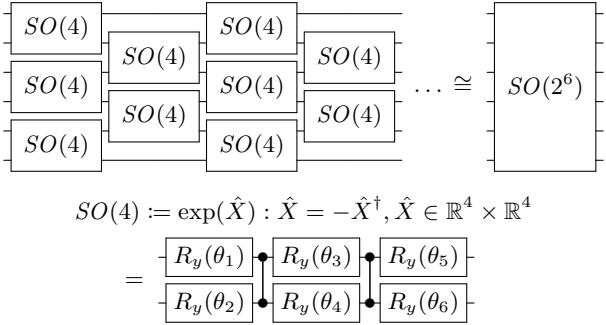

The imposition of specific symmetries which constrain the subgroup of
$\mathcal{SU}(2^{N})$ may or may not present considerable difficulties in
constructing gate fabrics of the type defined above. For an example
that does not introduce significant difficulty, consider the case where we
restrict the Hilbert space operators to have real value, i.e., a restriction to
$\mathcal{SO}(2^{N})$. In this case, one may simply
substitute $\mathcal{SU}(4) \rightarrow \mathcal{SO}(4)$ in the gate fabric of
Figure \ref{fig:SU4} to construct the desired gate fabric sketched in Figure
\ref{fig:SO4} for $\mathcal{SO}(2^{N})$. 

\begin{figure}[b]
\centering
\begin{equation*}
\begin{array}{l}
\Qcircuit @R=0.3em @C=0.3em @!R {
 & \multigate{1}{\mathit{H}(4)}
 & \qw
 & \multigate{1}{\mathit{H}(4)}
 & \qw
 & \qw \\
 & \ghost{\mathit{H}(4)}
 & \multigate{1}{\mathit{H}(4)}
 & \ghost{\mathit{H}(4)}
 & \multigate{1}{\mathit{H}(4)}
 & \qw \\
 & \multigate{1}{\mathit{H}(4)}
 & \ghost{\mathit{H}(4)}
 & \multigate{1}{\mathit{H}(4)}
 & \ghost{\mathit{H}(4)}
 & \qw \\
 & \ghost{\mathit{H}(4)}
 & \multigate{1}{\mathit{H}(4)}
 & \ghost{\mathit{H}(4)}
 & \multigate{1}{\mathit{H}(4)}
 & \qw \\
 & \multigate{1}{\mathit{H}(4)}
 & \ghost{\mathit{H}(4)}
 & \multigate{1}{\mathit{H}(4)}
 & \ghost{\mathit{H}(4)}
 & \qw \\
 & \ghost{\mathit{H}(4)}
 & \qw
 & \ghost{\mathit{H}(4)}
 & \qw
 & \qw \\
}
\end{array}
\ldots
\phantom{}
\not\cong
\phantom{}
\begin{array}{l}
\Qcircuit @R=0.3em @C=0.3em @!R {
 & \multigate{5}{\mathit{H}(2^6)}
 & \qw \\
 & \ghost{\mathit{H}(2^6)}
 & \qw \\
 & \ghost{\mathit{H}(2^6)}
 & \qw \\
 & \ghost{\mathit{H}(2^6)}
 & \qw \\
 & \ghost{\mathit{H}(2^6)}
 & \qw \\
 & \ghost{\mathit{H}(2^6)}
 & \qw \\
}
\end{array}
\end{equation*}
\begin{equation*}
\begin{array}{l}
\Qcircuit @R=0.3em @C=0.3em @!R {
 & \multigate{1}{H(4)}
 & \qw \\
 & \ghost{H(4)}
 & \qw \\
}
\end{array}
\coloneqq
\begin{array}{l}
\Qcircuit @R=0.3em @C=0.3em @!R {
 & \gate{R_{y} (+\pi / 4)}
 & \ctrl{1}
 & \gate{R_{y} (+\lambda / 2)}
 & \ctrl{1}
 & \gate{R_{y} (-\pi / 4)}
 & \qw \\
 & \gate{R_{y} (+\pi / 4)}
 & \ctrl{-1}
 & \gate{R_{y} (-\lambda/2)}
 & \ctrl{-1}
 & \gate{R_{y} (-\pi / 4)}
 & \qw \\
}
\end{array}
\end{equation*}
\begin{equation*}
=
\left [
\begin{array}{rrrr}
1 & & & \\
 & c & +s & \\
 & -s & c & \\
 & & & 1 \\
\end{array}
\right ]
:
\
\begin{array}{l}
c \coloneqq \cos(\lambda/2)
\\
s \coloneqq \sin(\lambda/2)
\\
\end{array}
\end{equation*}
\caption{Gate fabric attempt \emph{not} universal for the
Hamming-weight-preserving subgroup $\mathcal{H}(2^N)$ (sketched for $N=6)$.  The
gate fabric is a 2-local-nearest-neighbor tessellation of alternating even and
odd qubit-pair 1-parameter, 2-qubit Hamming-weight-preserving $\hat H(4)$ gates.
The gate fabric exactly commutes with the Hamming weight operator $\hat P \equiv
\sum_{p} (\hat I - \hat Z_p) / 2$, but the gate fabric does not span
$\mathcal{H}(2^N)$ for any depth.
}
\label{fig:H1}
\end{figure}
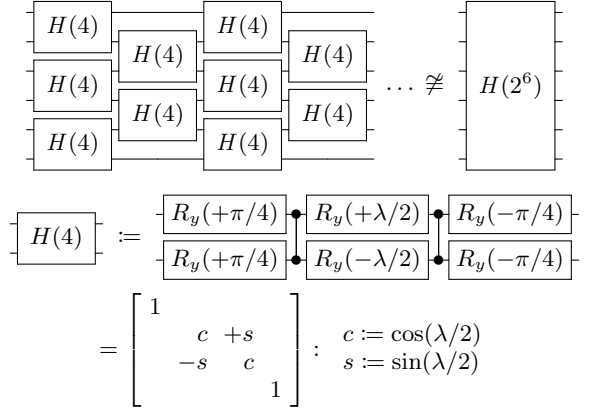

For an example that
does introduce significant difficulty, consider the case where we restrict
$\mathcal{SO}(2^{N})$ Hilbert space operators to preserve Hamming weight, i.e.,
to respect the commutation constraint $[\hat U, \hat P] = 0$ where $\hat P
\coloneqq \sum_{p} (\hat I - \hat Z_{p}) / 2$ is the Hamming weight or ``particle
counting'' operator. We denote this subgroup as $\mathcal{H}(2^N)$. Here, we
might be tempted to continue restricting the 2-qubit $\mathcal{SO}(4)$ gates to
preserve Hamming weight, mandating that we substitute $\mathcal{SO}(4)
\rightarrow \mathcal{H}(4)$, where $\mathcal{H}(4)$ implements a Givens rotation
between configurations $|01\rangle$ and $|10\rangle$ while acting as the
identity in $|00\rangle$ and $|11\rangle$. This is sketched in Figure
\ref{fig:H1}. However, a tessellation of 2-qubit Givens gates is not a gate
fabric for $\mathcal{H}(2^N)$, as it does not provide universality for this
subgroup. In fact, it can be shown that a tessellation of Givens gates amounts
to a one-particle rotation of the qubit creation and annihilation operators
$\hat p^{\pm} \coloneqq \sum_{q} V_{qp} \hat q^{\pm}$ for $V_{qp} \in
\mathcal{SO}(N)$ and $\hat q^{\pm} \coloneqq (\hat X_q \mp i \hat Y_q) /
2$, and thus after exactly $N$ layers and $N (N - 1) / 2$ gates the part of
Hilbert space reachable with the fabric does no longer increase anymore, and in fact 
the fabric is
classically simulable in polynomial time via techniques such as the match gate
formalism or direct implementation with classical photons and
beamsplitters. Note that $\mathcal{H}(2^N)$ has irreducible
representations of dimension up to $\binom{N}{\lfloor N / 2 \rfloor}$, so
failure to
reach universality can be shown by simply parameter counting. Speaking more practically,
this proposed gate fabric has very limited expressive power for most irreps of
$\mathcal{H}(2^{N})$, and does not provide a good approximation to most desired
actions within this space.

In fact, no gate fabric for $\mathcal{H} (2^N)$ is possible with 2-qubit gate
elements. One possible construction of a gate fabric for $\mathcal{H}(2^N)$ with
3-qubit gate elements is sketched in Figure \ref{fig:H2}. Note that this might
not be a minimal representation - we will see examples of fermionic systems
shortly where a much simpler gate element than the fully
explicitly universal $k$-minimal qubit gate provides a gate fabric.

\begin{figure}
\centering
\begin{equation*}
\label{eq:SO2N-Hamming}
\begin{array}{l}
\Qcircuit @R=0.3em @C=0.3em @!R {
 & \multigate{2}{\mathit{H}(8)}
 & \qw 
 & \qw 
 & \qw 
 & \qw \\
 & \ghost{\mathit{H}(8)}
 & \multigate{2}{\mathit{H}(8)}
 & \qw 
 & \qw \\
 & \ghost{\mathit{H}(8)}
 & \ghost{\mathit{H}(8)}
 & \multigate{2}{\mathit{H}(8)}
 & \qw \\
 & \multigate{2}{\mathit{H}(8)}
 & \ghost{\mathit{H}(8)}
 & \ghost{\mathit{H}(8)}
 & \qw \\
 & \ghost{\mathit{H}(8)}
 & \multigate{2}{\mathit{H}(8)}
 & \ghost{\mathit{H}(8)}
 & \qw \\
 & \ghost{\mathit{H}(8)}
 & \ghost{\mathit{H}(8)}
 & \multigate{2}{\mathit{H}(8)}
 & \qw \\
 & \multigate{2}{\mathit{H}(8)}
 & \ghost{\mathit{H}(8)}
 & \ghost{\mathit{H}(8)}
 & \qw \\
 & \ghost{\mathit{H}(8)}
 & \qw 
 & \ghost{\mathit{H}(8)}
 & \qw \\
 & \ghost{\mathit{H}(8)}
 & \qw 
 & \qw 
 & \qw \\
}
\end{array}
\phantom{}
\ldots
\cong
\phantom{}
\begin{array}{l}
\Qcircuit @R=0.3em @C=0.3em @!R {
 & \multigate{8}{\mathit{H}(2^9)}
 & \qw \\
 & \ghost{\mathit{H}(2^9)}
 & \qw \\
 & \ghost{\mathit{H}(2^9)}
 & \qw \\
 & \ghost{\mathit{H}(2^9)}
 & \qw \\
 & \ghost{\mathit{H}(2^9)}
 & \qw \\
 & \ghost{\mathit{H}(2^9)}
 & \qw \\
 & \ghost{\mathit{H}(2^9)}
 & \qw \\
 & \ghost{\mathit{H}(2^9)}
 & \qw \\
 & \ghost{\mathit{H}(2^9)}
 & \qw \\
}
\end{array}
\end{equation*}
\begin{equation*}
\mathit{H} (8)
\coloneqq
\left [
\begin{array}{rrrrrrrrr}
1 & & & & & & & & \\
 & \color{brickred}{u_{00}^{1}} & \color{brickred}{u_{01}^{1}} & &
\color{brickred}{u_{02}^{1}} & & & & \\
 & \color{brickred}{u_{10}^{1}} & \color{brickred}{u_{11}^{1}} & &
\color{brickred}{u_{12}^{1}} & & & & \\
 & & & \color{darkblue}{u_{00}^{2}} & & \color{darkblue}{u_{01}^{2}} &
\color{darkblue}{u_{02}^{2}} & \\
 & \color{brickred}{u_{20}^{1}} & \color{brickred}{u_{21}^{1}} & &
\color{brickred}{u_{22}^{1}} & & & & \\
 & & & \color{darkblue}{u_{10}^{2}} & & \color{darkblue}{u_{11}^{2}} &
\color{darkblue}{u_{12}^{2}} & \\
 & & & \color{darkblue}{u_{20}^{2}} & & \color{darkblue}{u_{21}^{2}} &
\color{darkblue}{u_{22}^{2}} & \\
& & & & & & & & 1 \\
\end{array}
\right ]
\end{equation*}
\begin{equation*}
\hat u^{d}
\coloneqq
\exp(\hat x^{d})
\in 
\mathcal{SO}(3)
:
\
\hat x^{d\dagger}
=
-
\hat x^{d}
:
\hat x^{d}
\in
\mathbb{R}^3
\times
\mathbb{R}^3
, 
\
d \in [1, 2]
\end{equation*}
\caption{Gate fabric  universal for the Hamming-weight-preserving subgroup
$\mathcal{H}(2^N)$ (sketched for $N=9)$.  The gate fabric is a
3-local-nearest-neighbor tessellation of cascading qubit-triple 6-parameter,
3-qubit Hamming-weight-preserving $\hat H(8)$ gates. Each $\hat H(8)$ gate is
composed of a 3-parameter $\mathcal{SO}(3)$ rotation in the $d$-Hamming-weight
subspace, where $d \in [1, 2]$ for a total of 6 parameters. The gate fabric exactly
commutes with the Hamming weight operator $\hat P \equiv \sum_{p} (\hat I - \hat
Z_p) / 2$ and spans $\mathcal{H}(2^N)$ at sufficient depth.
}
\label{fig:H2}
\end{figure}

\section{Additional Details on the Jordan-Wigner Mapping}
\label{appendix:jw_details}
This section is included to enumerate the expansion of same-spin number
operators and the $\hat S^2$ operator into Pauli operators in the Jordan-Wigner
mapping defined in the main text.

\subsection{Same-Spin Occupation and Substitution Operators}

\subsubsection{Same-Spin Occupation Number Operators}

The same-spin occupation number operators are
\begin{equation}
p^{\pm}
p^{\mp}
=
(
\hat I_{p}
\mp
\hat Z_{p}
) / 2 
\end{equation}
whereby $p^{+} p^{-}$ is a ``particle occupation number operator'' (counts $1$s). $p^{-}
p^{+}$ is a ``hole occupation number operator'' (counts $0$s).  Note that the
Jordan-Wigner strings cancel for these operators.

\subsubsection{Same-Spin Substitution Operators}

The same-spin one-particle substitution operator is
\begin{equation}
\left .
p^{\pm}
q^{\mp}
\right |_{p < q}
=
\pm
\hat P_{p}^{\pm}
\otimes
\hat Z_{p+1, q-1}^{\leftrightarrow}
\otimes
\hat P_{q}^{\mp}.
\end{equation}
Here $Z_{p+1,q-1}^{\leftrightarrow} \coloneqq \bigotimes_{r=p+1}^{r=q-1} \hat Z_{r}$.
For completeness
\begin{equation}
\left .
p^{\pm}
q^{\mp}
\right |_{p > q}
=
\pm
\hat P_{q}^{\mp}
\otimes
\hat Z_{q+1, p-1}^{\leftrightarrow}
\otimes
\hat P_{p}^{\pm}
\end{equation}
Technically, $p^{+} q^{-}$ is the ``one-particle substitution operator'' and
$p^{-} q^{+}$ is the ``one-hole substitution operator.''

With some algebra, one can show that,
\begin{equation}
\left .
p^{\pm}
q^{\mp}
+
q^{\pm}
p^{\mp}
\right |_{p < q}
\end{equation}
\[
=
\pm
\hat X_{p}
\otimes
\hat Z_{p+1, q-1}^{\leftrightarrow}
\otimes
\hat X_{q}
/ 2
\pm
\hat Y_{p}
\otimes
\hat Z_{p+1, q-1}^{\leftrightarrow}
\otimes
\hat Y_{q}
/ 2
\]
(The formula for $p > q$ is the same except for the indices on $Z_{p+1,
q-1}^{\leftrightarrow}$.)

Here, the Jordan-Wigner strings only cancel partially. However, the
$\alpha$-then-$\beta$ Jordan-Wigner ordering does provide the advantage that the
remaining $Z_{p,q}^{\leftrightarrow}$ strings are supported only on the intermediate
$\alpha$($\beta)$ spin orbital indices for $\alpha$($\beta$) substitution
operators.

\subsection{Quantum Number Operators}

\subsubsection{Alpha Number Operator}

The $\alpha$ number operator is
\begin{equation}
\hat N_{\alpha}
\coloneqq
\sum_{p}
p^{\dagger}
p
=
(M / 2) \hat I
-
\sum_{p} \hat Z_{p} /2.
\end{equation}
The eigenvalues of the $\alpha$ number operator are $N_{\alpha} \in [0, 1, \ldots
M]$ with degeneracy $\binom{M}{N_{\alpha}} 2^{M}$. The determinants are
eigenfunctions of the $\alpha$ number operator, with eigenvalues given by the
$\alpha$ population count,
\begin{equation}
\hat N_{\alpha} | \vec I \rangle
=
\mathrm{popcount} (\vec I_{\alpha} )
| \vec I \rangle
=
N_{\alpha} 
| \vec I \rangle
\end{equation}
and $\mathrm{popcount}(\vec I_{\alpha})$ counts the number of ones in $\vec I_{\alpha}$. 

\subsubsection{Beta Number Operator}

The $\beta$ number operator is,
\begin{equation}
\hat N_{\beta}
\coloneqq
\sum_{p}
\bar p^{\dagger}
\bar p
=
(M / 2) \hat I
-
\sum_{\bar p} \hat Z_{\bar p} /2
\end{equation}
The eigenvalues of the $\beta$ number operator are $N_{\beta} \in [0, 1, \ldots
M]$ with degeneracy $\binom{M}{N_{\beta}} 2^{M}$. The dets are
eigenfunctions of the $\beta$ number operator, with eigenvalues given by the
$\beta$ population count,
\begin{equation}
\hat N_{\beta} | \vec I \rangle
=
\mathrm{popcount} (\vec I_{\beta} )
| \vec I \rangle
=
N_{\beta} 
| \vec I \rangle
\end{equation}

\subsubsection{Total Spin Squared Operator}

The total spin squared operator is,
\begin{equation}
\hat S^2 = 
\hat S_{-}
\hat S_{+}
+
\hat S_{z}
+
\hat S_{z}^2
\end{equation}
with the spin lowering operator,
\begin{equation}
\hat S_{-}
\coloneqq
\sum_{p}
\bar p^{\dagger}
p
\end{equation}
the spin raising operator,
\begin{equation}
\hat S_{+}
\coloneqq
\sum_{p}
p^{\dagger}
\bar p
\end{equation}
and the $z$-spin,
\begin{equation}
\hat S_{z}
\coloneqq
\frac{1}{2}
\left [
\hat N_{\alpha}
-
\hat N_{\beta}
\right ].
\end{equation}
After some algebra, under the chosen Jordan Wigner mapping, this resolves to
\begin{equation}
\hat S^2 = 
\end{equation}
\[
=
\frac{3 M}{8}
\hat I
-
\frac{3}{8}
\sum_{p}
\hat Z_{p}
\otimes
\hat Z_{\bar p} 
\]
\[
+
\sum_{p < q}
\frac{1}{8}
\hat Z_{p}
\otimes
\hat Z_{q}
+
\sum_{p < q}
\frac{1}{8}
\hat Z_{\bar p}
\otimes
\hat Z_{\bar q}
\]
\[
-
\sum_{p < q}
\frac{1}{8}
\hat Z_{p}
\otimes
\hat Z_{\bar q}
-
\sum_{p < q}
\frac{1}{8}
\hat Z_{\bar p}
\otimes
\hat Z_{q}
\]
\[
-
\frac{1}{8}
\sum_{p<q}
\hat X_{p}
\otimes
\hat Z_{p+1,q-1}^{\leftrightarrow}
\otimes
\hat X_{q}
\otimes
\hat X_{\bar p}
\otimes
\hat Z_{\bar p+1,\bar q-1}^{\leftrightarrow}
\otimes
\hat X_{\bar q}
\]
\[
-
\frac{1}{8}
\sum_{p<q}
\hat X_{p}
\otimes
\hat Z_{p+1,q-1}^{\leftrightarrow}
\otimes
\hat X_{q}
\otimes
\hat Y_{\bar p}
\otimes
\hat Z_{\bar p+1,\bar q-1}^{\leftrightarrow}
\otimes
\hat Y_{\bar q}
\]
\[
-
\frac{1}{8}
\sum_{p<q}
\hat X_{p}
\otimes
\hat Z_{p+1,q-1}^{\leftrightarrow}
\otimes
\hat Y_{q}
\otimes
\hat X_{\bar p}
\otimes
\hat Z_{\bar p+1,\bar q-1}^{\leftrightarrow}
\otimes
\hat Y_{\bar q}
\]
\[
+
\frac{1}{8}
\sum_{p<q}
\hat X_{p}
\otimes
\hat Z_{p+1,q-1}^{\leftrightarrow}
\otimes
\hat Y_{q}
\otimes
\hat Y_{\bar p}
\otimes
\hat Z_{\bar p+1,\bar q-1}^{\leftrightarrow}
\otimes
\hat X_{\bar q}
\]
\[
+
\frac{1}{8}
\sum_{p<q}
\hat Y_{p}
\otimes
\hat Z_{p+1,q-1}^{\leftrightarrow}
\otimes
\hat X_{q}
\otimes
\hat X_{\bar p}
\otimes
\hat Z_{\bar p+1,\bar q-1}^{\leftrightarrow}
\otimes
\hat Y_{\bar q}
\]
\[
-
\frac{1}{8}
\sum_{p<q}
\hat Y_{p}
\otimes
\hat Z_{p+1,q-1}^{\leftrightarrow}
\otimes
\hat X_{q}
\otimes
\hat Y_{\bar p}
\otimes
\hat Z_{\bar p+1,\bar q-1}^{\leftrightarrow}
\otimes
\hat X_{\bar q}
\]
\[
-
\frac{1}{8}
\sum_{p<q}
\hat Y_{p}
\otimes
\hat Z_{p+1,q-1}^{\leftrightarrow}
\otimes
\hat Y_{q}
\otimes
\hat X_{\bar p}
\otimes
\hat Z_{\bar p+1,\bar q-1}^{\leftrightarrow}
\otimes
\hat X_{\bar q}
\]
\[
-
\frac{1}{8}
\sum_{p<q}
\hat Y_{p}
\otimes
\hat Z_{p+1,q-1}^{\leftrightarrow}
\otimes
\hat Y_{q}
\otimes
\hat Y_{\bar p}
\otimes
\hat Z_{\bar p+1,\bar q-1}^{\leftrightarrow}
\otimes
\hat Y_{\bar q}
\]
The eigenvalues of the $\hat S^2$ operator can be written as $S/2 (S/2+1)$ with $S \in [0, 1,
2, \ldots]$ (singlet, doublet, triplet, etc).

\section{The Specific Case of $M=2$ in $\mathcal{F}(2^{2M})$}
\label{appendix:two_mode_fermions}
This section explicitly enumerates, for the special case of $M=2$, the
characteristics of the Jordan-Wigner computational basis functions (representing
Fock space Slater determinants), $S^2$-pure configuration state function (CSF)
linear combinations thereof, and arbitrary linear combinations thereof which we
will refer to as full configuration interaction (FCI) states. This section is
useful to develop the beginnings of a picture for the arbitrary
$M$-spatial-orbital Fock space, and also directly leads to the FCI gate fabric
of the next full section.

\subsection{Slater Determinants / Jordan-Wigner Computational Basis States}

Table~\ref{tab:dets} enumerates the Slater determinants in the $M=2$ case along
with their corresponding Jordan-Wigner computational basis states, provides the
$N_{\alpha}$ and $N_{\beta}$ eigenvalues of each determinant (all determinants
are proper eigenstates of the particle number operators), identifies which
determinants are configuration state functions (CSFs) (only some determinants
are also CSFs), and if a CSF, provides the $S$ eigenvalue of the
determinant/CSF.

\begin{table}[ht!]
\renewcommand{\arraystretch}{1.5}
\centering
\caption{Enumeration of characteristics of $M=2$ Fock space in Jordan-Wigner
representation. First column: Base-2 qubit occupation string (i.e., qubit
computational
basis state). Second column: Base-10 qubit occupation string (i.e.,
base-10 index for vector and matrix quantities). 
Third column: Slater determinantal
configuration represented by this qubit computational basis state.
Fourth column: number of
$\alpha$ electrons in this configuration (always a proper eigenstate).
Fifth column: number of
$\beta$ electrons in this configuration (always a proper eigenstate).
Sixth column: is this configuration a valid configuration state function (CSF),
i.e., a proper eigenstate of $\hat S^2$?
Seventh column: if yes to the previous question, $S$ eigenvalue for this
simultaneous Slater determinant/CSF ($S=0$ - singlet, $S=1$ -
doublet, $S=2$ - triplet, \ldots).}
\label{tab:dets}
\begin{tabular}{rrcrrrr}
\hline 
\hline 
Base 2 & Base 10 & Determinant & $N_{\alpha}$ & $N_{\beta}$ &
Is CSF? & $S$ \\
\hline
$|0000\rangle$ & $|\# 0\rangle$ & 
\raisebox{6pt}{{\Qcircuit @R=0.6em @C=0.7em @! {
 & \qw
 & \qw
 & \qw \\
 & \qw
 & \qw
 & \qw \\
}}}
& 0 & 0 & Y & 0 \\
\hline 
$|0001\rangle$ & $|\# 1\rangle$ & 
\raisebox{6pt}{{\Qcircuit @R=0.3em @C=0.45em @! {
 & \qw
 & \qw
 & \qw \\
 & \ctrl{0}
 & \qw
 & \qw \\
}}}
& 1 & 0 & Y & 1 \\
\hline 
$|0010\rangle$ & $|\# 2\rangle$ & 
\raisebox{6pt}{{\Qcircuit @R=0.3em @C=0.3em @! {
 & \qw
 & \qw
 & \qw \\
 & \qw
 & \ctrlo{0}
 & \qw \\
}}}
& 0 & 1 & Y & 1 \\
\hline 
$|0011\rangle$ & $|\# 3\rangle$ & 
\raisebox{6pt}{{\Qcircuit @R=0.3em @C=0.3em @! {
 & \qw
 & \qw
 & \qw \\
 & \ctrl{0}
 & \ctrlo{0}
 & \qw \\
}}}
& 1 & 1 & Y & 0 \\
\hline 
$|0100\rangle$ & $|\# 4\rangle$ & 
\raisebox{6pt}{{\Qcircuit @R=0.3em @C=0.45em @! {
 & \ctrl{0}
 & \qw
 & \qw \\
 & \qw
 & \qw
 & \qw \\
}}}
& 1 & 0 & Y & 1 \\
\hline 
$|0101\rangle$ & $|\# 5\rangle$ & 
\raisebox{6pt}{{\Qcircuit @R=0.3em @C=0.45em @! {
 & \ctrl{0}
 & \qw
 & \qw \\
 & \ctrl{0}
 & \qw
 & \qw \\
}}}
& 2 & 0 & Y & 2 \\
\hline 
$|0110\rangle$ & $|\# 6\rangle$ & 
\raisebox{6pt}{{\Qcircuit @R=0.3em @C=0.3em @! {
 & \ctrl{0}
 & \qw
 & \qw \\
 & \qw
 & \ctrlo{0}
 & \qw \\
}}}
& 1 & 1 & N &   \\
\hline 
$|0111\rangle$ & $|\# 7\rangle$ & 
\raisebox{6pt}{{\Qcircuit @R=0.3em @C=0.3em @! {
 & \ctrl{0}
 & \qw
 & \qw \\
 & \ctrl{0}
 & \ctrlo{0}
 & \qw \\
}}}
& 2 & 1 & Y & 1 \\
\hline 
$|1000\rangle$ & $|\# 8\rangle$ & 
\raisebox{6pt}{{\Qcircuit @R=0.3em @C=0.3em @! {
 & \qw
 & \ctrlo{0}
 & \qw \\
 & \qw
 & \qw
 & \qw \\
}}}
& 0 & 1 & Y & 1 \\
\hline 
$|1001\rangle$ & $|\# 9\rangle$ & 
\raisebox{6pt}{{\Qcircuit @R=0.3em @C=0.3em @! {
 & \qw
 & \ctrlo{0}
 & \qw \\
 & \ctrl{0}
 & \qw
 & \qw \\
}}}
& 1 & 1 & N &   \\
\hline 
$|1010\rangle$ & $|\#10\rangle$ & 
\raisebox{6pt}{{\Qcircuit @R=0.3em @C=0.3em @! {
 & \qw
 & \ctrlo{0}
 & \qw \\
 & \qw
 & \ctrlo{0}
 & \qw \\
}}}
& 0 & 2 & Y & 2 \\
\hline 
$|1011\rangle$ & $|\#11\rangle$ & 
\raisebox{6pt}{{\Qcircuit @R=0.3em @C=0.3em @! {
 & \qw
 & \ctrlo{0}
 & \qw \\
 & \ctrl{0}
 & \ctrlo{0}
 & \qw \\
}}}
& 1 & 2 & Y & 1 \\
\hline 
$|1100\rangle$ & $|\#12\rangle$ & 
\raisebox{6pt}{{\Qcircuit @R=0.3em @C=0.3em @! {
 & \ctrl{0}
 & \ctrlo{0}
 & \qw \\
 & \qw
 & \qw
 & \qw \\
}}}
& 1 & 1 & Y & 0 \\
\hline 
$|1101\rangle$ & $|\#13\rangle$ & 
\raisebox{6pt}{{\Qcircuit @R=0.3em @C=0.3em @! {
 & \ctrl{0}
 & \ctrlo{0}
 & \qw \\
 & \ctrl{0}
 & \qw
 & \qw \\
}}}
& 2 & 1 & Y & 1 \\
\hline 
$|1110\rangle$ & $|\#14\rangle$ & 
\raisebox{6pt}{{\Qcircuit @R=0.3em @C=0.3em @! {
 & \ctrl{0}
 & \ctrlo{0}
 & \qw \\
 & \qw
 & \ctrlo{0}
 & \qw \\
}}}
& 1 & 2 & Y & 1 \\
\hline 
$|1111\rangle$ & $|\#15\rangle$ & 
\raisebox{6pt}{{\Qcircuit @R=0.3em @C=0.3em @! {
 & \ctrl{0}
 & \ctrlo{0}
 & \qw \\
 & \ctrl{0}
 & \ctrlo{0}
 & \qw \\
}}}
& 2 & 2 & Y & 0 \\
\hline 
\hline 
\end{tabular}
\end{table}

\subsection{Quantum Number Operators}

The $\hat N_{\alpha}$ and $\hat N_{\beta}$ operators are diagonal (as is always
true in the Jordan-Wigner representation), and their diagonal values are
depicted in the fourth and fifth columns of Table \ref{tab:dets}, respectively.

The $\hat S^2$ operator is not diagonal (as is generally true in the
Jordan-Winger representation). Instead, only 14 out of the 16 rows/columns of
this operator are diagonal, and their diagonal entries are given in the seventh
column of Table \ref{tab:dets}. The non-diagonal contributions arise from the
non-CSF determinants
\[
|0110\rangle
\equiv
\raisebox{5pt}{\Qcircuit @R=0.3em @C=0.3em @! {
 & \ctrl{0}
 & \qw
 & \qw \\
 & \qw
 & \ctrlo{0}
 & \qw \\
}}
\equiv
|\#6\rangle
\]
and,
\[
|1001\rangle
\equiv
\raisebox{5pt}{\Qcircuit @R=0.3em @C=0.3em @! {
 & \qw
 & \ctrlo{0}
 & \qw \\
 & \ctrl{0}
 & \qw
 & \qw \\
}}
\equiv
|\#9\rangle
\]
These two determinants form the seniority-2 coupling set (the set of
determinants with 2$\times$ non-spin-paired electrons). In this restricted
basis, the $\hat S^2$ operator is,
\[
\hat S^2 
=
\left [
\begin{array}{rr}
 1 & -1 \\
-1 &  1 \\
\end{array}
\right ]
\]

\subsection{Configuration State Functions (CSFs)}

Configuration state functions (CSFs) are defined as sparse linear combinations
of Slater determinants that provide proper eigenstates of $\hat S^2$.  The 14
spin-pure Slater determinants discussed above are also CSFs, with corresponding
eigenvalues $S$.

In the seniority-2 coupling set of the non-spin-pure Slater determinants
$|\#6\rangle$ and $|\#9\rangle$, the eigenvectors of the $\hat S^2$ operator are,
\[
\hat V
\equiv
\frac{1}{\sqrt{2}}
\left [
\begin{array}{cc}
1 & 1 \\
1 & -1 \\
\end{array}
\right ]
\]
and the corresponding eigenvalues are,
\[
\hat \Lambda
\equiv
\left [
\begin{array}{c}
0 \\
2 \\
\end{array}
\right ]
.
\]
E.g., the $+$ combination yields an $S=0$ singlet CSF, while the $-$ combination
yields an $S=2$ triplet CSF.

Thus the symmetry-adapted configuration state functions (CSFs) for this seniority
coupling set are,
\[
| \Phi_{Z=2}^{S=0} \rangle
\equiv
\frac{1}{\sqrt{2}}
\left (
|0110\rangle
+
|1001\rangle
\right )
\]
and,
\[
| \Phi_{Z=2}^{S=2} \rangle
\equiv
\frac{1}{\sqrt{2}}
\left (
|0110\rangle
-
|1001\rangle
\right )
\]
Therefore, we have a complete real, orthonormal set of $16$ CSFs for
$\mathcal{F}(2^{2 * 2})$: 5 singlets, 8 doublets, and 3 triplets. These CSFs are
proper eigenfunctions of $\hat N_{\alpha}$, $\hat N_{\beta}$, and $\hat S^2$.

\subsection{Quantum Number Irreps}

Valid solutions to the time-dependent or time-independent Schr\"odinger equation
for spin-$1/2$ fermions governed by spin-free Hamiltonian operators must be
definite simultaneous eigenstates of the quantum number operators $(\hat
N_{\alpha}, \hat N_{\beta}, \hat S^2$) with definite target eigenvalues
$(N_{\alpha}, N_{\beta}, S)$. We refer to the set of valid simultaneous
eigenstates for a given set of target quantum numbers $(N_{\alpha},
N_{\beta}, S)$ as a quantum number irrep.

Table~\ref{tab:csfs} enumerates the dimensionality and our particular convention
for the CSF basis for each definite $(N_{\alpha}, N_{\beta}, S)$ irrep of the
$M=2$ Fock space.  An arbitrary special orthogonal rotation within each irrep
would also provide a faithful representation of the basis for that irrep.

\begin{table}[ht!]
\centering
\caption{CSF Irreps for $M=2$ Fock space. $D$ refers to the irrep dimension. The
listed elements are our particular convention for the CSF basis functions of
each irrep.}
\label{tab:csfs}
\begin{tabular}{rrrrr}
\hline
\hline
$N_{\alpha}$ & $N_{\beta}$ & $S$ & $D$ & Elements \\
\hline
0 & 0 & 0 & 1 & $|\#0\rangle$ \\
2 & 2 & 0 & 1 & $|\#15\rangle$ \\
2 & 0 & 2 & 1 & $|\#5\rangle$ \\
0 & 2 & 2 & 1 & $|\#10\rangle$ \\
1 & 1 & 2 & 1 & $|\Phi_{Z=2}^{S=2}\rangle$ \\
1 & 0 & 1 & 2 & $|\#1\rangle$, $|\#4\rangle$ \\
0 & 1 & 1 & 2 & $|\#2\rangle$, $|\#8\rangle$ \\
1 & 2 & 1 & 2 & $|\#11\rangle$, $|\#14\rangle$ \\
2 & 1 & 1 & 2 & $|\#7\rangle$, $|\#13\rangle$ \\
1 & 1 & 0 & 3 & $|\#3\rangle$, $|\#12\rangle$, $|\Phi_{Z=2}^{S=0}\rangle$ \\
\hline
\hline
\end{tabular}
\end{table}

\subsection{Full Configuration Interaction (FCI) States}

The restriction of physically valid solutions of the time-dependent or
time-independent Schr\"odinger equation to a given target quantum number irrep
severely constrains, but does not exactly determine the valid solution for most
irreps. For instance, in the $(N_{\alpha}=1, N_{\beta}=0, S=1)$ irrep, the
15-parameter generic solution,
\[
| \Psi \rangle
\stackrel{?}{=}
\sum_{I=0}^{I=15}
c_{I}
| \#I \rangle
:
\sum_{I=0}^{I=15}
|c_{I} |^2
=
1
,
c_{I} \in \mathbb{R}
\]
is invalid because it does not respect the quantum number symmetries, but the
1-parameter solution,
\[
| \Psi \rangle
=
\sum_{I \in <1, 4>}
c_{I}
| \#I \rangle
:
\sum_{I \in <1, 4>}
|c_{I} |^2
=
1
,
c_{I} \in \mathbb{R}
\]
is valid due to the fact that the dimension of the target irrep is $D = 2 > 1$. 

We generically refer to states which exactly lie within a given target quantum
number irrep, but where the remaining flexibility in the state is determined by
solving an auxiliary equation such as the time-dependent or time-independent
Schr\"odinger equation, as ``full configuration interaction'' (FCI) states. The
motivation for this naming is the set of states that emerge from exactly
diagonalizing the spin-free electronic Hamiltonian within a given quantum number
irrep, i.e., the classical FCI method, though the usage within this work should
be understood to be generalized to solving any linear auxiliary equation
governed by a spin-free operator which is simultaneously diagonalized by the
three quantum number operators.

The question that arises at this point is how to construct special orthogonal
operators that respect the quantum number symmetry but have the power to move
from an arbitrary quantum-number-pure trial state to an FCI state within the
same quantum number irrep. The simple
answer is to construct complete special orthogonal operators acting on the CSF
basis of each irrep, with the property that these operators commute with all
three quantum number operators. This leads to the construction of $4\times$ 1-parameter
$\mathcal{SO}(2)$ operators (simple Givens rotation matrices) acting within the
$4\times$ $S=1$ doublet irreps, and $1\times$ 3-parameter $\mathcal{SO}(3)$
operator acting within the $(N_{\alpha}=1, N_{\beta}=1, S=0)$ irrep. This seems
to imply that the parameter dimension of $\mathcal{F}(2^{2 * 2})$ is $7$. However,
further analysis reveals that to preserve $\hat S^2$ symmetry, the same operator
must be applied in the 
$(N_{\alpha} = 1, N_{\beta} = 0, S = 1)$ and 
$(N_{\alpha} = 0, N_{\beta} = 1, S = 1)$ irreps
and that the same operator must be applied in the
$(N_{\alpha} = 1, N_{\beta} = 2, S = 1)$ and 
$(N_{\alpha} = 2, N_{\beta} = 1, S = 1)$ irreps. This is related to the fact
that the spin-free Schr\"odinger equation is invariant under permutation of the
$\alpha$ and $\beta$ labels in the working equations. This reduces the total
number of parameter of $\mathcal{F}(2^{2 * 2})$ to $5$, and yields the highly
structured special orthogonal operators that will be encountered as $M=2$ FCI
gate operators in the next section.

\section{Gate Fabric for $\mathcal{F}(2^{2M})$ via $M=2$ FCI Gates}

\label{appendix:other_qnp_gate_fabrics}

\begin{figure*}[htp!]
\centering
\begin{equation*}
\begin{array}{l}
\Qcircuit @R=0.3em @C=0.3em @!R {
 & \multigate{3}{F}
 & \qw
 & \qw \\
 & \ghost{F}
 & \qw
 & \qw \\
 & \ghost{F}
 & \multigate{3}{F}
 & \qw \\
 & \ghost{F}
 & \ghost{F}
 & \qw \\
 & \multigate{3}{F}
 & \ghost{F}
 & \qw \\
 & \ghost{F}
 & \ghost{F}
 & \qw \\
 & \ghost{F}
 & \qw
 & \qw \\
 & \ghost{F}
 & \qw
 & \qw \\
}
\end{array}
\ldots
\cong
\begin{array}{l}
\Qcircuit @R=0.3em @C=0.3em @!R {
 & \multigate{7}{\mathcal{F}(2^{2 * 4})}
 & \qw \\
 & \ghost{\mathcal{F}(2^{2 * 4})}
 & \qw \\
 & \ghost{\mathcal{F}(2^{2 * 4})}
 & \qw \\
 & \ghost{\mathcal{F}(2^{2 * 4})}
 & \qw \\
 & \ghost{\mathcal{F}(2^{2 * 4})}
 & \qw \\
 & \ghost{\mathcal{F}(2^{2 * 4})}
 & \qw \\
 & \ghost{\mathcal{F}(2^{2 * 4})}
 & \qw \\
 & \ghost{\mathcal{F}(2^{2 * 4})}
 & \qw \\
}
\end{array}
\end{equation*}
\begin{equation*}
\begin{array}{l}
\Qcircuit @R=0.3em @C=0.3em @!R {
 & \multigate{3}{F}
 & \qw \\
 & \ghost{F}
 & \qw \\
 & \ghost{F}
 & \qw \\
 & \ghost{F}
 & \qw \\
}
\end{array}
\coloneqq
\left [
\begin{array}{rrrrrrrrrrrrrrrr}
 1 & & & & & & & & & & & & & & &  \\
 & \color{brickred}{c_{1\mathrm{p}}} & & & \color{brickred}{s_{1\mathrm{p}}} & & & & & & & & & & & \\
 & & \color{brickred}{c_{1\mathrm{p}}} & & & & & & \color{brickred}{s_{1\mathrm{p}}} & & & & & & &  \\
 & & & \color{darkgreen}{u_{00}} & & & \color{darkgreen}{u_{01}} & & & \color{darkgreen}{u_{02}} & & & \color{darkgreen}{u_{03}} & & &  \\
 & \color{brickred}{-s_{1\mathrm{p}}} & & & \color{brickred}{c_{1\mathrm{p}}} & & & & & & & & & & &  \\
 & & & & & 1 & & & & & & & & & &  \\
 & & & \color{darkgreen}{u_{10}} & & & \color{darkgreen}{u_{11}} & & & \color{darkgreen}{u_{12}} & & & \color{darkgreen}{u_{13}} & & &  \\
 & & & & & & & \color{darkblue}{c_{1\mathrm{h}}} & & & & & & \color{darkblue}{+s_{1\mathrm{h}}} & &  \\
 & & \color{brickred}{-s_{1\mathrm{p}}} & & & & & & \color{brickred}{c_{1\mathrm{p}}} & & & & & & &  \\
 & & & \color{darkgreen}{u_{20}} & & & \color{darkgreen}{u_{21}} & & & \color{darkgreen}{u_{22}} & & & \color{darkgreen}{u_{23}} & & &  \\
 & & & & & & & & & & 1 & & & & &  \\
 & & & & & & & & & & & \color{darkblue}{c_{1\mathrm{h}}} & & & \color{darkblue}{+s_{1\mathrm{h}}} &  \\
 & & & \color{darkgreen}{u_{30}} & & & \color{darkgreen}{u_{31}} & & & \color{darkgreen}{u_{32}} & & & \color{darkgreen}{u_{33}} & & &  \\
 & & & & & & & \color{darkblue}{-s_{1\mathrm{h}}} & & & & & & \color{darkblue}{c_{1\mathrm{h}}} & &  \\
 & & & & & & & & & & & \color{darkblue}{-s_{1\mathrm{h}}} & & & \color{darkblue}{c_{1\mathrm{h}}} &  \\
 & & & & & & & & & & & & & & & 1 \\
\end{array}
\right ] \in \mathcal{F} (2^{2*2})
\end{equation*}
\begin{equation*}
\left [
\begin{array}{rrrr}
u_{00} & u_{01} & u_{02} & u_{03} \\
u_{10} & u_{11} & u_{12} & u_{13} \\
u_{20} & u_{21} & u_{22} & u_{23} \\
u_{30} & u_{31} & u_{32} & u_{33} \\
\end{array}
\right ]
\coloneqq
\left [
\begin{array}{rrrr}
1 & & & \\
& & & 1 \\
& 1 / \sqrt{2} & 1 / \sqrt{2} & \\
& 1 / \sqrt{2} & -1 / \sqrt{2} & \\
\end{array}
\right ]^{\dagger}
\left [
\begin{array}{rrrr}
v_{00} & v_{01} & v_{02} & \\
v_{10} & v_{11} & v_{12} & \\
v_{20} & v_{21} & v_{22} & \\
 & & & 1 \\
\end{array}
\right ]
\left [
\begin{array}{rrrr}
1 & & & \\
& & & 1 \\
& 1 / \sqrt{2} & 1 / \sqrt{2} & \\
& 1 / \sqrt{2} & -1 / \sqrt{2} & \\
\end{array}
\right ] 
\end{equation*}
\begin{equation*}
\begin{array}{ll}
c_{1\mathrm{p}} \coloneqq \cos(\theta_{1\mathrm{p}}), & 
s_{1\mathrm{p}} \coloneqq \sin(\theta_{1\mathrm{p}}) \\
c_{1\mathrm{h}} \coloneqq \cos(\theta_{1\mathrm{h}}), & 
s_{1\mathrm{h}} \coloneqq \sin(\theta_{1\mathrm{h}}) \\
\end{array}
\end{equation*}
\begin{equation*}
\hat v
\coloneqq
\exp(\hat x)
:
\
\hat x
=
-\hat x^\dagger
,
\
\hat x 
\in
\mathbb{R}^3
\times
\mathbb{R}^3
\end{equation*}
\caption{Gate fabric hypothesized to be universal for $\mathcal{F}(2^{2M})$
(sketched for $M=4$).  The spin orbitals in Jordan-Wigner representation are
physically ordered in ``interleaved'' ordering with even(odd) qubit indices
denoting $\alpha$($\beta$) spin orbitals. The Jordan-Wigner strings are defined
in ``$\alpha$-then-$\beta$'' order as defined in the main text. 
The gate fabric
is a 4-local-nearest-neighbor- tessellation of
alternating even and odd spatial-orbital-pair 5-parameter, 4-qubit $\hat F$
gates, constructed to be universal for $M=2$ FCI. 
Each $\hat F$ gate consists of:
(A) a spin-adapted 1-parameter $\mathcal{SO}(2)$
rotation in the 1-particle ($N_{\alpha} + N_{\beta} = 1$) $S=0$ block denoted in red (parameter
$\theta_{1\mathrm{p}}$).
(B) a spin-adapted 1-parameter $\mathcal{SO}(2)$ rotation in the 1-hole
($N_{\alpha} + N_{\beta} = 3$) $S=0$ block denoted in blue (parameter
$\theta_{1\mathrm{h}}$).
(C) a spin-adapted 3-parameter $\mathcal{SO}(3)$ rotation in the ($N_{\alpha} =
1$, $N_{\beta} = 0$, $S = 0$) irrep denoted in green (3 parameters in the unique
upper triangle of $\hat x$).
The decomposition of $\hat u$ into a transformation from the determinant basis
of the Jordan-Wigner computational basis to the CSF basis, followed by the
application of an $\mathcal{SO}(3)$ rotation in the $S=0$ CSFs, followed by
backtransformation to the determinant basis is depicted below the definition of
$\hat F$. 
The factoring of $\hat F$ into a product over a set of $5\times$ representative
$1$-parameter $4$-qubit gates, and the explicit decomposition of these gates
into physically realizable forms in the standard 2-qubit gate library is
discussed later in the Appendix.
}
\label{fig:F2}
\end{figure*}
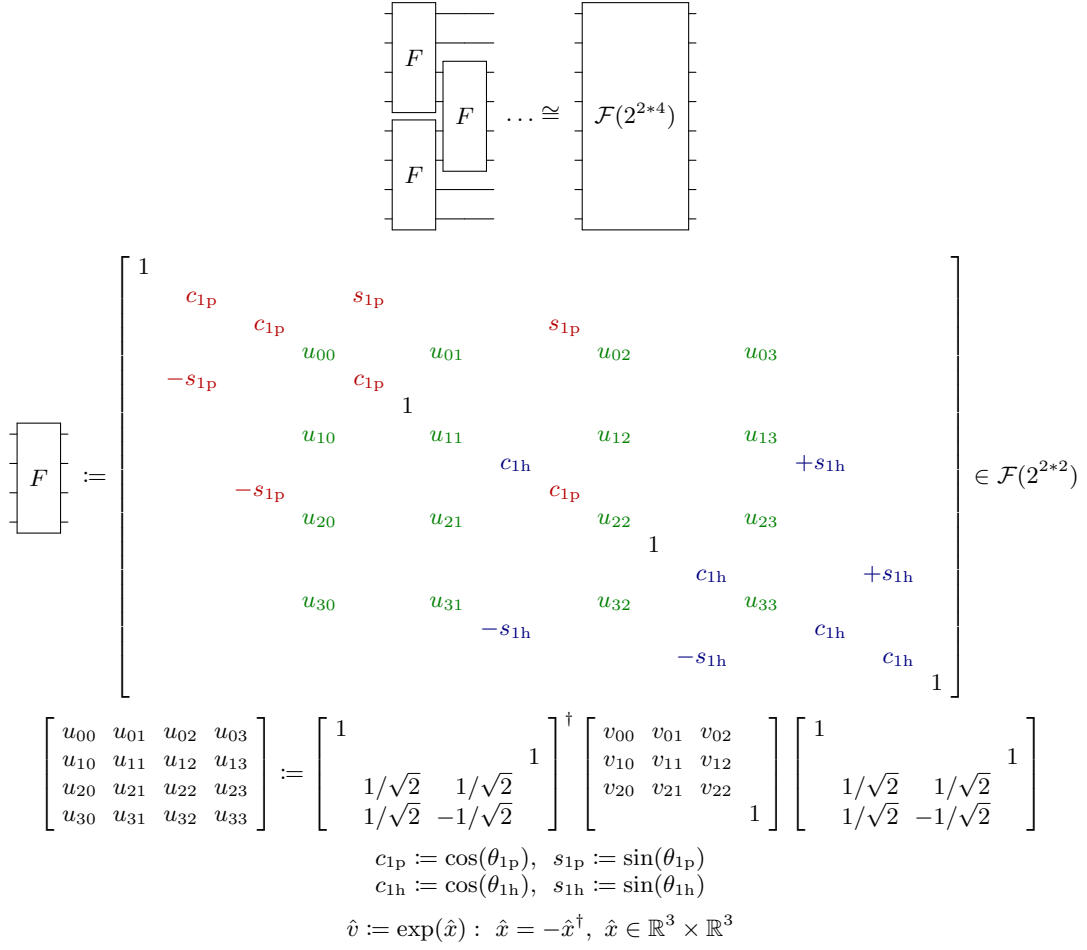

An early iteration of the gate fabric described in the main text was
developed by constructing a gate fabric comprising a 5-parameter 4-qubit $\hat F$ gate
universal for $M=2$ FCI as detailed in Figure \ref{fig:F2}. A fabric of
these $\hat F$ gates was found to exactly preserve quantum number symmetry, to
provide universality for $\mathcal{F} (2^{2M})$ for sufficient parameter depth,
and to yield an expressive approximate representation at intermediate depths.
The representation power and numerical convergence was found to be similar
between $\hat F$ gate fabrics and the $\hat Q$ gate fabrics, and the latter is
conceptually simpler, so we have elected to focus on the latter in the main
text.
In the following we describe this alternative gate fabric and
additional variants and refer to and use the concepts and notation introduced in
Appendices~\ref{appendix:jw_details} and \ref{appendix:two_mode_fermions}.

\subsection{Decomposition of $\hat F$ into Simple Gate Elements}

There are many different possible implementations of $\hat F$ into products of simpler
(e.g., 1-parameter) gate elements. However, the block diagonal nature and
configuration constituency of $\hat F$ suggests the following pragmatic choice,
leading to a decomposition with a simple decomposition all the way down to a
standard two-qubit gate library. The text refers to the $\hat F$ gate matrix in
the second line of Figure~\ref{fig:F2}:

The red block (matrix entries $c_{1p} \coloneqq \cos(\theta_{1p}/2)$ and $s_{1p}
\coloneqq \sin(\theta_{1p}/2)$) corresponds to a Givens rotation between the one
particle $(N_{\alpha} = 1, N_{\beta} = 0, S=1)$ CSFs $
|0001\rangle
\eqqcolon
\raisebox{5pt}{\Qcircuit @R=0.3em @C=0.3em @! {
 & \qw
 & \qw
 & \qw \\
 & \ctrl{0}
 & \qw
 & \qw \\
}}
=
|\#1\rangle
$
and
$
|0100\rangle
\eqqcolon
\raisebox{5pt}{\Qcircuit @R=0.3em @C=0.3em @! {
 & \ctrl{0}
 & \qw
 & \qw \\
 & \qw
 & \qw
 & \qw \\
}}
=
|\#4\rangle
$
and the same Givens rotation between the $(N_{\alpha} = 0, N_{\beta} = 1, S=1)$
CSFs $|\#2\rangle$ and $|\#8\rangle$ (to preserve $\hat S^2$
symmetry). We call this operation the $\mathrm{QNP_{1p}}$ gate
(quantum-number-preserving 1-particle gate).

The blue block (matrix entries $c_{1h} \coloneqq \cos(\theta_{1h}/2)$ and
$s_{1h} \coloneqq \sin(\theta_{1h}/2)$) implements a Givens rotation between the
one hole $(N_{\alpha} = 1, N_{\beta} = 2, S=1)$ CSFs $
|1011\rangle
\eqqcolon
\raisebox{5pt}{\Qcircuit @R=0.3em @C=0.3em @! {
 & \qw
 & \ctrlo{0}
 & \qw \\
 & \ctrl{0}
 & \ctrlo{0}
 & \qw \\
}}
=
|\#11\rangle
$ 
and 
$
|1110\rangle
\coloneqq
\raisebox{5pt}{\Qcircuit @R=0.3em @C=0.3em @! {
 & \ctrl{0}
 & \ctrlo{0}
 & \qw \\
 & \qw
 & \ctrlo{0}
 & \qw \\
}}
=
|\#14\rangle
$
and the same Givens rotation between the $(N_{\alpha} = 2, N_{\beta} = 1, S=1)$
CSFs $|\#7\rangle$ and $|\#13\rangle$ (to preserve $\hat S^2$ symmetry).
We call this operation the $\mathrm{QNP_{1h}}(\theta_{1h})$ gate
(quantum-number-preserving 1-hole gate).

The green block implements an $\mathcal{SO}(3)$ rotation between the three
$(N_{\alpha} = 1, N_{\beta} = 1, S=0)$ CSFs,
$
|0011\rangle
\coloneqq
\raisebox{5pt}{\Qcircuit @R=0.3em @C=0.3em @! {
 & \qw
 & \qw
 & \qw \\
 & \ctrl{0}
 & \ctrlo{0}
 & \qw \\
}}
\eqqcolon
|\#3\rangle
$
,
$
|1100\rangle
\eqqcolon
\raisebox{5pt}{\Qcircuit @R=0.3em @C=0.3em @! {
 & \ctrl{0}
 & \ctrlo{0}
 & \qw \\
 & \qw
 & \qw
 & \qw \\
}}
=
|\#12\rangle
$
,
and
$
(
|0110\rangle
+
|1001\rangle
) / \sqrt{2}
\eqqcolon
\left (
\raisebox{5pt}{\Qcircuit @R=0.3em @C=0.3em @! {
 & \ctrl{0}
 & \qw
 & \qw \\
 & \qw
 & \ctrlo{0}
 & \qw \\
}}
+
\raisebox{5pt}{\Qcircuit @R=0.3em @C=0.3em @! {
 & \qw
 & \ctrlo{0}
 & \qw \\
 & \ctrl{0}
 & \qw
 & \qw \\
}}
\right ) / \sqrt{2}
=
(
|\#6\rangle
+
|\#9\rangle
) / \sqrt{2}
$.
There are three natural rotation gates (i.e., Euler-angle-like rotation gates)
in this subspace:
First, the $\mathrm{QNP_{PX}}$ gate (quantum-number-preserving
pair exchange gate) implements a Givens rotation between the two closed shell
CSFs
$
|0011\rangle
\eqqcolon
\raisebox{5pt}{\Qcircuit @R=0.3em @C=0.3em @! {
 & \qw
 & \qw
 & \qw \\
 & \ctrl{0}
 & \ctrlo{0}
 & \qw \\
}}
\coloneqq
|3\rangle
$
and
$
|1100\rangle
\eqqcolon
\raisebox{5pt}{\Qcircuit @R=0.3em @C=0.3em @! {
 & \ctrl{0}
 & \ctrlo{0}
 & \qw \\
 & \qw
 & \qw
 & \qw \\
}}
=
|\#12\rangle
$.
Second and third the $\mathrm{QNP_{PBU}}$ and $\mathrm{QNP_{PBL}}$ (quantum number
preserving pair-break upper/lower gates) rotate between the upper,
respectively, lower closed shell CSF and the open-shell singlet CSF $
(
|0110\rangle
+
|1001\rangle
) / \sqrt{2}
$.

Explicit decompositions of the five gates 
$\mathrm{QNP_{1p}}$,
$\mathrm{QNP_{1h}}$,
$\mathrm{QNP_{PX}}$,
$\mathrm{QNP_{PBU}}$, and
$\mathrm{QNP_{PBL}}$ 
to elementary 2-qubit gate operations is provided in Appendix~\ref{sec:qnp_decompositions}.

\subsection{Simplifications of the $\hat F$ Gate Fabric}

A natural question at this point is whether there exist gate fabrics for
$\mathcal{F}(2^{2M})$ which are simpler than the $\hat F$-gate fabric described
above. E.g., a simpler gate fabric might have fewer parameters per gate element,
and/or fewer QNP product gates per gate element, while still preserving quantum
number symmetry and numerical efficiency. For one explicit example, it is clear
that the $3\times$ QNP product gates in $(N_{\alpha} = 1, N_{\beta} = 1, S=0)$
irrep are redundant, as $\mathrm{QNP_{PX}}$ and $\mathrm{QNP_{PBU}}$ (or
$\mathrm{QNP_{PBL}}$) are sufficient to attain any desired action in the irrep.
Further, repeated application of pairs of $\mathrm{QNP_{PX}}$ and
$\mathrm{QNP_{PBU}}$ (or $\mathrm{QNP_{PBL}}$) gates is sufficient to attain any
desired operator in the irrep. So we can already reduce from a 5-parameter
$\hat F$ gate fabric to a 4-parameter modified $\hat F'$ gate fabric. 

Next, we can consider the 1-hole and 1-particle spaces. Depending on the target
irrep, only one of these rotations is generally needed, e.g., for most irreps, a
gate fabric of 
$\mathrm{QNP_{1p}}$,
$\mathrm{QNP_{PX}}$,
and,
$\mathrm{QNP_{PBU}}$ is universal. So, for most irreps, a 3-parameter modified
$\hat F''$ gate fabric is sufficient.
A technical detail here is that
the choice of $\mathrm{QNP_{1p}}$ vs. $\mathrm{QNP_{1h}}$ required for
universality is contingent on whether there are more particles or holes in the
desired irrep of $\mathcal{F}(2^{M})$ for extreme edge case irreps. 

As we show in the main text, it is possible
to reduce $\hat F''$ even further to a 2-parameter $\hat Q$ gate fabric, where the
$\hat Q$ fabric symmetrizes the rotations between the 1-particle and 1-hole
irreps, and additional mixes rotations between the 1-particle/hole irreps and
the $(N_{\alpha} = 1, N_{\beta} = 1, S = 0)$ irrep. 
To that end we consider an alternative quantum number preserving gate which is already
well-known in the literature, the spatial orbital rotation gate, which we describe in the following section.

\subsection{Orbital Rotations}
A well-known operation in both classical and quantum electronic structure
methods is the spin-adapted spatial orbital rotation gate, which implements,
\begin{equation}
| \phi_{p}' \rangle
\equiv
\sum_{q}
V_{qp}
| \phi_{q} \rangle
\end{equation}
for $V_{pq} \in \mathcal{SO}(M)$, and for the particular case of $M=2$ adjacent
spatial orbitals. If we take $V_{pq}$ to be a $2 \times 2$ special orthogonal
matrix, i.e., a Givens rotation matrix with parameter $\varphi$, then this
1-parameter, 4-qubit $\mathrm{QNP_{OR}}$ gate (qantum-number-preserving orbital
rotation gate) is a special case of the 5-parameter, 4-qubit $\hat F$ gate from Figure~\ref{fig:F2} with,
$c \coloneqq c_{p1} = c_{h1} = \cos(\theta/2)$, $s \coloneqq s_{p1} = s_{h1} = \sin(\theta/2)$, and
\begin{equation}
    u_{ij} = 
    \left [
\begin{array}{rrrr}
c^2 & c s & c s & +s^2 \\
-c s & c^2 & -s^2 & cs \\
-c s & -s^2 & c^2 & cs \\
+s^2 & -c s & -c s & c^2 \\
\end{array}
\right]_{i,j} .
\end{equation}
This gate can be viewed as a simultaneous and symmetrical application of the
$\mathrm{QNP_{1p}}$ and $\mathrm{QNP_{1h}}$ gates which also acts in a direct
product manner in the $(N_{\alpha} = 1, N_{\beta} = 1, S = 0)$ irrep. The
explicit action of the $\mathrm{QNP_{OR}}$ gate is depicted in Figure
\ref{fig:O2}.

\begin{figure*}[htp!]
\centering
\begin{equation*}
\begin{array}{l}
\Qcircuit @R=0.3em @C=0.3em @!R {
 & \multigate{3}{\mathrm{QNP_{OR}} (\varphi)}
 & \qw \\
 & \ghost{\mathrm{QNP_{OR}} (\varphi)}
 & \qw \\
 & \ghost{\mathrm{QNP_{OR}} (\varphi)}
 & \qw \\
 & \ghost{\mathrm{QNP_{OR}} (\varphi)}
 & \qw \\
}
\end{array}
\coloneqq
\left [
\begin{array}{rrrrrrrrrrrrrrrr}
 1 & & & & & & & & & & & & & & &  \\
 & \color{brickred}{c} & & & \color{brickred}{+s} & & & & & & & & & & & \\
 & & \color{brickred}{c} & & & & & & \color{brickred}{-s} & & & & & & &  \\
 & & & \color{darkgreen}{c^2} & & & \color{darkgreen}{+cs} & & & \color{darkgreen}{+cs} & & & \color{darkgreen}{+s^2} & & &  \\
 & \color{brickred}{-s} & & & \color{brickred}{c} & & & & & & & & & & &  \\
 & & & & & 1 & & & & & & & & & &  \\
 & & & \color{darkgreen}{-cs} & & & \color{darkgreen}{c^2} & & & \color{darkgreen}{-s^2} & & & \color{darkgreen}{+cs} & & &  \\
 & & & & & & & \color{darkblue}{c} & & & & & & \color{darkblue}{+s} & &  \\
 & & \color{brickred}{-s} & & & & & & \color{brickred}{c} & & & & & & &  \\
 & & & \color{darkgreen}{-cs} & & & \color{darkgreen}{-s^2} & & & \color{darkgreen}{c^2} & & & \color{darkgreen}{+cs} & & &  \\
 & & & & & & & & & & 1 & & & & &  \\
 & & & & & & & & & & & \color{darkblue}{c} & & & \color{darkblue}{+s} &  \\
 & & & \color{darkgreen}{+s^2} & & & \color{darkgreen}{-cs} & & & \color{darkgreen}{-cs} & & & \color{darkgreen}{c^2} & & &  \\
 & & & & & & & \color{darkblue}{-s} & & & & & & \color{darkblue}{c} & &  \\
 & & & & & & & & & & & \color{darkblue}{-s} & & & \color{darkblue}{c} &  \\
 & & & & & & & & & & & & & & & 1 \\
\end{array}
\right ] \in \mathcal{F} (2^{2*2})
\end{equation*}
\begin{equation*}
\begin{array}{ll}
c \coloneqq \cos(\varphi), & 
s \coloneqq \sin(\varphi) \\
\end{array}
\end{equation*}
\caption{Spin-adapted spatial orbital rotation gate between two adjacent spatial
orbitals. The parameter $\varphi$ is the argument of the Givens rotation between
orbitals $|\phi_{0}\rangle$ and $|\phi_{1}\rangle$, with the same Givens rotation
applied in the $\alpha$ and $\beta$ spaces. 
}
\label{fig:O2}
\end{figure*}

It is well-known that a fabric of $\binom{M}{2}$ $\mathrm{QNP_{OR}}$ gates
arranged in a rectangular or triangular gate fabric pattern can exactly implement
an arbitrary orbital rotation within $M$ spatial orbitals, with a classically
tractable relationship between the $V_{pq}$ orbital rotation matrix and the
parameters $\{ \phi_{d} \}$ of the fabric being possible through the QR decomposition
of the orbital rotation matrix.

\subsection{Variants of the QNP fabrics}
As with the gate fabric in the main text it can be interesting to prepend the parametrized gate elements 
$\hat F'$ or $\hat F''$ with a fixed gate like the $\hat \Pi$ gate as this may improve trainability of the fabric
or even expressiveness at intermediate depths.
In the main text we have explored the options $\hat \Pi \in \{\hat I, \mathrm{QNP_{OR}} (\pi)\}$.
Another natural option, inspired by the concept of fermionic swap networks, would be to take $\hat \Pi$ to be an orbital wise fermionic swap gate.
This gate is also quantum number preserving and we introduce it in the end of the following section.

\begin{widetext}
\section{Explicit decompositions of the quantum number preserving gates}\label{sec:qnp_decompositions}

\usetikzlibrary{calc}
\pgfdeclarelayer{bg}
\pgfsetlayers{bg,main}

Here we provide explicit decompositions of all the quantum number preserving gates introduced in the main text,
namely the gates: $\mathrm{QNP_{1p}}$, $\mathrm{QNP_{1h}}$, $\mathrm{QNP_{PX}}$, $\mathrm{QNP_{PBU}}$, $\mathrm{QNP_{PBL}}$, and $\mathrm{QNP_{OR}}$ as well as one additional gate $\mathrm{OFSWAP}$.
We call these gates number preserving gates because for any gate $G$ from the above list it holds that
\begin{equation}
\left [
\hat G(\theta),
\hat N_{\alpha}
\right ]
=
0
, \
\left [
\hat G(\theta),
\hat N_{\beta}
\right ]
=
0
, \
\left [
\hat G(\theta),
\hat S^2
\right ]
=
0.
\end{equation}
The decompositions of $\mathrm{QNP_{1p}}(\theta)$ and $\mathrm{QNP_{1h}}(\theta)$ are given in terms of decompositions
of two gates, each acting on just the alpha or beta space such that $\mathrm{QNP_{1p}}(\theta) = \mathrm{QNP_{A0B1}}(\theta) \,  \mathrm{QNP_{A1B0}}(\theta)$ and $\mathrm{QNP_{1h}}(\theta) = \mathrm{QNP_{A2B1}}(\theta) \, \mathrm{QNP_{A1B2}}(\theta)$.
None of these four gates is individually quantum number preserving.

\subsection{Pair exchange gate $\mathrm{QNP_{PX}}(\theta)$}
For the $\mathrm{QNP_{PX}}(\theta)$ gate we present the following decomposition in terms of standard gates and controlled $Y$ rotations:
\begin{center}
\begin{tikzpicture}[scale=0.7, transform shape]
\begin{scope}
  \node[draw, fill=white, minimum height=height("$A$")+3cm, text width=2.7cm-10mm, inner sep=2mm, align=center] (lhs) at (-2.1500000000000004, -1.5) {$$\mathrm{QNP_{PX}}(\theta)$$}; 
  \begin{pgfonlayer}{bg}
    \draw ($(lhs.west |- 0,0) + (-0.2,0)$) -- ($(lhs.east |- 0,0) + (0.2,0)$);
    \draw ($(lhs.west |- 0,-1) + (-0.2,0)$) -- ($(lhs.east |- 0,-1) + (0.2,0)$);
    \draw ($(lhs.west |- 0,-2) + (-0.2,0)$) -- ($(lhs.east |- 0,-2) + (0.2,0)$);
    \draw ($(lhs.west |- 0,-3) + (-0.2,0)$) -- ($(lhs.east |- 0,-3) + (0.2,0)$);
  \end{pgfonlayer}
  \node at (lhs.center-|-0.4,0) {{$=$}};  \draw[] (0.5, -1) -- +(0, 1) node[pos=0, circle, fill=black, inner sep=0pt,minimum size=3pt] {{}} node[pos=1] (gate0) {$\oplus$}; 
  \draw[] (0.5, -3) -- +(0, 1) node[pos=0, circle, fill=black, inner sep=0pt,minimum size=3pt] {{}} node[pos=1] (gate1) {$\oplus$}; 
  \node[draw, fill=white] (gate2) at (1.5, 0) {$X$}; 
  \draw[] (1.5, -3) -- +(0, 2) node[pos=0, circle, fill=black, inner sep=0pt,minimum size=3pt] {{}} node[pos=1] (gate3) {$\oplus$}; 
  \draw[] (2.9, 0) -- +(0, -3) node[pos=0, circle, fill=black, inner sep=0pt,minimum size=3pt] {{}} node[pos=1, draw, fill=white] (gate4) {$RY(\theta/4)$}; 
  \draw[] (4.3, 0) -- +(0, -2) node[pos=0, circle, fill=black, inner sep=0pt,minimum size=3pt] {{}} node[pos=1] (gate5) {$\oplus$}; 
  \draw[] (5.7, -2) -- +(0, -1) node[pos=0, circle, fill=black, inner sep=0pt,minimum size=3pt] {{}} node[pos=1, draw, fill=white] (gate6) {$RY(\theta/4)$}; 
  \draw[] (7.1, 0) -- +(0, -2) node[pos=0, circle, fill=black, inner sep=0pt,minimum size=3pt] {{}} node[pos=1] (gate7) {$\oplus$}; 
  \draw[] (8.5, -2) -- +(0, -1) node[pos=0, circle, fill=black, inner sep=0pt,minimum size=3pt] {{}} node[pos=1, draw, fill=white] (gate8) {$RY(-\theta/4)$}; 
  \draw[] (9.9, -1) -- +(0, -2) node[pos=0, circle, fill=black, inner sep=0pt,minimum size=3pt] {{}} node[pos=1, circle, fill=black, inner sep=0pt,minimum size=3pt] {{}}; 
  \draw[] (11.3, 0) -- +(0, -3) node[pos=0, circle, fill=black, inner sep=0pt,minimum size=3pt] {{}} node[pos=1, draw, fill=white] (gate10) {$RY(-\theta/4)$}; 
  \draw[] (12.700000000000001, 0) -- +(0, -2) node[pos=0, circle, fill=black, inner sep=0pt,minimum size=3pt] {{}} node[pos=1] (gate11) {$\oplus$}; 
  \draw[] (14.100000000000001, -2) -- +(0, -1) node[pos=0, circle, fill=black, inner sep=0pt,minimum size=3pt] {{}} node[pos=1, draw, fill=white] (gate12) {$RY(-\theta/4)$}; 
  \draw[] (15.500000000000002, 0) -- +(0, -2) node[pos=0, circle, fill=black, inner sep=0pt,minimum size=3pt] {{}} node[pos=1] (gate13) {$\oplus$}; 
  \draw[] (16.9, -2) -- +(0, -1) node[pos=0, circle, fill=black, inner sep=0pt,minimum size=3pt] {{}} node[pos=1, draw, fill=white] (gate14) {$RY(\theta/4)$}; 
  \node[draw, fill=white] (gate15) at (16.5, 0) {$X$}; 
  \draw[] (18.3, -1) -- +(0, -2) node[pos=0, circle, fill=black, inner sep=0pt,minimum size=3pt] {{}} node[pos=1, circle, fill=black, inner sep=0pt,minimum size=3pt] {{}}; 
  \draw[] (19.3, -3) -- +(0, 2) node[pos=0, circle, fill=black, inner sep=0pt,minimum size=3pt] {{}} node[pos=1] (gate17) {$\oplus$}; 
  \draw[] (20.3, -3) -- +(0, 1) node[pos=0, circle, fill=black, inner sep=0pt,minimum size=3pt] {{}} node[pos=1] (gate18) {$\oplus$}; 
  \draw[] (20.3, -1) -- +(0, 1) node[pos=0, circle, fill=black, inner sep=0pt,minimum size=3pt] {{}} node[pos=1] (gate19) {$\oplus$}; 
\begin{pgfonlayer}{bg}
  \draw (0, 0) -- (20.8, 0);
  \draw (0, -1) -- (20.8, -1);
  \draw (0, -2) -- (20.8, -2);
  \draw (0, -3) -- (20.8, -3);
\end{pgfonlayer}
\end{scope}
\end{tikzpicture}
\end{center}
Due to cancellations when expanding out the controlled $Y$ rotations, a decomposition in terms of only standard gates has only slightly higher depth ($16\to 18$) and requires less two-qubit gates ($18\to 14$) even if the controlled $Y$ rotation is a native operation:
\begin{center}
\begin{tikzpicture}[scale=0.7, transform shape]
\begin{scope}
  \node[draw, fill=white, minimum height=height("$A$")+3cm, text width=2.7cm-10mm, inner sep=2mm, align=center] (lhs) at (-2.1500000000000004, -1.5) {$$\mathrm{QNP_{PX}}(\theta)$$}; 
  \begin{pgfonlayer}{bg}
    \draw ($(lhs.west |- 0,0) + (-0.2,0)$) -- ($(lhs.east |- 0,0) + (0.2,0)$);
    \draw ($(lhs.west |- 0,-1) + (-0.2,0)$) -- ($(lhs.east |- 0,-1) + (0.2,0)$);
    \draw ($(lhs.west |- 0,-2) + (-0.2,0)$) -- ($(lhs.east |- 0,-2) + (0.2,0)$);
    \draw ($(lhs.west |- 0,-3) + (-0.2,0)$) -- ($(lhs.east |- 0,-3) + (0.2,0)$);
  \end{pgfonlayer}
  \node at (lhs.center-|-0.4,0) {{$=$}};  \draw[] (0.5, -1) -- +(0, 1) node[pos=0, circle, fill=black, inner sep=0pt,minimum size=3pt] {{}} node[pos=1] (gate0) {$\oplus$}; 
  \draw[] (1.5, -3) -- +(0, 2) node[pos=0, circle, fill=black, inner sep=0pt,minimum size=3pt] {{}} node[pos=1] (gate1) {$\oplus$}; 
  \draw[] (2.5, 0) -- +(0, -1) node[pos=0, circle, fill=black, inner sep=0pt,minimum size=3pt] {{}} node[pos=1, circle, fill=black, inner sep=0pt,minimum size=3pt] {{}}; 
  \node[draw, fill=white] (gate3) at (2.5, -3) {$H$}; 
  \draw[] (3.5, -3) -- +(0, 1) node[pos=0, circle, fill=black, inner sep=0pt,minimum size=3pt] {{}} node[pos=1] (gate4) {$\oplus$}; 
  \node[draw, fill=white] (gate5) at (4.9, -3) {$RY(-\theta/8)$}; 
  \node[draw, fill=white] (gate6) at (4.9, -2) {$RY(\theta/8)$}; 
  \draw[] (6.3, 0) -- +(0, -3) node[pos=0, circle, fill=black, inner sep=0pt,minimum size=3pt] {{}} node[pos=1, circle, fill=black, inner sep=0pt,minimum size=3pt] {{}}; 
  \node[draw, fill=white] (gate8) at (7.7, -3) {$RY(-\theta/8)$}; 
  \draw[] (7.3, 0) -- +(0, -2) node[pos=0, circle, fill=black, inner sep=0pt,minimum size=3pt] {{}} node[pos=1] (gate9) {$\oplus$}; 
  \node[draw, fill=white] (gate10) at (8.7, -2) {$RY(\theta/8)$}; 
  \draw[] (10.1, -1) -- +(0, -1) node[pos=0, circle, fill=black, inner sep=0pt,minimum size=3pt] {{}} node[pos=1] (gate11) {$\oplus$}; 
  \draw[] (11.1, -1) -- +(0, -2) node[pos=0, circle, fill=black, inner sep=0pt,minimum size=3pt] {{}} node[pos=1] (gate12) {$\oplus$}; 
  \node[draw, fill=white] (gate13) at (12.5, -3) {$RY(\theta/8)$}; 
  \node[draw, fill=white] (gate14) at (12.5, -2) {$RY(-\theta/8)$}; 
  \draw[] (13.9, 0) -- +(0, -2) node[pos=0, circle, fill=black, inner sep=0pt,minimum size=3pt] {{}} node[pos=1] (gate15) {$\oplus$}; 
  \draw[] (14.9, 0) -- +(0, -3) node[pos=0, circle, fill=black, inner sep=0pt,minimum size=3pt] {{}} node[pos=1, circle, fill=black, inner sep=0pt,minimum size=3pt] {{}}; 
  \node[draw, fill=white] (gate17) at (16.3, -2) {$RY(-\theta/8)$}; 
  \node[draw, fill=white] (gate18) at (16.3, -3) {$RY(\theta/8)$}; 
  \draw[] (17.7, -3) -- +(0, 1) node[pos=0, circle, fill=black, inner sep=0pt,minimum size=3pt] {{}} node[pos=1] (gate19) {$\oplus$}; 
  \draw[] (18.7, -1) -- +(0, -2) node[pos=0, circle, fill=black, inner sep=0pt,minimum size=3pt] {{}} node[pos=1] (gate20) {$\oplus$}; 
  \node[draw, fill=white] (gate21) at (19.7, -3) {$H$}; 
  \draw[] (20.7, -3) -- +(0, 2) node[pos=0, circle, fill=black, inner sep=0pt,minimum size=3pt] {{}} node[pos=1] (gate22) {$\oplus$}; 
  \draw[] (21.7, -1) -- +(0, 1) node[pos=0, circle, fill=black, inner sep=0pt,minimum size=3pt] {{}} node[pos=1] (gate23) {$\oplus$}; 
\begin{pgfonlayer}{bg}
  \draw (0, 0) -- (22.2, 0);
  \draw (0, -1) -- (22.2, -1);
  \draw (0, -2) -- (22.2, -2);
  \draw (0, -3) -- (22.2, -3);
\end{pgfonlayer}
\end{scope}
\begin{scope}[xshift=-22.2cm, yshift=-4.5cm]
\begin{pgfonlayer}{bg}
  \draw (22.2, 0) -- (22.2, 0);
  \draw (22.2, -1) -- (22.2, -1);
  \draw (22.2, -2) -- (22.2, -2);
  \draw (22.2, -3) -- (22.2, -3);
\end{pgfonlayer}
\end{scope}
\end{tikzpicture}
\end{center}


\subsection{Two orbital Givens rotation gate $\mathrm{QNP_{OR}}$}
We describe the construction of the Givens rotation gate in more detail.
A Givens rotation is generally a rotation in a two dimensional subspace of the form
\begin{equation}
\left [
\begin{array}{r}
\phi_{0}' (\vec r_1) \\
\phi_{1}' (\vec r_1) \\
\end{array}
\right ]
\coloneqq
\left [
\begin{array}{rr}
c & +s \\
-s & c \\
\end{array}
\right ]
\left [
\begin{array}{r}
\phi_{0} (\vec r_1) \\
\phi_{1} (\vec r_1) \\
\end{array}
\right ]
\end{equation}
where $c \coloneqq \cos(\theta/2)$ and $s \coloneqq \sin(\theta/2)$ for a continuous
parameter $\theta$.
Under the Jordan-Wigner mapping a Givens rotation between the orbital bases can be
implemented as as pair of parallel Givens gates as follows:
\begin{equation}
\begin{array}{l}
\Qcircuit @R=0.3em @C=0.3em @!R {
\lstick{|0_{\alpha}\rangle}
& \multigate{3}{\mathrm{QNP_{OR}}(\theta)}
 & \qw \\
\lstick{|0_{\beta}\rangle}
 & \ghost{\mathrm{QNP_{OR}}(\theta)}
 & \qw \\
\lstick{|1_{\alpha}\rangle}
 & \ghost{\mathrm{QNP_{OR}}(\theta)}
 & \qw \\
\lstick{|1_{\beta}\rangle}
 & \ghost{\mathrm{QNP_{OR}}(\theta)}
 & \qw \\
}
\end{array}
\qquad
\coloneqq
\qquad
\qquad
\begin{array}{l}
\Qcircuit @R=0.3em @C=0.3em @!R {
\lstick{|0_{\alpha}\rangle}
 & \gate{G(\theta)}
 & \qw
 & \qw \\
\lstick{|0_{\beta}\rangle}
 & \qw \qwx[1] \qwx[-1]
 & \gate{G(\theta)}
 & \qw \\
\lstick{|1_{\alpha}\rangle}
 & \gate{G(\theta)}
 & \qw \qwx[1] \qwx[-1]
 & \qw \\
\lstick{|1_{\beta}\rangle}
 & \qw
 & \gate{G(\theta)}
 & \qw \\
}
\end{array}
\end{equation}
In the two qubit space, the Givens rotation gate $G(\theta)$ has the action
\begin{equation}
\hat G (\theta)
\coloneqq
\left [
\begin{array}{rrrr}
1 & & & \\
 & c & +s & \\
 & -s & c & \\
& & & 1 \\
\end{array}
\right ]
\end{equation}
and can be decomposed into elementary gates as follows:
\begin{equation}
\raisebox{-0.075cm}{\Qcircuit @R=0.3em @C=0.3em @!R {
 & \multigate{1}{G(\theta)}
 & \qw \\
 & \ghost{G(\theta)}
 & \qw \\
}}
\raisebox{-0.35cm}{\;=\;}
\Qcircuit @R=0.3em @C=0.3em @!R {
 & \gate{H}
 & \ctrl{1}
 & \gate{R_{y} (\theta / 2)}
 & \ctrl{1}
 & \gate{H}
 & \qw \\
 & \gate{H}
 & \control \qw
 & \gate{R_{y} (-\theta / 2)}
 & \control \qw
 & \gate{H}
 & \qw \\
}
\raisebox{-0.35cm}{\;=\;}
\Qcircuit @R=0.3em @C=0.3em @!R {
& \targ
& \ctrl{1}
& \targ
& \qw
\\
& \ctrl{-1}
& \gate{R_{y} (\theta)}
& \ctrl{-1} 
& \qw
\\
}
\raisebox{-0.35cm}{\;=\;}
\Qcircuit @R=0.3em @C=0.3em @!R {
& \targ
& \qw
& \ctrl{1}
& \qw
& \ctrl{1}
& \targ
& \qw
\\
& \ctrl{-1}
& \gate{R_{y} (\theta/2)}
& \targ
& \gate{R_{y} (-\theta/2)}
& \targ 
& \ctrl{-1}
& \qw
\\
}
\end{equation}
In the 4-qubit Hilbert space, the two orbital Givens rotation gate $\mathrm{QNP_{OR}}(\theta)$ has the action
\begin{equation}
\mathrm{QNP_{OR}}(\theta) 
\coloneqq
\left [
\begin{array}{rrrrrrrrrrrrrrrr}
 1 & & & & & & & & & & & & & & &  \\
 & \color{brickred}{c} & & & \color{brickred}{+s} & & & & & & & & & & & \\
 & & \color{brickred}{c} & & & & & & \color{brickred}{-s} & & & & & & &  \\
 & & & \color{darkgreen}{c^2} & & & \color{darkgreen}{+cs} & & & \color{darkgreen}{+cs} & & & \color{darkgreen}{+s^2} & & &  \\
 & \color{brickred}{-s} & & & \color{brickred}{c} & & & & & & & & & & &  \\
 & & & & & 1 & & & & & & & & & &  \\
 & & & \color{darkgreen}{-cs} & & & \color{darkgreen}{c^2} & & & \color{darkgreen}{-s^2} & & & \color{darkgreen}{+cs} & & &  \\
 & & & & & & & \color{darkblue}{c} & & & & & & \color{darkblue}{+s} & &  \\
 & & \color{brickred}{-s} & & & & & & \color{brickred}{c} & & & & & & &  \\
 & & & \color{darkgreen}{-cs} & & & \color{darkgreen}{-s^2} & & & \color{darkgreen}{c^2} & & & \color{darkgreen}{+cs} & & &  \\
 & & & & & & & & & & 1 & & & & &  \\
 & & & & & & & & & & & \color{darkblue}{c} & & & \color{darkblue}{+s} &  \\
 & & & \color{darkgreen}{+s^2} & & & \color{darkgreen}{-cs} & & & \color{darkgreen}{-cs} & & & \color{darkgreen}{c^2} & & &  \\
 & & & & & & & \color{darkblue}{-s} & & & & & & \color{darkblue}{c} & &  \\
 & & & & & & & & & & & \color{darkblue}{-s} & & & \color{darkblue}{c} &  \\
 & & & & & & & & & & & & & & & 1 \\
\end{array}
\right ] .
\end{equation}
When applied to two neighboring spatial orbitals, this gate also preserves all three
quantum numbers and has the following decomposition with gate depth $5$ and just $4$ CNOT gates:

\begin{center}
\begin{tikzpicture}[scale=0.7, transform shape]
\begin{scope}
  \node[draw, fill=white, minimum height=height("$A$")+3cm, text width=2.8cm-10mm, inner sep=2mm, align=center] (lhs) at (-2.2, -1.5) {$$\mathrm{QNP_{OR}}(\theta)$$}; 
  \begin{pgfonlayer}{bg}
    \draw ($(lhs.west |- 0,0) + (-0.2,0)$) -- ($(lhs.east |- 0,0) + (0.2,0)$);
    \draw ($(lhs.west |- 0,-1) + (-0.2,0)$) -- ($(lhs.east |- 0,-1) + (0.2,0)$);
    \draw ($(lhs.west |- 0,-2) + (-0.2,0)$) -- ($(lhs.east |- 0,-2) + (0.2,0)$);
    \draw ($(lhs.west |- 0,-3) + (-0.2,0)$) -- ($(lhs.east |- 0,-3) + (0.2,0)$);
  \end{pgfonlayer}
  \node at (lhs.center-|-0.4,0) {{$=$}};  \node[draw, fill=white] (gate0) at (0.5, 0) {$H$}; 
  \node[draw, fill=white] (gate1) at (0.5, -1) {$H$}; 
  \draw[] (1.5, 0) -- +(0, -2) node[pos=0, circle, fill=black, inner sep=0pt,minimum size=3pt] {{}} node[pos=1] (gate2) {$\oplus$}; 
  \draw[] (2.5, -1) -- +(0, -2) node[pos=0, circle, fill=black, inner sep=0pt,minimum size=3pt] {{}} node[pos=1] (gate3) {$\oplus$}; 
  \node[draw, fill=white] (gate4) at (2.9, 0) {$RY(\theta/2)$}; 
  \node[draw, fill=white] (gate5) at (3.9, -2) {$RY(\theta/2)$}; 
  \node[draw, fill=white] (gate6) at (3.9, -1) {$RY(\theta/2)$}; 
  \node[draw, fill=white] (gate7) at (3.9, -3) {$RY(\theta/2)$}; 
  \draw[] (5.3, 0) -- +(0, -2) node[pos=0, circle, fill=black, inner sep=0pt,minimum size=3pt] {{}} node[pos=1] (gate8) {$\oplus$}; 
  \node[draw, fill=white] (gate9) at (6.3, 0) {$H$}; 
  \draw[] (6.3, -1) -- +(0, -2) node[pos=0, circle, fill=black, inner sep=0pt,minimum size=3pt] {{}} node[pos=1] (gate10) {$\oplus$}; 
  \node[draw, fill=white] (gate11) at (7.3, -1) {$H$}; 
\begin{pgfonlayer}{bg}
  \draw (0, 0) -- (7.8, 0);
  \draw (0, -1) -- (7.8, -1);
  \draw (0, -2) -- (7.8, -2);
  \draw (0, -3) -- (7.8, -3);
\end{pgfonlayer}
\end{scope}
\end{tikzpicture}
\end{center}

An alternative decomposition into controlled $Y$ rotation gates is: 
\begin{center}
\begin{tikzpicture}[scale=0.7, transform shape]
\begin{scope}
  \node[draw, fill=white, minimum height=height("$A$")+3cm, text width=2.8cm-10mm, inner sep=2mm, align=center] (lhs) at (-2.2, -1.5) {$$\mathrm{QNP_{OR}}(\theta)$$}; 
  \begin{pgfonlayer}{bg}
    \draw ($(lhs.west |- 0,0) + (-0.2,0)$) -- ($(lhs.east |- 0,0) + (0.2,0)$);
    \draw ($(lhs.west |- 0,-1) + (-0.2,0)$) -- ($(lhs.east |- 0,-1) + (0.2,0)$);
    \draw ($(lhs.west |- 0,-2) + (-0.2,0)$) -- ($(lhs.east |- 0,-2) + (0.2,0)$);
    \draw ($(lhs.west |- 0,-3) + (-0.2,0)$) -- ($(lhs.east |- 0,-3) + (0.2,0)$);
  \end{pgfonlayer}
  \node at (lhs.center-|-0.4,0) {{$=$}};  \draw[] (0.5, -2) -- +(0, 2) node[pos=0, circle, fill=black, inner sep=0pt,minimum size=3pt] {{}} node[pos=1] (gate0) {$\oplus$}; 
  \draw[] (1.9, 0) -- +(0, -2) node[pos=0, circle, fill=black, inner sep=0pt,minimum size=3pt] {{}} node[pos=1, draw, fill=white] (gate1) {$RY(\theta)$}; 
  \draw[] (3.3, -2) -- +(0, 2) node[pos=0, circle, fill=black, inner sep=0pt,minimum size=3pt] {{}} node[pos=1] (gate2) {$\oplus$}; 
  \draw[] (4.3, -3) -- +(0, 2) node[pos=0, circle, fill=black, inner sep=0pt,minimum size=3pt] {{}} node[pos=1] (gate3) {$\oplus$}; 
  \draw[] (5.7, -1) -- +(0, -2) node[pos=0, circle, fill=black, inner sep=0pt,minimum size=3pt] {{}} node[pos=1, draw, fill=white] (gate4) {$RY(\theta)$}; 
  \draw[] (7.1, -3) -- +(0, 2) node[pos=0, circle, fill=black, inner sep=0pt,minimum size=3pt] {{}} node[pos=1] (gate5) {$\oplus$}; 
\begin{pgfonlayer}{bg}
  \draw (0, 0) -- (7.6, 0);
  \draw (0, -1) -- (7.6, -1);
  \draw (0, -2) -- (7.6, -2);
  \draw (0, -3) -- (7.6, -3);
\end{pgfonlayer}
\end{scope}
\end{tikzpicture}
\end{center}
Note that if these gates are native, the $2$-qubit gate count is raised to $6$ while reducing the depth to $3$.
Expanding out the controlled $Y$ rotations yields:
\begin{center}
\begin{tikzpicture}[scale=0.7, transform shape]
\begin{scope}
  \node[draw, fill=white, minimum height=height("$A$")+3cm, text width=2.8cm-10mm, inner sep=2mm, align=center] (lhs) at (-2.2, -1.5) {$$\mathrm{QNP_{OR}}(\theta)$$}; 
  \begin{pgfonlayer}{bg}
    \draw ($(lhs.west |- 0,0) + (-0.2,0)$) -- ($(lhs.east |- 0,0) + (0.2,0)$);
    \draw ($(lhs.west |- 0,-1) + (-0.2,0)$) -- ($(lhs.east |- 0,-1) + (0.2,0)$);
    \draw ($(lhs.west |- 0,-2) + (-0.2,0)$) -- ($(lhs.east |- 0,-2) + (0.2,0)$);
    \draw ($(lhs.west |- 0,-3) + (-0.2,0)$) -- ($(lhs.east |- 0,-3) + (0.2,0)$);
  \end{pgfonlayer}
  \node at (lhs.center-|-0.4,0) {{$=$}};  \draw[] (0.5, -2) -- +(0, 2) node[pos=0, circle, fill=black, inner sep=0pt,minimum size=3pt] {{}} node[pos=1] (gate0) {$\oplus$}; 
  \draw[] (1.5, -3) -- +(0, 2) node[pos=0, circle, fill=black, inner sep=0pt,minimum size=3pt] {{}} node[pos=1] (gate1) {$\oplus$}; 
  \node[draw, fill=white] (gate2) at (2.9, -2) {$RY(\theta/2)$}; 
  \node[draw, fill=white] (gate3) at (2.9, -3) {$RY(\theta/2)$}; 
  \draw[] (4.3, 0) -- +(0, -2) node[pos=0, circle, fill=black, inner sep=0pt,minimum size=3pt] {{}} node[pos=1] (gate4) {$\oplus$}; 
  \draw[] (5.3, -1) -- +(0, -2) node[pos=0, circle, fill=black, inner sep=0pt,minimum size=3pt] {{}} node[pos=1] (gate5) {$\oplus$}; 
  \node[draw, fill=white] (gate6) at (6.7, -2) {$RY(-\theta/2)$}; 
  \node[draw, fill=white] (gate7) at (6.7, -3) {$RY(-\theta/2)$}; 
  \draw[] (8.1, 0) -- +(0, -2) node[pos=0, circle, fill=black, inner sep=0pt,minimum size=3pt] {{}} node[pos=1] (gate8) {$\oplus$}; 
  \draw[] (9.1, -2) -- +(0, 2) node[pos=0, circle, fill=black, inner sep=0pt,minimum size=3pt] {{}} node[pos=1] (gate9) {$\oplus$}; 
  \draw[] (10.1, -1) -- +(0, -2) node[pos=0, circle, fill=black, inner sep=0pt,minimum size=3pt] {{}} node[pos=1] (gate10) {$\oplus$}; 
  \draw[] (11.1, -3) -- +(0, 2) node[pos=0, circle, fill=black, inner sep=0pt,minimum size=3pt] {{}} node[pos=1] (gate11) {$\oplus$}; 
\begin{pgfonlayer}{bg}
  \draw (0, 0) -- (11.6, 0);
  \draw (0, -1) -- (11.6, -1);
  \draw (0, -2) -- (11.6, -2);
  \draw (0, -3) -- (11.6, -3);
\end{pgfonlayer}
\end{scope}
\end{tikzpicture}
\end{center}
Of course the two Givens rotations of the gate commute and can be performed at the same time, giving a gate depth of $6$ and a CNOT count of $8$.
Depending on the preceding and following gates, it may furthermore be favourable to substitute the doubled CNOT gates with a single CNOT and a SWAP gate.

\subsection{Single particle and single hole gates}
In the following we abbreviate $R\coloneqq RY(\theta/8)$. The $\mathrm{QNP_{A1B0}}(\theta)$ gate can be decomposed as follows:
\begin{center}
\begin{tikzpicture}[scale=0.7, transform shape]
\begin{scope}
  \node[draw, fill=white, minimum height=height("$A$")+3cm, text width=2.9cm-10mm, inner sep=2mm, align=center] (lhs) at (-2.25, -1.5) {$$\mathrm{QNP_{A1B0}}(\theta)$$}; 
  \begin{pgfonlayer}{bg}
    \draw ($(lhs.west |- 0,0) + (-0.2,0)$) -- ($(lhs.east |- 0,0) + (0.2,0)$);
    \draw ($(lhs.west |- 0,-1) + (-0.2,0)$) -- ($(lhs.east |- 0,-1) + (0.2,0)$);
    \draw ($(lhs.west |- 0,-2) + (-0.2,0)$) -- ($(lhs.east |- 0,-2) + (0.2,0)$);
    \draw ($(lhs.west |- 0,-3) + (-0.2,0)$) -- ($(lhs.east |- 0,-3) + (0.2,0)$);
  \end{pgfonlayer}
  \node at (lhs.center-|-0.4,0) {{$=$}};  \draw[] (0.5, -2) -- +(0, 2) node[pos=0, circle, fill=black, inner sep=0pt,minimum size=3pt] {{}} node[pos=1] (gate0) {$\oplus$}; 
  \node[draw, fill=white, minimum height=height("$A$")+0cm, text width=1cm, align=center] (gate1) at (1.5, -2.0) {$R$}; 
  \draw[] (2.5, -3) -- +(0, 1) node[pos=0, circle, fill=black, inner sep=0pt,minimum size=3pt] {{}} node[pos=1] (gate2) {$\oplus$}; 
  \node[draw, fill=white, minimum height=height("$A$")+0cm, text width=1cm, align=center] (gate3) at (3.5, -2.0) {$R$}; 
  \draw[] (4.5, -1) -- +(0, -1) node[pos=0, circle, fill=black, inner sep=0pt,minimum size=3pt] {{}} node[pos=1] (gate4) {$\oplus$}; 
  \node[draw, fill=white, minimum height=height("$A$")+0cm, text width=1cm, align=center] (gate5) at (5.5, -2.0) {$R$}; 
  \draw[] (5.5, -1) -- +(0, 1) node[pos=0, circle, fill=black, inner sep=0pt,minimum size=3pt] {{}} node[pos=1, circle, fill=black, inner sep=0pt,minimum size=3pt] {{}}; 
  \draw[] (6.5, -3) -- +(0, 1) node[pos=0, circle, fill=black, inner sep=0pt,minimum size=3pt] {{}} node[pos=1] (gate7) {$\oplus$}; 
  \node[draw, fill=white, minimum height=height("$A$")+0cm, text width=1cm, align=center] (gate8) at (7.5, -2.0) {$R$}; 
  \draw[] (8.5, 0) -- +(0, -2) node[pos=0, circle, fill=black, inner sep=0pt,minimum size=3pt] {{}} node[pos=1, circle, fill=black, inner sep=0pt,minimum size=3pt] {{}}; 
  \node[draw, fill=white, minimum height=height("$A$")+0cm, text width=1cm, align=center] (gate10) at (9.5, -2.0) {$R^\dagger$}; 
  \draw[] (10.5, -3) -- +(0, 1) node[pos=0, circle, fill=black, inner sep=0pt,minimum size=3pt] {{}} node[pos=1] (gate11) {$\oplus$}; 
  \node[draw, fill=white, minimum height=height("$A$")+0cm, text width=1cm, align=center] (gate12) at (11.5, -2.0) {$R^\dagger$}; 
  \draw[] (12.5, -1) -- +(0, -1) node[pos=0, circle, fill=black, inner sep=0pt,minimum size=3pt] {{}} node[pos=1] (gate13) {$\oplus$}; 
  \node[draw, fill=white, minimum height=height("$A$")+0cm, text width=1cm, align=center] (gate14) at (13.5, -2.0) {$R^\dagger$}; 
  \draw[] (14.5, -3) -- +(0, 1) node[pos=0, circle, fill=black, inner sep=0pt,minimum size=3pt] {{}} node[pos=1] (gate15) {$\oplus$}; 
  \node[draw, fill=white, minimum height=height("$A$")+0cm, text width=1cm, align=center] (gate16) at (15.5, -2.0) {$R^\dagger$}; 
  \draw[] (16.5, 0) -- +(0, -2) node[pos=0, circle, fill=black, inner sep=0pt,minimum size=3pt] {{}} node[pos=1, circle, fill=black, inner sep=0pt,minimum size=3pt] {{}}; 
  \draw[] (17.5, -2) -- +(0, 2) node[pos=0, circle, fill=black, inner sep=0pt,minimum size=3pt] {{}} node[pos=1] (gate18) {$\oplus$}; 
\begin{pgfonlayer}{bg}
  \draw (0, 0) -- (18.0, 0);
  \draw (0, -1) -- (18.0, -1);
  \draw (0, -2) -- (18.0, -2);
  \draw (0, -3) -- (18.0, -3);
\end{pgfonlayer}
\end{scope}
\end{tikzpicture}
\end{center}


The $\mathrm{QNP_{A0B1}}(\theta)$ gate can be decomposed as follows:
\begin{center}
\begin{tikzpicture}[scale=0.7, transform shape]
\begin{scope}
  \node[draw, fill=white, minimum height=height("$A$")+3cm, text width=2.9cm-10mm, inner sep=2mm, align=center] (lhs) at (-2.25, -1.5) {$$\mathrm{QNP_{A0B1}}(\theta)$$}; 
  \begin{pgfonlayer}{bg}
    \draw ($(lhs.west |- 0,0) + (-0.2,0)$) -- ($(lhs.east |- 0,0) + (0.2,0)$);
    \draw ($(lhs.west |- 0,-1) + (-0.2,0)$) -- ($(lhs.east |- 0,-1) + (0.2,0)$);
    \draw ($(lhs.west |- 0,-2) + (-0.2,0)$) -- ($(lhs.east |- 0,-2) + (0.2,0)$);
    \draw ($(lhs.west |- 0,-3) + (-0.2,0)$) -- ($(lhs.east |- 0,-3) + (0.2,0)$);
  \end{pgfonlayer}
  \node at (lhs.center-|-0.4,0) {{$=$}};  \draw[] (0.5, -3) -- +(0, 2) node[pos=0, circle, fill=black, inner sep=0pt,minimum size=3pt] {{}} node[pos=1] (gate0) {$\oplus$}; 
  \node[draw, fill=white, minimum height=height("$A$")+0cm, text width=1cm, align=center] (gate1) at (1.5, -3.0) {$R$}; 
  \draw[] (2.5, -2) -- +(0, -1) node[pos=0, circle, fill=black, inner sep=0pt,minimum size=3pt] {{}} node[pos=1] (gate2) {$\oplus$}; 
  \node[draw, fill=white, minimum height=height("$A$")+0cm, text width=1cm, align=center] (gate3) at (3.5, -3.0) {$R$}; 
  \draw[] (4.5, 0) -- +(0, -3) node[pos=0, circle, fill=black, inner sep=0pt,minimum size=3pt] {{}} node[pos=1] (gate4) {$\oplus$}; 
  \node[draw, fill=white, minimum height=height("$A$")+0cm, text width=1cm, align=center] (gate5) at (5.5, -3.0) {$R$}; 
  \draw[] (5.5, 0) -- +(0, -1) node[pos=0, circle, fill=black, inner sep=0pt,minimum size=3pt] {{}} node[pos=1, circle, fill=black, inner sep=0pt,minimum size=3pt] {{}}; 
  \draw[] (6.5, -2) -- +(0, -1) node[pos=0, circle, fill=black, inner sep=0pt,minimum size=3pt] {{}} node[pos=1] (gate7) {$\oplus$}; 
  \node[draw, fill=white, minimum height=height("$A$")+0cm, text width=1cm, align=center] (gate8) at (7.5, -3.0) {$R$}; 
  \draw[] (8.5, -1) -- +(0, -2) node[pos=0, circle, fill=black, inner sep=0pt,minimum size=3pt] {{}} node[pos=1, circle, fill=black, inner sep=0pt,minimum size=3pt] {{}}; 
  \node[draw, fill=white, minimum height=height("$A$")+0cm, text width=1cm, align=center] (gate10) at (9.5, -3.0) {$R^\dagger$}; 
  \draw[] (10.5, -2) -- +(0, -1) node[pos=0, circle, fill=black, inner sep=0pt,minimum size=3pt] {{}} node[pos=1] (gate11) {$\oplus$}; 
  \node[draw, fill=white, minimum height=height("$A$")+0cm, text width=1cm, align=center] (gate12) at (11.5, -3.0) {$R^\dagger$}; 
  \draw[] (12.5, 0) -- +(0, -3) node[pos=0, circle, fill=black, inner sep=0pt,minimum size=3pt] {{}} node[pos=1] (gate13) {$\oplus$}; 
  \node[draw, fill=white, minimum height=height("$A$")+0cm, text width=1cm, align=center] (gate14) at (13.5, -3.0) {$R^\dagger$}; 
  \draw[] (14.5, -2) -- +(0, -1) node[pos=0, circle, fill=black, inner sep=0pt,minimum size=3pt] {{}} node[pos=1] (gate15) {$\oplus$}; 
  \node[draw, fill=white, minimum height=height("$A$")+0cm, text width=1cm, align=center] (gate16) at (15.5, -3.0) {$R^\dagger$}; 
  \draw[] (16.5, -1) -- +(0, -2) node[pos=0, circle, fill=black, inner sep=0pt,minimum size=3pt] {{}} node[pos=1, circle, fill=black, inner sep=0pt,minimum size=3pt] {{}}; 
  \draw[] (17.5, -3) -- +(0, 2) node[pos=0, circle, fill=black, inner sep=0pt,minimum size=3pt] {{}} node[pos=1] (gate18) {$\oplus$}; 
\begin{pgfonlayer}{bg}
  \draw (0, 0) -- (18.0, 0);
  \draw (0, -1) -- (18.0, -1);
  \draw (0, -2) -- (18.0, -2);
  \draw (0, -3) -- (18.0, -3);
\end{pgfonlayer}
\end{scope}
\end{tikzpicture}
\end{center}


The $\mathrm{QNP_{A2B1}}(\theta)$ gate can be decomposed as follows:
\begin{center}
\begin{tikzpicture}[scale=0.7, transform shape]
\begin{scope}
  \node[draw, fill=white, minimum height=height("$A$")+3cm, text width=2.9cm-10mm, inner sep=2mm, align=center] (lhs) at (-2.25, -1.5) {$$\mathrm{QNP_{A2B1}}(\theta)$$}; 
  \begin{pgfonlayer}{bg}
    \draw ($(lhs.west |- 0,0) + (-0.2,0)$) -- ($(lhs.east |- 0,0) + (0.2,0)$);
    \draw ($(lhs.west |- 0,-1) + (-0.2,0)$) -- ($(lhs.east |- 0,-1) + (0.2,0)$);
    \draw ($(lhs.west |- 0,-2) + (-0.2,0)$) -- ($(lhs.east |- 0,-2) + (0.2,0)$);
    \draw ($(lhs.west |- 0,-3) + (-0.2,0)$) -- ($(lhs.east |- 0,-3) + (0.2,0)$);
  \end{pgfonlayer}
  \node at (lhs.center-|-0.4,0) {{$=$}};  \draw[] (0.5, -3) -- +(0, 2) node[pos=0, circle, fill=black, inner sep=0pt,minimum size=3pt] {{}} node[pos=1] (gate0) {$\oplus$}; 
  \node[draw, fill=white, minimum height=height("$A$")+0cm, text width=1cm, align=center] (gate1) at (1.5, -3.0) {$R$}; 
  \draw[] (2.5, -2) -- +(0, -1) node[pos=0, circle, fill=black, inner sep=0pt,minimum size=3pt] {{}} node[pos=1] (gate2) {$\oplus$}; 
  \node[draw, fill=white, minimum height=height("$A$")+0cm, text width=1cm, align=center] (gate3) at (3.5, -3.0) {$R^\dagger$}; 
  \draw[] (4.5, 0) -- +(0, -3) node[pos=0, circle, fill=black, inner sep=0pt,minimum size=3pt] {{}} node[pos=1] (gate4) {$\oplus$}; 
  \node[draw, fill=white, minimum height=height("$A$")+0cm, text width=1cm, align=center] (gate5) at (5.5, -3.0) {$R$}; 
  \draw[] (6.5, -2) -- +(0, -1) node[pos=0, circle, fill=black, inner sep=0pt,minimum size=3pt] {{}} node[pos=1] (gate6) {$\oplus$}; 
  \node[draw, fill=white, minimum height=height("$A$")+0cm, text width=1cm, align=center] (gate7) at (7.5, -3.0) {$R^\dagger$}; 
  \draw[] (8.5, -1) -- +(0, -2) node[pos=0, circle, fill=black, inner sep=0pt,minimum size=3pt] {{}} node[pos=1, circle, fill=black, inner sep=0pt,minimum size=3pt] {{}}; 
  \draw[] (9.5, 0) -- +(0, -1) node[pos=0, circle, fill=black, inner sep=0pt,minimum size=3pt] {{}} node[pos=1, circle, fill=black, inner sep=0pt,minimum size=3pt] {{}}; 
  \node[draw, fill=white, minimum height=height("$A$")+0cm, text width=1cm, align=center] (gate10) at (9.5, -3.0) {$R$}; 
  \draw[] (10.5, -2) -- +(0, -1) node[pos=0, circle, fill=black, inner sep=0pt,minimum size=3pt] {{}} node[pos=1] (gate11) {$\oplus$}; 
  \node[draw, fill=white, minimum height=height("$A$")+0cm, text width=1cm, align=center] (gate12) at (11.5, -3.0) {$R^\dagger$}; 
  \draw[] (12.5, 0) -- +(0, -3) node[pos=0, circle, fill=black, inner sep=0pt,minimum size=3pt] {{}} node[pos=1] (gate13) {$\oplus$}; 
  \node[draw, fill=white, minimum height=height("$A$")+0cm, text width=1cm, align=center] (gate14) at (13.5, -3.0) {$R$}; 
  \draw[] (14.5, -2) -- +(0, -1) node[pos=0, circle, fill=black, inner sep=0pt,minimum size=3pt] {{}} node[pos=1] (gate15) {$\oplus$}; 
  \node[draw, fill=white, minimum height=height("$A$")+0cm, text width=1cm, align=center] (gate16) at (15.5, -3.0) {$R^\dagger$}; 
  \draw[] (16.5, -1) -- +(0, -2) node[pos=0, circle, fill=black, inner sep=0pt,minimum size=3pt] {{}} node[pos=1, circle, fill=black, inner sep=0pt,minimum size=3pt] {{}}; 
  \draw[] (17.5, -3) -- +(0, 2) node[pos=0, circle, fill=black, inner sep=0pt,minimum size=3pt] {{}} node[pos=1] (gate18) {$\oplus$}; 
\begin{pgfonlayer}{bg}
  \draw (0, 0) -- (18.0, 0);
  \draw (0, -1) -- (18.0, -1);
  \draw (0, -2) -- (18.0, -2);
  \draw (0, -3) -- (18.0, -3);
\end{pgfonlayer}
\end{scope}
\end{tikzpicture}
\end{center}


Finally, the $\mathrm{QNP_{A1B2}}(\theta)$ gate can be decomposed as follows:
\begin{center}
\begin{tikzpicture}[scale=0.7, transform shape]
\begin{scope}
  \node[draw, fill=white, minimum height=height("$A$")+3cm, text width=2.9cm-10mm, inner sep=2mm, align=center] (lhs) at (-2.25, -1.5) {$$\mathrm{QNP_{A1B2}}(\theta)$$}; 
  \begin{pgfonlayer}{bg}
    \draw ($(lhs.west |- 0,0) + (-0.2,0)$) -- ($(lhs.east |- 0,0) + (0.2,0)$);
    \draw ($(lhs.west |- 0,-1) + (-0.2,0)$) -- ($(lhs.east |- 0,-1) + (0.2,0)$);
    \draw ($(lhs.west |- 0,-2) + (-0.2,0)$) -- ($(lhs.east |- 0,-2) + (0.2,0)$);
    \draw ($(lhs.west |- 0,-3) + (-0.2,0)$) -- ($(lhs.east |- 0,-3) + (0.2,0)$);
  \end{pgfonlayer}
  \node at (lhs.center-|-0.4,0) {{$=$}};  \draw[] (0.5, -2) -- +(0, 2) node[pos=0, circle, fill=black, inner sep=0pt,minimum size=3pt] {{}} node[pos=1] (gate0) {$\oplus$}; 
  \node[draw, fill=white, minimum height=height("$A$")+0cm, text width=1cm, align=center] (gate1) at (1.5, -2.0) {$R$}; 
  \draw[] (2.5, -3) -- +(0, 1) node[pos=0, circle, fill=black, inner sep=0pt,minimum size=3pt] {{}} node[pos=1] (gate2) {$\oplus$}; 
  \node[draw, fill=white, minimum height=height("$A$")+0cm, text width=1cm, align=center] (gate3) at (3.5, -2.0) {$R^\dagger$}; 
  \draw[] (4.5, -1) -- +(0, -1) node[pos=0, circle, fill=black, inner sep=0pt,minimum size=3pt] {{}} node[pos=1] (gate4) {$\oplus$}; 
  \node[draw, fill=white, minimum height=height("$A$")+0cm, text width=1cm, align=center] (gate5) at (5.5, -2.0) {$R$}; 
  \draw[] (6.5, -3) -- +(0, 1) node[pos=0, circle, fill=black, inner sep=0pt,minimum size=3pt] {{}} node[pos=1] (gate6) {$\oplus$}; 
  \node[draw, fill=white, minimum height=height("$A$")+0cm, text width=1cm, align=center] (gate7) at (7.5, -2.0) {$R^\dagger$}; 
  \draw[] (8.5, 0) -- +(0, -2) node[pos=0, circle, fill=black, inner sep=0pt,minimum size=3pt] {{}} node[pos=1, circle, fill=black, inner sep=0pt,minimum size=3pt] {{}}; 
  \draw[] (9.5, -1) -- +(0, 1) node[pos=0, circle, fill=black, inner sep=0pt,minimum size=3pt] {{}} node[pos=1, circle, fill=black, inner sep=0pt,minimum size=3pt] {{}}; 
  \node[draw, fill=white, minimum height=height("$A$")+0cm, text width=1cm, align=center] (gate10) at (9.5, -2.0) {$R$}; 
  \draw[] (10.5, -3) -- +(0, 1) node[pos=0, circle, fill=black, inner sep=0pt,minimum size=3pt] {{}} node[pos=1] (gate11) {$\oplus$}; 
  \node[draw, fill=white, minimum height=height("$A$")+0cm, text width=1cm, align=center] (gate12) at (11.5, -2.0) {$R^\dagger$}; 
  \draw[] (12.5, -1) -- +(0, -1) node[pos=0, circle, fill=black, inner sep=0pt,minimum size=3pt] {{}} node[pos=1] (gate13) {$\oplus$}; 
  \node[draw, fill=white, minimum height=height("$A$")+0cm, text width=1cm, align=center] (gate14) at (13.5, -2.0) {$R$}; 
  \draw[] (14.5, -3) -- +(0, 1) node[pos=0, circle, fill=black, inner sep=0pt,minimum size=3pt] {{}} node[pos=1] (gate15) {$\oplus$}; 
  \node[draw, fill=white, minimum height=height("$A$")+0cm, text width=1cm, align=center] (gate16) at (15.5, -2.0) {$R^\dagger$}; 
  \draw[] (16.5, 0) -- +(0, -2) node[pos=0, circle, fill=black, inner sep=0pt,minimum size=3pt] {{}} node[pos=1, circle, fill=black, inner sep=0pt,minimum size=3pt] {{}}; 
  \draw[] (17.5, -2) -- +(0, 2) node[pos=0, circle, fill=black, inner sep=0pt,minimum size=3pt] {{}} node[pos=1] (gate18) {$\oplus$}; 
\begin{pgfonlayer}{bg}
  \draw (0, 0) -- (18.0, 0);
  \draw (0, -1) -- (18.0, -1);
  \draw (0, -2) -- (18.0, -2);
  \draw (0, -3) -- (18.0, -3);
\end{pgfonlayer}
\end{scope}
\end{tikzpicture}
\end{center}


\subsection{Pair breaking gates}
For the pair breaking gates we present decompositions into standard gates and controlled $Y$ rotations.
The pair break low gate $\mathrm{QNP_{PBL}}$ has the decomposion:
\begin{center}
\begin{tikzpicture}[scale=0.7, transform shape]
\begin{scope}
  \node[draw, fill=white, minimum height=height("$A$")+3cm, text width=2.8cm-10mm, inner sep=2mm, align=center] (lhs) at (-2.2, -1.5) {$$\mathrm{QNP_{PBL}}(\theta)$$}; 
  \begin{pgfonlayer}{bg}
    \draw ($(lhs.west |- 0,0) + (-0.2,0)$) -- ($(lhs.east |- 0,0) + (0.2,0)$);
    \draw ($(lhs.west |- 0,-1) + (-0.2,0)$) -- ($(lhs.east |- 0,-1) + (0.2,0)$);
    \draw ($(lhs.west |- 0,-2) + (-0.2,0)$) -- ($(lhs.east |- 0,-2) + (0.2,0)$);
    \draw ($(lhs.west |- 0,-3) + (-0.2,0)$) -- ($(lhs.east |- 0,-3) + (0.2,0)$);
  \end{pgfonlayer}
  \node at (lhs.center-|-0.4,0) {{$=$}};  \draw[] (0.5, -3) -- +(0, 3) node[pos=0, circle, fill=black, inner sep=0pt,minimum size=3pt] {{}} node[pos=1] (gate0) {$\oplus$}; 
  \draw[] (1.5, -2) -- +(0, 1) node[pos=0, circle, fill=black, inner sep=0pt,minimum size=3pt] {{}} node[pos=1] (gate1) {$\oplus$}; 
  \node[draw, fill=white] (gate2) at (1.5, 0) {$X$}; 
  \draw[] (2.5, -2) -- +(0, -1) node[pos=0, circle, fill=black, inner sep=0pt,minimum size=3pt] {{}} node[pos=1] (gate3) {$\oplus$}; 
  \draw[] (3.9, 0) -- +(0, -2) node[pos=0, circle, fill=black, inner sep=0pt,minimum size=3pt] {{}} node[pos=1, draw, fill=white] (gate4) {$RY(\pi/8)$}; 
  \draw[] (5.3, 0) -- +(0, -1) node[pos=0, circle, fill=black, inner sep=0pt,minimum size=3pt] {{}} node[pos=1] (gate5) {$\oplus$}; 
  \draw[] (6.7, -1) -- +(0, -1) node[pos=0, circle, fill=black, inner sep=0pt,minimum size=3pt] {{}} node[pos=1, draw, fill=white] (gate6) {$RY(\pi/8)$}; 
  \draw[] (8.1, 0) -- +(0, -1) node[pos=0, circle, fill=black, inner sep=0pt,minimum size=3pt] {{}} node[pos=1] (gate7) {$\oplus$}; 
  \draw[] (9.5, -1) -- +(0, -1) node[pos=0, circle, fill=black, inner sep=0pt,minimum size=3pt] {{}} node[pos=1, draw, fill=white] (gate8) {$RY(-\pi/8)$}; 
  \draw[] (10.9, -3) -- +(0, 1) node[pos=0, circle, fill=black, inner sep=0pt,minimum size=3pt] {{}} node[pos=1, circle, fill=black, inner sep=0pt,minimum size=3pt] {{}}; 
  \draw[] (12.3, 0) -- +(0, -2) node[pos=0, circle, fill=black, inner sep=0pt,minimum size=3pt] {{}} node[pos=1, draw, fill=white] (gate10) {$RY(-\pi/8)$}; 
  \draw[] (13.700000000000001, 0) -- +(0, -1) node[pos=0, circle, fill=black, inner sep=0pt,minimum size=3pt] {{}} node[pos=1] (gate11) {$\oplus$}; 
  \draw[] (15.100000000000001, -1) -- +(0, -1) node[pos=0, circle, fill=black, inner sep=0pt,minimum size=3pt] {{}} node[pos=1, draw, fill=white] (gate12) {$RY(-\pi/8)$}; 
  \draw[] (16.5, 0) -- +(0, -1) node[pos=0, circle, fill=black, inner sep=0pt,minimum size=3pt] {{}} node[pos=1] (gate13) {$\oplus$}; 
  \draw[] (17.9, -1) -- +(0, -1) node[pos=0, circle, fill=black, inner sep=0pt,minimum size=3pt] {{}} node[pos=1, draw, fill=white] (gate14) {$RY(\pi/8)$}; 
\begin{pgfonlayer}{bg}
  \draw (0, 0) -- (18.8, 0);
  \draw (0, -1) -- (18.8, -1);
  \draw (0, -2) -- (18.8, -2);
  \draw (0, -3) -- (18.8, -3);
\end{pgfonlayer}
\end{scope}
\begin{scope}[xshift=-18.8cm, yshift=-4.5cm]
  \draw[] (19.3, -3) -- +(0, 1) node[pos=0, circle, fill=black, inner sep=0pt,minimum size=3pt] {{}} node[pos=1, circle, fill=black, inner sep=0pt,minimum size=3pt] {{}}; 
  \draw[] (20.3, -2) -- +(0, -1) node[pos=0, circle, fill=black, inner sep=0pt,minimum size=3pt] {{}} node[pos=1] (gate16) {$\oplus$}; 
  \draw[] (21.3, -2) -- +(0, 2) node[pos=0, circle, fill=black, inner sep=0pt,minimum size=3pt] {{}} node[pos=1] (gate17) {$\oplus$}; 
  \node[draw, fill=white] (gate18) at (22.3, 0) {$X$}; 
  \draw[] (22.3, -1) -- +(0, -1) node[pos=0, circle, fill=black, inner sep=0pt,minimum size=3pt] {{}} node[pos=1] (gate19) {$\oplus$}; 
  \draw[] (23.7, 0) -- +(0, -1) node[pos=0, circle, fill=black, inner sep=0pt,minimum size=3pt] {{}} node[pos=1, draw, fill=white] (gate20) {$RY(\theta/4)$}; 
  \draw[] (25.1, 0) -- +(0, -3) node[pos=0, circle, fill=black, inner sep=0pt,minimum size=3pt] {{}} node[pos=1] (gate21) {$\oplus$}; 
  \draw[] (26.5, -3) -- +(0, 2) node[pos=0, circle, fill=black, inner sep=0pt,minimum size=3pt] {{}} node[pos=1, draw, fill=white] (gate22) {$RY(\theta/4)$}; 
  \draw[] (27.900000000000002, 0) -- +(0, -3) node[pos=0, circle, fill=black, inner sep=0pt,minimum size=3pt] {{}} node[pos=1] (gate23) {$\oplus$}; 
  \draw[] (29.3, -3) -- +(0, 2) node[pos=0, circle, fill=black, inner sep=0pt,minimum size=3pt] {{}} node[pos=1, draw, fill=white] (gate24) {$RY(-\theta/4)$}; 
  \draw[] (30.700000000000003, -2) -- +(0, 1) node[pos=0, circle, fill=black, inner sep=0pt,minimum size=3pt] {{}} node[pos=1, circle, fill=black, inner sep=0pt,minimum size=3pt] {{}}; 
  \draw[] (32.1, 0) -- +(0, -1) node[pos=0, circle, fill=black, inner sep=0pt,minimum size=3pt] {{}} node[pos=1, draw, fill=white] (gate26) {$RY(-\theta/4)$}; 
  \draw[] (33.5, 0) -- +(0, -3) node[pos=0, circle, fill=black, inner sep=0pt,minimum size=3pt] {{}} node[pos=1] (gate27) {$\oplus$}; 
  \draw[] (34.9, -3) -- +(0, 2) node[pos=0, circle, fill=black, inner sep=0pt,minimum size=3pt] {{}} node[pos=1, draw, fill=white] (gate28) {$RY(-\theta/4)$}; 
  \draw[] (36.3, 0) -- +(0, -3) node[pos=0, circle, fill=black, inner sep=0pt,minimum size=3pt] {{}} node[pos=1] (gate29) {$\oplus$}; 
  \draw[] (37.699999999999996, -3) -- +(0, 2) node[pos=0, circle, fill=black, inner sep=0pt,minimum size=3pt] {{}} node[pos=1, draw, fill=white] (gate30) {$RY(\theta/4)$}; 
\begin{pgfonlayer}{bg}
  \draw (18.8, 0) -- (38.599999999999994, 0);
  \draw (18.8, -1) -- (38.599999999999994, -1);
  \draw (18.8, -2) -- (38.599999999999994, -2);
  \draw (18.8, -3) -- (38.599999999999994, -3);
\end{pgfonlayer}
\end{scope}
\begin{scope}[xshift=-38.599999999999994cm, yshift=-9.0cm]
  \node[draw, fill=white] (gate31) at (39.099999999999994, 0) {$X$}; 
  \draw[] (39.099999999999994, -2) -- +(0, 1) node[pos=0, circle, fill=black, inner sep=0pt,minimum size=3pt] {{}} node[pos=1, circle, fill=black, inner sep=0pt,minimum size=3pt] {{}}; 
  \draw[] (40.099999999999994, -1) -- +(0, -1) node[pos=0, circle, fill=black, inner sep=0pt,minimum size=3pt] {{}} node[pos=1] (gate33) {$\oplus$}; 
  \draw[] (41.099999999999994, -2) -- +(0, 2) node[pos=0, circle, fill=black, inner sep=0pt,minimum size=3pt] {{}} node[pos=1] (gate34) {$\oplus$}; 
  \draw[] (42.099999999999994, -2) -- +(0, -1) node[pos=0, circle, fill=black, inner sep=0pt,minimum size=3pt] {{}} node[pos=1] (gate35) {$\oplus$}; 
  \draw[] (43.49999999999999, 0) -- +(0, -2) node[pos=0, circle, fill=black, inner sep=0pt,minimum size=3pt] {{}} node[pos=1, draw, fill=white] (gate36) {$RY(-\pi/8)$}; 
  \draw[] (44.89999999999999, 0) -- +(0, -1) node[pos=0, circle, fill=black, inner sep=0pt,minimum size=3pt] {{}} node[pos=1] (gate37) {$\oplus$}; 
  \draw[] (46.29999999999999, -1) -- +(0, -1) node[pos=0, circle, fill=black, inner sep=0pt,minimum size=3pt] {{}} node[pos=1, draw, fill=white] (gate38) {$RY(-\pi/8)$}; 
  \draw[] (47.69999999999999, 0) -- +(0, -1) node[pos=0, circle, fill=black, inner sep=0pt,minimum size=3pt] {{}} node[pos=1] (gate39) {$\oplus$}; 
  \draw[] (49.09999999999999, -1) -- +(0, -1) node[pos=0, circle, fill=black, inner sep=0pt,minimum size=3pt] {{}} node[pos=1, draw, fill=white] (gate40) {$RY(\pi/8)$}; 
  \draw[] (50.499999999999986, -3) -- +(0, 1) node[pos=0, circle, fill=black, inner sep=0pt,minimum size=3pt] {{}} node[pos=1, circle, fill=black, inner sep=0pt,minimum size=3pt] {{}}; 
  \draw[] (51.899999999999984, 0) -- +(0, -2) node[pos=0, circle, fill=black, inner sep=0pt,minimum size=3pt] {{}} node[pos=1, draw, fill=white] (gate42) {$RY(\pi/8)$}; 
  \draw[] (53.29999999999998, 0) -- +(0, -1) node[pos=0, circle, fill=black, inner sep=0pt,minimum size=3pt] {{}} node[pos=1] (gate43) {$\oplus$}; 
  \draw[] (54.69999999999998, -1) -- +(0, -1) node[pos=0, circle, fill=black, inner sep=0pt,minimum size=3pt] {{}} node[pos=1, draw, fill=white] (gate44) {$RY(\pi/8)$}; 
  \draw[] (56.09999999999998, 0) -- +(0, -1) node[pos=0, circle, fill=black, inner sep=0pt,minimum size=3pt] {{}} node[pos=1] (gate45) {$\oplus$}; 
  \draw[] (57.49999999999998, -1) -- +(0, -1) node[pos=0, circle, fill=black, inner sep=0pt,minimum size=3pt] {{}} node[pos=1, draw, fill=white] (gate46) {$RY(-\pi/8)$}; 
\begin{pgfonlayer}{bg}
  \draw (38.599999999999994, 0) -- (58.39999999999998, 0);
  \draw (38.599999999999994, -1) -- (58.39999999999998, -1);
  \draw (38.599999999999994, -2) -- (58.39999999999998, -2);
  \draw (38.599999999999994, -3) -- (58.39999999999998, -3);
\end{pgfonlayer}
\end{scope}
\begin{scope}[xshift=-58.39999999999998cm, yshift=-13.5cm]
  \node[draw, fill=white] (gate47) at (58.89999999999998, 0) {$X$}; 
  \draw[] (58.89999999999998, -3) -- +(0, 1) node[pos=0, circle, fill=black, inner sep=0pt,minimum size=3pt] {{}} node[pos=1, circle, fill=black, inner sep=0pt,minimum size=3pt] {{}}; 
  \draw[] (59.89999999999998, -2) -- +(0, -1) node[pos=0, circle, fill=black, inner sep=0pt,minimum size=3pt] {{}} node[pos=1] (gate49) {$\oplus$}; 
  \draw[] (60.89999999999998, -2) -- +(0, 1) node[pos=0, circle, fill=black, inner sep=0pt,minimum size=3pt] {{}} node[pos=1] (gate50) {$\oplus$}; 
  \draw[] (61.89999999999998, -3) -- +(0, 3) node[pos=0, circle, fill=black, inner sep=0pt,minimum size=3pt] {{}} node[pos=1] (gate51) {$\oplus$}; 
\begin{pgfonlayer}{bg}
  \draw (58.39999999999998, 0) -- (62.39999999999998, 0);
  \draw (58.39999999999998, -1) -- (62.39999999999998, -1);
  \draw (58.39999999999998, -2) -- (62.39999999999998, -2);
  \draw (58.39999999999998, -3) -- (62.39999999999998, -3);
\end{pgfonlayer}
\end{scope}
\end{tikzpicture}
\end{center}

While the Pair break up gate $\mathrm{QNP_{PBU}}$ has the decomposition:
\begin{center}
\begin{tikzpicture}[scale=0.7, transform shape]
\begin{scope}
  \node[draw, fill=white, minimum height=height("$A$")+3cm, text width=2.8cm-10mm, inner sep=2mm, align=center] (lhs) at (-2.2, -1.5) {$$\mathrm{QNP_{PBU}}(\theta)$$}; 
  \begin{pgfonlayer}{bg}
    \draw ($(lhs.west |- 0,0) + (-0.2,0)$) -- ($(lhs.east |- 0,0) + (0.2,0)$);
    \draw ($(lhs.west |- 0,-1) + (-0.2,0)$) -- ($(lhs.east |- 0,-1) + (0.2,0)$);
    \draw ($(lhs.west |- 0,-2) + (-0.2,0)$) -- ($(lhs.east |- 0,-2) + (0.2,0)$);
    \draw ($(lhs.west |- 0,-3) + (-0.2,0)$) -- ($(lhs.east |- 0,-3) + (0.2,0)$);
  \end{pgfonlayer}
  \node at (lhs.center-|-0.4,0) {{$=$}};  \draw[] (0.5, -3) -- +(0, 3) node[pos=0, circle, fill=black, inner sep=0pt,minimum size=3pt] {{}} node[pos=1] (gate0) {$\oplus$}; 
  \draw[] (1.5, -2) -- +(0, 1) node[pos=0, circle, fill=black, inner sep=0pt,minimum size=3pt] {{}} node[pos=1] (gate1) {$\oplus$}; 
  \node[draw, fill=white] (gate2) at (1.5, 0) {$X$}; 
  \draw[] (2.5, -2) -- +(0, -1) node[pos=0, circle, fill=black, inner sep=0pt,minimum size=3pt] {{}} node[pos=1] (gate3) {$\oplus$}; 
  \draw[] (3.9, 0) -- +(0, -2) node[pos=0, circle, fill=black, inner sep=0pt,minimum size=3pt] {{}} node[pos=1, draw, fill=white] (gate4) {$RY(\pi/8)$}; 
  \draw[] (5.3, 0) -- +(0, -1) node[pos=0, circle, fill=black, inner sep=0pt,minimum size=3pt] {{}} node[pos=1] (gate5) {$\oplus$}; 
  \draw[] (6.7, -1) -- +(0, -1) node[pos=0, circle, fill=black, inner sep=0pt,minimum size=3pt] {{}} node[pos=1, draw, fill=white] (gate6) {$RY(\pi/8)$}; 
  \draw[] (8.1, 0) -- +(0, -1) node[pos=0, circle, fill=black, inner sep=0pt,minimum size=3pt] {{}} node[pos=1] (gate7) {$\oplus$}; 
  \draw[] (9.5, -1) -- +(0, -1) node[pos=0, circle, fill=black, inner sep=0pt,minimum size=3pt] {{}} node[pos=1, draw, fill=white] (gate8) {$RY(-\pi/8)$}; 
  \draw[] (10.9, -3) -- +(0, 1) node[pos=0, circle, fill=black, inner sep=0pt,minimum size=3pt] {{}} node[pos=1, circle, fill=black, inner sep=0pt,minimum size=3pt] {{}}; 
  \draw[] (12.3, 0) -- +(0, -2) node[pos=0, circle, fill=black, inner sep=0pt,minimum size=3pt] {{}} node[pos=1, draw, fill=white] (gate10) {$RY(-\pi/8)$}; 
  \draw[] (13.700000000000001, 0) -- +(0, -1) node[pos=0, circle, fill=black, inner sep=0pt,minimum size=3pt] {{}} node[pos=1] (gate11) {$\oplus$}; 
  \draw[] (15.100000000000001, -1) -- +(0, -1) node[pos=0, circle, fill=black, inner sep=0pt,minimum size=3pt] {{}} node[pos=1, draw, fill=white] (gate12) {$RY(-\pi/8)$}; 
  \draw[] (16.5, 0) -- +(0, -1) node[pos=0, circle, fill=black, inner sep=0pt,minimum size=3pt] {{}} node[pos=1] (gate13) {$\oplus$}; 
  \draw[] (17.9, -1) -- +(0, -1) node[pos=0, circle, fill=black, inner sep=0pt,minimum size=3pt] {{}} node[pos=1, draw, fill=white] (gate14) {$RY(\pi/8)$}; 
\begin{pgfonlayer}{bg}
  \draw (0, 0) -- (18.8, 0);
  \draw (0, -1) -- (18.8, -1);
  \draw (0, -2) -- (18.8, -2);
  \draw (0, -3) -- (18.8, -3);
\end{pgfonlayer}
\end{scope}
\begin{scope}[xshift=-18.8cm, yshift=-4.5cm]
  \draw[] (19.3, -3) -- +(0, 1) node[pos=0, circle, fill=black, inner sep=0pt,minimum size=3pt] {{}} node[pos=1, circle, fill=black, inner sep=0pt,minimum size=3pt] {{}}; 
  \draw[] (20.3, -2) -- +(0, -1) node[pos=0, circle, fill=black, inner sep=0pt,minimum size=3pt] {{}} node[pos=1] (gate16) {$\oplus$}; 
  \draw[] (21.3, -3) -- +(0, 3) node[pos=0, circle, fill=black, inner sep=0pt,minimum size=3pt] {{}} node[pos=1] (gate17) {$\oplus$}; 
  \draw[] (22.3, -3) -- +(0, 2) node[pos=0, circle, fill=black, inner sep=0pt,minimum size=3pt] {{}} node[pos=1] (gate18) {$\oplus$}; 
  \draw[] (23.7, 0) -- +(0, -3) node[pos=0, circle, fill=black, inner sep=0pt,minimum size=3pt] {{}} node[pos=1, draw, fill=white] (gate19) {$RY(\theta/4)$}; 
  \draw[] (25.1, 0) -- +(0, -2) node[pos=0, circle, fill=black, inner sep=0pt,minimum size=3pt] {{}} node[pos=1] (gate20) {$\oplus$}; 
  \draw[] (26.5, -2) -- +(0, -1) node[pos=0, circle, fill=black, inner sep=0pt,minimum size=3pt] {{}} node[pos=1, draw, fill=white] (gate21) {$RY(-\theta/4)$}; 
  \draw[] (27.900000000000002, 0) -- +(0, -2) node[pos=0, circle, fill=black, inner sep=0pt,minimum size=3pt] {{}} node[pos=1] (gate22) {$\oplus$}; 
  \draw[] (29.3, -2) -- +(0, -1) node[pos=0, circle, fill=black, inner sep=0pt,minimum size=3pt] {{}} node[pos=1, draw, fill=white] (gate23) {$RY(\theta/4)$}; 
  \draw[] (30.700000000000003, -1) -- +(0, -2) node[pos=0, circle, fill=black, inner sep=0pt,minimum size=3pt] {{}} node[pos=1, circle, fill=black, inner sep=0pt,minimum size=3pt] {{}}; 
  \draw[] (32.1, 0) -- +(0, -3) node[pos=0, circle, fill=black, inner sep=0pt,minimum size=3pt] {{}} node[pos=1, draw, fill=white] (gate25) {$RY(\theta/4)$}; 
  \draw[] (33.5, 0) -- +(0, -2) node[pos=0, circle, fill=black, inner sep=0pt,minimum size=3pt] {{}} node[pos=1] (gate26) {$\oplus$}; 
  \draw[] (34.9, -2) -- +(0, -1) node[pos=0, circle, fill=black, inner sep=0pt,minimum size=3pt] {{}} node[pos=1, draw, fill=white] (gate27) {$RY(-\theta/4)$}; 
  \draw[] (36.3, 0) -- +(0, -2) node[pos=0, circle, fill=black, inner sep=0pt,minimum size=3pt] {{}} node[pos=1] (gate28) {$\oplus$}; 
  \draw[] (37.699999999999996, -2) -- +(0, -1) node[pos=0, circle, fill=black, inner sep=0pt,minimum size=3pt] {{}} node[pos=1, draw, fill=white] (gate29) {$RY(\theta/4)$}; 
\begin{pgfonlayer}{bg}
  \draw (18.8, 0) -- (38.599999999999994, 0);
  \draw (18.8, -1) -- (38.599999999999994, -1);
  \draw (18.8, -2) -- (38.599999999999994, -2);
  \draw (18.8, -3) -- (38.599999999999994, -3);
\end{pgfonlayer}
\end{scope}
\begin{scope}[xshift=-38.599999999999994cm, yshift=-9.0cm]
  \draw[] (39.099999999999994, -1) -- +(0, -2) node[pos=0, circle, fill=black, inner sep=0pt,minimum size=3pt] {{}} node[pos=1, circle, fill=black, inner sep=0pt,minimum size=3pt] {{}}; 
  \draw[] (40.099999999999994, -3) -- +(0, 2) node[pos=0, circle, fill=black, inner sep=0pt,minimum size=3pt] {{}} node[pos=1] (gate31) {$\oplus$}; 
  \draw[] (41.099999999999994, -3) -- +(0, 3) node[pos=0, circle, fill=black, inner sep=0pt,minimum size=3pt] {{}} node[pos=1] (gate32) {$\oplus$}; 
  \draw[] (42.099999999999994, -2) -- +(0, -1) node[pos=0, circle, fill=black, inner sep=0pt,minimum size=3pt] {{}} node[pos=1] (gate33) {$\oplus$}; 
  \draw[] (43.49999999999999, 0) -- +(0, -2) node[pos=0, circle, fill=black, inner sep=0pt,minimum size=3pt] {{}} node[pos=1, draw, fill=white] (gate34) {$RY(-\pi/8)$}; 
  \draw[] (44.89999999999999, 0) -- +(0, -1) node[pos=0, circle, fill=black, inner sep=0pt,minimum size=3pt] {{}} node[pos=1] (gate35) {$\oplus$}; 
  \draw[] (46.29999999999999, -1) -- +(0, -1) node[pos=0, circle, fill=black, inner sep=0pt,minimum size=3pt] {{}} node[pos=1, draw, fill=white] (gate36) {$RY(-\pi/8)$}; 
  \draw[] (47.69999999999999, 0) -- +(0, -1) node[pos=0, circle, fill=black, inner sep=0pt,minimum size=3pt] {{}} node[pos=1] (gate37) {$\oplus$}; 
  \draw[] (49.09999999999999, -1) -- +(0, -1) node[pos=0, circle, fill=black, inner sep=0pt,minimum size=3pt] {{}} node[pos=1, draw, fill=white] (gate38) {$RY(\pi/8)$}; 
  \draw[] (50.499999999999986, -3) -- +(0, 1) node[pos=0, circle, fill=black, inner sep=0pt,minimum size=3pt] {{}} node[pos=1, circle, fill=black, inner sep=0pt,minimum size=3pt] {{}}; 
  \draw[] (51.899999999999984, 0) -- +(0, -2) node[pos=0, circle, fill=black, inner sep=0pt,minimum size=3pt] {{}} node[pos=1, draw, fill=white] (gate40) {$RY(\pi/8)$}; 
  \draw[] (53.29999999999998, 0) -- +(0, -1) node[pos=0, circle, fill=black, inner sep=0pt,minimum size=3pt] {{}} node[pos=1] (gate41) {$\oplus$}; 
  \draw[] (54.69999999999998, -1) -- +(0, -1) node[pos=0, circle, fill=black, inner sep=0pt,minimum size=3pt] {{}} node[pos=1, draw, fill=white] (gate42) {$RY(\pi/8)$}; 
  \draw[] (56.09999999999998, 0) -- +(0, -1) node[pos=0, circle, fill=black, inner sep=0pt,minimum size=3pt] {{}} node[pos=1] (gate43) {$\oplus$}; 
  \draw[] (57.49999999999998, -1) -- +(0, -1) node[pos=0, circle, fill=black, inner sep=0pt,minimum size=3pt] {{}} node[pos=1, draw, fill=white] (gate44) {$RY(-\pi/8)$}; 
\begin{pgfonlayer}{bg}
  \draw (38.599999999999994, 0) -- (58.39999999999998, 0);
  \draw (38.599999999999994, -1) -- (58.39999999999998, -1);
  \draw (38.599999999999994, -2) -- (58.39999999999998, -2);
  \draw (38.599999999999994, -3) -- (58.39999999999998, -3);
\end{pgfonlayer}
\end{scope}
\begin{scope}[xshift=-58.39999999999998cm, yshift=-13.5cm]
  \draw[] (58.89999999999998, -3) -- +(0, 1) node[pos=0, circle, fill=black, inner sep=0pt,minimum size=3pt] {{}} node[pos=1, circle, fill=black, inner sep=0pt,minimum size=3pt] {{}}; 
  \draw[] (59.89999999999998, -2) -- +(0, -1) node[pos=0, circle, fill=black, inner sep=0pt,minimum size=3pt] {{}} node[pos=1] (gate46) {$\oplus$}; 
  \draw[] (60.89999999999998, -2) -- +(0, 1) node[pos=0, circle, fill=black, inner sep=0pt,minimum size=3pt] {{}} node[pos=1] (gate47) {$\oplus$}; 
  \draw[] (61.89999999999998, -3) -- +(0, 3) node[pos=0, circle, fill=black, inner sep=0pt,minimum size=3pt] {{}} node[pos=1] (gate48) {$\oplus$}; 
  \node[draw, fill=white] (gate49) at (62.89999999999998, 0) {$X$}; 
\begin{pgfonlayer}{bg}
  \draw (58.39999999999998, 0) -- (63.39999999999998, 0);
  \draw (58.39999999999998, -1) -- (63.39999999999998, -1);
  \draw (58.39999999999998, -2) -- (63.39999999999998, -2);
  \draw (58.39999999999998, -3) -- (63.39999999999998, -3);
\end{pgfonlayer}
\end{scope}
\end{tikzpicture}
\end{center}

\subsection{Fermionic orbital swap gate}
Finally, the $\mathrm{OFSWAP}$ gate is an orbital wise fermionic swap gate,
i.e., two fermionic swap gates (a $\mathrm{SWAP}$ gate followed by a controlled
$\mathrm{Z}$ gate) with action 
\begin{equation}
\mathrm{FSWAP} = 
\left[ 
\begin{array}{rrrr}
 +1& & & \\
  & &+1& \\
  &+1& & \\
  & & &-1\\
 \end{array}
\right],
\end{equation}
where the $-1$ in the lower right corner takes into account the sign of the
fermionic anti-communtation relations, applied between the alpha and beta wires
respectively.  Curiously $\mathrm{OFSWAP}$ is only up to phases representable by
an orbital rotation as we have $\mathrm{OFSWAP} \, \hat Z_0 \, \hat Z_{\bar
0} = \mathrm{QNP_{2OGR}}(\pi)$.

\end{widetext}

\section{Generalized Parameter-Shift Rules}

In order to compute the derivative of expectation values with respect to quantum gate parameters, the so-called parameter-shift rule has been established as a tool to avoid finite difference derivatives, which become unstable under the influence of noise from both, measurements and circuit imperfections \cite{Li_Sun_17}.
In addition to the original concept, multiple efforts have been made to analyze and generalize the parameter-shift rule \cite{Mitarai_Fujii_18, Schuld_Killoran_19, Mari_Killoran_20,Banchi_Crooks_21,Meyer_Eisert_20}.
In this appendix we introduce the concept of tuning the shift angle in parameter-shift rules for an algorithmic advantage (Section \ref{sec:tuning}), a new four-term parameter-shift rule for gates with three distinct eigenvalues (Section \ref{sec:four-term-rule}) and exclude a further straightforward generalization of this type of parameter-shift rules (Section \ref{sec:generalization}).
We also compare our new four-term rule to the one recently presented in \cite{Kottmann_Guzik_20} (Section \ref{sec:relation_to_Kottmann}) and extend the variance minimization strategy from \cite{Mari_Killoran_20} to both four-term rules.
This four term shift rule is applicable to all of the quantum number preserving gates introduced above, except for the spin adapted $\mathrm{QNP_{OR}}$ gate, which can be analytically differentiated by differentiating the individual $G$ gates and using chain rule.

\subsection{Shift tuning}\label{sec:tuning}
We briefly recap the derivation of the standard parameter-shift rule without fixing the shift angle, leading to a free parameter in the rule.
Consider a parametrized gate of the form
\begin{align}
    U(\theta)\coloneqq\exp\left(-i\frac{\theta}{2}P\right)
\end{align}
where $P^2=\mathbf{1}$, as is the case for example for Pauli rotation gates.
In a circuit with an arbitrary number of parameters, let's single out the parameter of the gate $U$ above and write our cost function of interest as
\begin{align}\label{eq:expval_short}
    f(\theta)=\bra{\psi(\theta)}B_0\ket{\psi(\theta)}\eqqcolon\bra{\phi}U(\theta)^\dagger B U(\theta)\ket{\phi},
\end{align}
where the part of the circuit preparing $\ket{\psi}(\theta)$ from some initial state applied before the gate $U$ has been absorbed into $\ket{\phi}$ and the part after $U$ is absorbed in to $B$.
Then the derivative is, by the product rule, given by
\begin{align}
    \frac{\partial}{\partial\theta}f(\theta)=\bra{\phi}U(\theta)^\dagger \left(-\frac{i}{2}[B,P]\right)U(\theta)\ket{\phi}.
\end{align}

Now look at the conjugation of $B$ by $U$ at an arbitrary shift angle $\pm\alpha$:
\begin{align}
    \mathcal{U}(\pm\alpha)&(B)\nonumber\\
    \coloneqq&U(\pm\alpha)^\dagger B U(\pm\alpha)\nonumber\\
    =&U(\pm\alpha)^\dagger B\left(\cos\left(\frac{\alpha}{2}\right)\mathbf{1}\mp i\sin\left(\frac{\alpha}{2}\right)P\right)\\
    =&\cos\left(\frac{\alpha}{2}\right)^2B+\sin\left(\frac{\alpha}{2}\right)^2PBP\mp\frac{i}{2}\sin(\alpha)[B,P]\nonumber 
\end{align}
Subtracting $\mathcal{U}(-\alpha)(B)$ from $\mathcal{U}(\alpha)(B)$ and excluding multiples of $\pi$ as values for $\alpha$, we obtain the generalized two-term parameter-shift rule
\begin{align}\label{eq:tuning_result}
    &\mathcal{U}(\alpha)(B)-\mathcal{U}(-\alpha)(B)=-i\sin(\alpha)[B,P]\\
\Rightarrow\quad&\frac{\partial}{\partial\theta}f(\theta)=\frac{1}{2\sin(\alpha)}\big(f(\theta+\alpha)-f(\theta-\alpha)\big)
\end{align}
where the original parameter-shift rule corresponds to choosing $\alpha=\pi/2$.
We note that the concept of shift-tuning was independently discovered in \cite{Mari_Killoran_20} and introduced in the quantum computing software package PennyLane \cite{PennyLane}.

\subsubsection{Reducing the gate count}
In particular, the general form of Eq. \eqref{eq:tuning_result} allows us -- provided that $\theta$ is not a multiple of $\pi$ -- to choose $\alpha=-\theta$, making the first of the cost function evaluations $f(0)$ and therefore reducing the gate count because $U(0)=\mathbf{1}$ can be skipped in the circuit.
This may lead to an additional gate count reduction if the neighbouring gates on both sides of $U$ can be merged, which is true for example in circuits for the Quantum Approximate Optimization Algorithm (QAOA).

\subsection{Four-term parameter-shift rule}\label{sec:four-term-rule}
Here we derive a four-term parameter-shift rule for gates that do not fulfil the two-term rule, e.g.~controlled rotation gates like $CR_Z(\theta)$ or many of our QNP gates with one parameter.

To this end, consider a gate 
\begin{align} \label{eq:four_term_shift_gate}
    U(\varphi)\coloneqq\exp\left(-i\frac{\varphi}{2}Q\right)
\end{align}
with  $Q^3=Q$ but not necessarily $Q^2=\mathbf{1}$, as is true for any gate with spectrum $\{-1,0,1\}$.
Then the exponential series can be rewritten as
\begin{align}
    U(\varphi)=\mathbf{1}+\left(\cos\left(\frac{\varphi}{2}\right)-1\right)Q^2-i\sin\left(\frac{\varphi}{2}\right)Q
\end{align}
and a computation similar to the one above leads to
\begin{align}
    &\mathcal{U}(\alpha)(B)-\mathcal{U}(-\alpha)(B)\nonumber\\
    =&- 2i\sin\left(\frac{\alpha}{2}\right)[B,Q]\label{eq:2point_op_diff}\\
    &- 2i\sin\left(\frac{\alpha}{2}\right)\left(\cos\left(\frac{\alpha}{2}\right)-1\right)[Q,QBQ].\nonumber
\end{align}
We can then obtain the commutator by linearly combining this difference with itself for a second angle $\pm\beta$, so that
\begin{align}
-\frac{i}{2}[B,Q]&=
    d_1\big(\mathcal{U}(\alpha)(B)-\mathcal{U}(-\alpha)(B)\big)\\
    &-d_2\big(\mathcal{U}(\beta)(B)-\mathcal{U}(-\beta)(B)\big)
\end{align}
which holds true if the angles $\alpha$, $\beta$ and the prefactors $d_{1,2}$ satisfy
\begin{align}\label{eq:4point_condition_1}
    \frac{1}{4}&=d_1\sin\left(\frac{\alpha}{2}\right)-d_2\sin\left(\frac{\beta}{2}\right)\\
    \frac{1}{2}&=d_1\sin\left(\alpha\right)-d_2\sin\left(\beta\right).\label{eq:4point_condition_2}
\end{align}
Therefore, we get the four-term parameter-shift rule
\begin{align}\label{eq:four-term-rule}
    \frac{\partial}{\partial\varphi}f(\varphi)&=d_1\big(f(\varphi+\alpha)-f(\varphi-\alpha)\big)\\
&-d_2\big(f(\varphi+\beta)-f(\varphi-\beta)\big)\nonumber
\end{align}
where we again can choose $\alpha$ or $\beta$ such that one of the function evaluations skips the gate $U$.
A particularly symmetric solution of Eqns.~\eqref{eq:4point_condition_1} and \eqref{eq:4point_condition_2} is
\begin{align}
    d_1 = \frac{1}{2},\quad d_2=\frac{\sqrt{2}-1}{4},\quad \alpha=\frac{\pi}{2},\quad\beta=\pi.
\end{align}

In general, any gate for which the spectrum of the generator is $\{-a+c, c, a+c\}$ obeys the four-term parameter-shift rule as the shift $c$ can be absorbed into a global phase that does not contribute to the gradient and $a$ can be absorbed into the variational parameter of the gate.

As an example, the four-term rule is applicable to (multi-)controlled Pauli rotations $CR_P(\varphi)$ for which $Q$ is the zero matrix except for the Pauli operator $P$ on the target qubit.
For multiple control qubits and our QNP gates, this will lead to less circuit evaluations using the chain rule and applying the two-term rule to the gate decomposition.

In order to find out whether a $n$-qubit single-parameter gate $U$ satisfies the four-term rule, one can  compute
\begin{align}
    Q=\left.\frac{\partial}{\partial\varphi}U(\varphi)\right\rvert_{\varphi=0} ,\quad \overline{Q}\coloneqq Q-\frac{1}{2^n}\operatorname{tr}(Q)
\end{align}
and test if there is an $a\in\mathbb{R}$ such that $\overline{Q}^3=a^2\,\overline{Q}$,
which is a sufficient condition, as the only thing we needed for the four term rule to apply was this assumptions about the generator spectrum.

\subsubsection{Relation to other four-term rule}\label{sec:relation_to_Kottmann}
Previous work showed the existence of a four-term parameter-shift rule \cite{Kottmann_Guzik_20} for gates of the form \eqref{eq:four_term_shift_gate}, which is implemented with only one shift angle but requires the two additional gates
\begin{align}
    V_\pm = \exp\left(\mp\frac{i\alpha \pi}{4}P_0\right) \;\;\text{with}\;\; P_0=\mathbf{1}-Q^2.\nonumber
\end{align}
There are four relevant aspects when comparing this rule to the one in \eqref{eq:four-term-rule}:
First, our four-term rule does not require any additional gates like $V_\pm$, which add overhead to the gradient evaluation circuits.
While the authors bound the additional cost by the cost of the differentiated gate itself, it might more crucially be non-trivial to construct $V_\pm$ for gates that do not have an obvious fermionic representation like the gates considered in \cite{Kottmann_Guzik_20}.

Second, the shift tuning technique for gate count reduction in \ref{sec:tuning} can easily be extended to both, our four-term rule the rule derived in \cite{Kottmann_Guzik_20}, provided one has access to the parametrized versions of $V_\pm$.
As the construction of $V_\pm$ for fermion-based gates is based on rotations, this access can be assumed for these gates whenever $V_\pm$ can be implemented.

Third, it was shown in \cite{Kottmann_Guzik_20} that their four-term rule reduces to a standard \emph{two}-term rule up to the insertions of the $V_\pm$ operators whenever both the circuit of interest and the measured observables are purely real-valued.
This is the case for virtually all molecular Hamiltonians and most of the circuits proposed for quantum chemistry problems -- including the fabrics in this work -- such that gradients of highly complex gates may be computed with just two circuit executions including the gates $V_\pm$ using the rule in \cite{Kottmann_Guzik_20}.

Fourth, the variances of the derivative estimators given by the two rules can be minimized to the same value by choosing the shift angles optimally, as shown in Sec.~\ref{sec:four_term_variance}.
This means that for a given budget of circuit executions, the quality of the estimated derivative is the same, even though the number of distinct circuits differs.

In summary, the specialized two-term parameter-shift rule in \cite{Kottmann_Guzik_20} is preferable if the following three criteria hold:
Firstly, the circuit and observable need to be real-valued.
Secondly, the auxiliary gates $V_\pm$ have to be available.
Thirdly, the computation must happen on a simulation level in which the number of \emph{distinct} circuits instead of the measurement budget is relevant, so that the reduction from four to two terms provides an advantage which is larger than the overhead of adding $V_\pm$.
In all other scenarios the four-term rule Eqn.~\eqref{eq:four-term-rule} with the optimal parameters in Eqns.~(\ref{eq:c1_four_point_opt}-\ref{eq:c2_four_point_opt}) requires slightly fewer gates and the same number of circuit executions, making it preferable in particular on quantum computers.

\subsubsection{Impossibility of some further shift rules}\label{sec:generalization}
One may wonder whether a three shift rule is possible for gates whose generators have just three distinct eigenvalues and whether shift rules exist for gates with more distinct eigenvalues.
We present some insights on these questions in the following.

During the derivation of the four-term parameter-shift rule we chose to first linearly combine $\mathcal{U}(\pm\alpha)(B)$ and $\mathcal{U}(\pm\beta)(B)$ with the same prefactors, respectively.
Alternatively one may try to combine $\mathcal{U}(\alpha_i)(B)$ at three shift angles $\{\alpha_i\}_{i\in\{1,2,3\}}$ linearly and demand the result to fulfil
\begin{align}
    \sum_{i=1}^3 d_i \mathcal{U}(\alpha_i)(B)\overset{!}{=}-\frac{i}{2}[B,Q].
\end{align}
This leads to the system of equations
\begin{align}
    0&=d_1\!\left[c_1-c_3\right] + d_2 \left[c_2-c_3\right]\nonumber\\
    0&=d_1\!\left[s_1^2-s_3^2\right] + d_2 \left[s_2^2-s_3^2\right]\nonumber\\
    1&=2d_1\!\left[s_1-s_3\right] + 2d_2 \left[s_2-s_3\right]\nonumber\\
    1&=d_1 \left[\sin\left(\alpha_1\right)-\sin\left(\alpha_3\right)\right] + d_2 \left[\sin\left(\alpha_2\right)-\sin\left(\alpha_3\right)\right]\nonumber
\end{align}
with $c_i=\cos\left(\frac{\alpha_i}{2}\right)$ and $s_i=\sin\left(\frac{\alpha_i}{2}\right)$,
which we conjecture to not have a solution.

Considering the generalization of the (standard) two-term shift rule to the four-term rule in \eqref{eq:four-term-rule} and their requirement on the gate generator, i.e.~$Q^2=\mathbf{1}$ and $Q^3=Q$, it seems a natural question whether further generalization is possible to gates that, e.g., fulfil $Q^5=Q$.
We show next that this is not the case.

Consider the generalized condition $Q^m=Q^n$, $m\neq n$ for the generator of a $d$-dimensional one-parameter gate.
We recall that we may absorb shifts and scaling prefactors of the spectrum of $Q$ into a global phase gate and the variational parameter, respectively, which may be used to obtain gates satisfying the generalized condition $Q^m=Q^n$.
In the eigenbasis of the Hermitian matrix $Q$, this condition becomes $\lambda_i^m=\lambda_i^n\;\forall 1\leq i\leq d$, which only ever is solved by $-1$, $0$ and $1$ over $\mathbb{R}$ (in which the spectrum of $Q$ must be contained) with the additional condition $m-n~\mod2~=0$ for $\lambda_i=-1$.
This means that $Q$ already satisfies $Q^3=Q$, allowing for the four-term rule to be applied.

Consequently, a direct generalization of the four-term rule is not possible.
Note that this does not exclude the existence of other schemes to compute the derivative of an expectation value w.r.t. parametrized states that are based on linear combinations of shifted expectation values.

\subsection{Minimizing the variance}

\label{sec:four_term_variance}

If we approximate the physical variance of the expectation value, $V$, to be independent of $\theta$, the variance of measuring $f$ at a given parameter for sufficiently many measurements $N$ is $V/N$.
The resulting variance of the two-term shift rule derivative for a budget of $N$ measurements is 
\begin{align}
    \sigma^2 = \frac{V}{N\sin^2\alpha},
\end{align}
where we chose the optimal allocation of $N/2$ measurements to each of the two terms in the shift rule.
We may optimize the shift angle in the two-term rule w.r.t.~this variance which yields the standard choice $\pi/2$ for the shift because
\begin{align}
    \underset{\alpha}{\text{argmin}}\left\{ \frac{V}{N\sin^2\alpha}\right\} = \frac{\pi}{2}.
\end{align}

The variance can be reduced further by introducing a multiplicative bias to the estimator, as presented in \cite{Mari_Killoran_20};
The optimal choice of the prefactor depends on the value and the variance of the derivative and is given by
\begin{align}\label{eq:scaling_bias}
    \lambda^* = \left(1+\frac{V}{N (\partial_\theta f)^2 }\right)^{-1}.
\end{align}
Note that $\lambda^*$ has to be estimated because $V$ and $\partial_\theta f$ are not known exactly.
The optimal choice of the shift parameter remains $\frac{\pi}{2}$.

For the four-term rule in Eqn.~\eqref{eq:four-term-rule}, the optimal shot allocation is proportional to the prefactors $d_{1,2}$ and leads to the variance
\begin{align}
    \sigma^2 = 4(d_1+d_2)^2\frac{V}{N}.
\end{align}

As for the two-term parameter-shift rule, we may minimize this variance w.r.t. $\alpha$ and $\beta$ via $d_1$ and $d_2$, which are given via Eqns.~(\ref{eq:4point_condition_1},\ref{eq:4point_condition_2}) by
\begin{align}\label{eq:c1_four_point}
    d_2 &= \frac{1}{4\sin(\beta/2)}\frac{\cos(\alpha/2)-1}{\cos(\beta/2)-\cos(\alpha/2)}\\\label{eq:c2_four_point}
    d_1 &= \frac{1}{\sin(\alpha)}\left(\frac{1}{2}+c_2\sin(\beta)\right).
\end{align}
This results in
\begin{align}\label{eq:c1_four_point_opt}
    d_1&=\frac{\sqrt{2}+1}{4\sqrt{2}},\;\alpha=\frac{\pi}{2}, \\\label{eq:c2_four_point_opt}
    d_2&=\frac{\sqrt{2}-1}{4\sqrt{2}},\;\beta=\frac{3\pi}{2}
\end{align}
and three equivalent solutions based on the symmetries of Eqns.~(\ref{eq:4point_condition_1},\ref{eq:4point_condition_2}).

The variance then is $\sigma^2=V/N$ as for the optimal two-term rule and again it may be further reduced by introducing a bias via a multiplicative prefactor $\lambda$, with the same optimal $\lambda^*$ as before.

For both, the four-term rule and the specialized two-term rule in \cite{Kottmann_Guzik_20}, the minimal variance is $\sigma^2=V/N$ as well, as the prefactors are equally large and sum to $1$.

In conclusion, under the constant variance assumption, the variance for all discussed two- and four-term parameter shift rules is the same at a given measurement budget, showing that they are equally expensive on a quantum device, for which the number of measurements instead of the number of distinct circuits is relevant.

\begin{widetext}

\section{Additional numerical results}

\label{sec:additional numerics}

\subsection{Computational Basis State Amplitudes}

In Figure~\ref{fig:experiments}b of the main text the individual ordering of each trace of computational basis states is ordered individually, which allows to view the shape of each tail but restricts comparability between single amplitudes.
In Figure~\ref{fig:probablities} only the computational basis states of the true ground state and of one optimized VQE state at 110 parameters are plotted in consistent ordering.
This allows for direct comparison between the amplitudes of the VQE and of FCI and demonstrates how our fabric finds a good approximation to most amplitudes while having far too few parameters to reproduce all amplitudes exactly.
\begin{figure*}[h]
\centering
\includegraphics[width=1.0\textwidth]{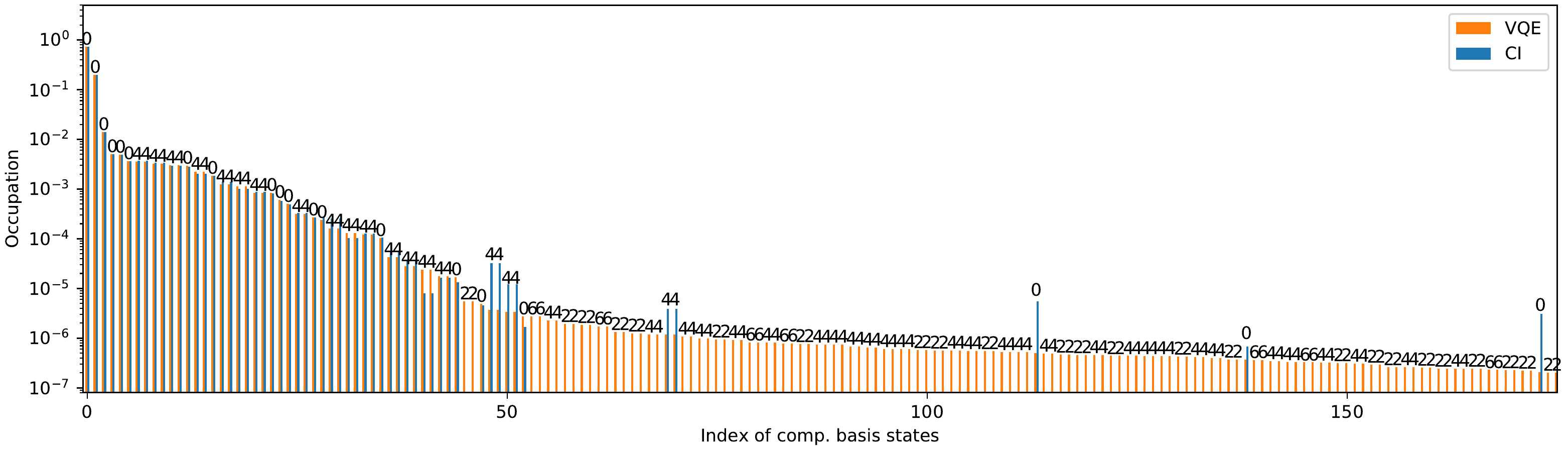}
\caption{A cutout from Figure \ref{fig:experiments}b for a VQE with 110
parameters (orange curve) and the blue shaded FCI probabilities at a vertical cut
off of $10^{-7}$ with consistent ordering between the FCI and VQE computational basis states.
The numbers above the columns indicate the seniority of the computational basis state at
the respective index, e.g. the first column at index 0 is the Hartree Fock
determinant with seniority 0.}
\label{fig:probablities}
\end{figure*}

\subsection{Numerical universality demonstration for Haar random states}

The test cases in real molecular systems in the main text are somewhat
complicated by the specifics of the electronic structure Hamiltonian and
especially by the spatial point group symmetry of the test molecules. One
notable artifact is that some of the left-most gates in our gate fabrics in real
molecules are ``dead,'' as they perform orbital rotations and diagonal pair
exchanges in the occupied or virtual subspaces of the Hartree-Fock starting
state.  The point group symmetry also seems to adversely affect the numerical
convergence behavior of the VQE gate fabric parameter optimization, e.g.,
suggesting the $\hat \Pi$-gate pre-mixing initialization adopted in the main
text. Noticeably better convergence behavior was observed when the molecules
were perturbed from $D_{2\mathrm{h}}$ symmetry to $C_1$ by random Gaussian
perturbations in XYZ coordinates. 

This section is included to demonstrate the numerical universality properties of
our proposed gate fabric for the artificial case of Haar random statevectors.
Specifically, for a number of test case irreps $(M, N_{\alpha}, N_{\beta}, S)$,
we form the full CSF basis, and then generate Haar random statevectors
$|A\rangle$ and $|B\rangle$ within this irrep of $\mathcal{F}(2^{2M})$ by
Gaussian random sampling and normalization of the statevector in the CSF basis,
and then backtransformation to the standard Jordan-Wigner computational basis.
We then optimize the VQE gate fabric parameters of the VQE entangler circuit
$\hat U$ to maximize $|\langle A | \hat U | B \rangle|^2$ via L-BFGS with
noise-free analytical gradients. Note that we do not perform $\hat \Pi$-based
pre-mixing convergence enhancement in this section. 

The results are shown for the half-filled cases for $M=4$ and $M=6$ in Figure
\ref{fig:haar-random}. The top panels show the bulk convergence properties with
respect to circuit depth $D$ / number of parameters $N_{\mathrm{parameter}}$
(roughly linearly proportional). The general finding here is roughly geometric
convergence at low parameter depths, followed by a sharp drop to near the
machine epsilon as the number of parameters crosses over the number of CSFs in
the irrep, indicating the onset of universality in the action of the VQE
entangler circuit. Quantum number symmetries are preserved to at least the
machine epsilon for all intermediate and final parameter values. The lower
panels show the numerical convergence behavior of the L-BFGS optimization
procedure for each point in the top panel. There are several salient features in
these plots: (1) The earliest convergence behavior appears to be roughly
geometric, and self-similar between gate fabrics with different numbers of
parameters (2) fabrics with smaller numbers of parameters deviate earlier from
this geometric convergence and eventually ``flatline'' at their non-universal
terminal values (3) some minor plateaus are observed in the convergence behavior
for small numbers of parameters (4) there is a distinct phase change as
universality is crossed, with circuits with larger numbers of parameters than
needed for universality exhibiting strongly geometric convergence behavior all
the way to the machine epsilon.

Such tests are assuredly artificial, but are free from the external artifacts
present in the molecular test cases, and serve to more-strongly indicate that
the gate fabric developed in Figure \ref{fig:F} are universal and
quantum-number-symmetry-preserving for $\mathcal{F}(2^{2M})$.

\begin{figure*}[h]
\centering
\includegraphics[width=0.45\textwidth]{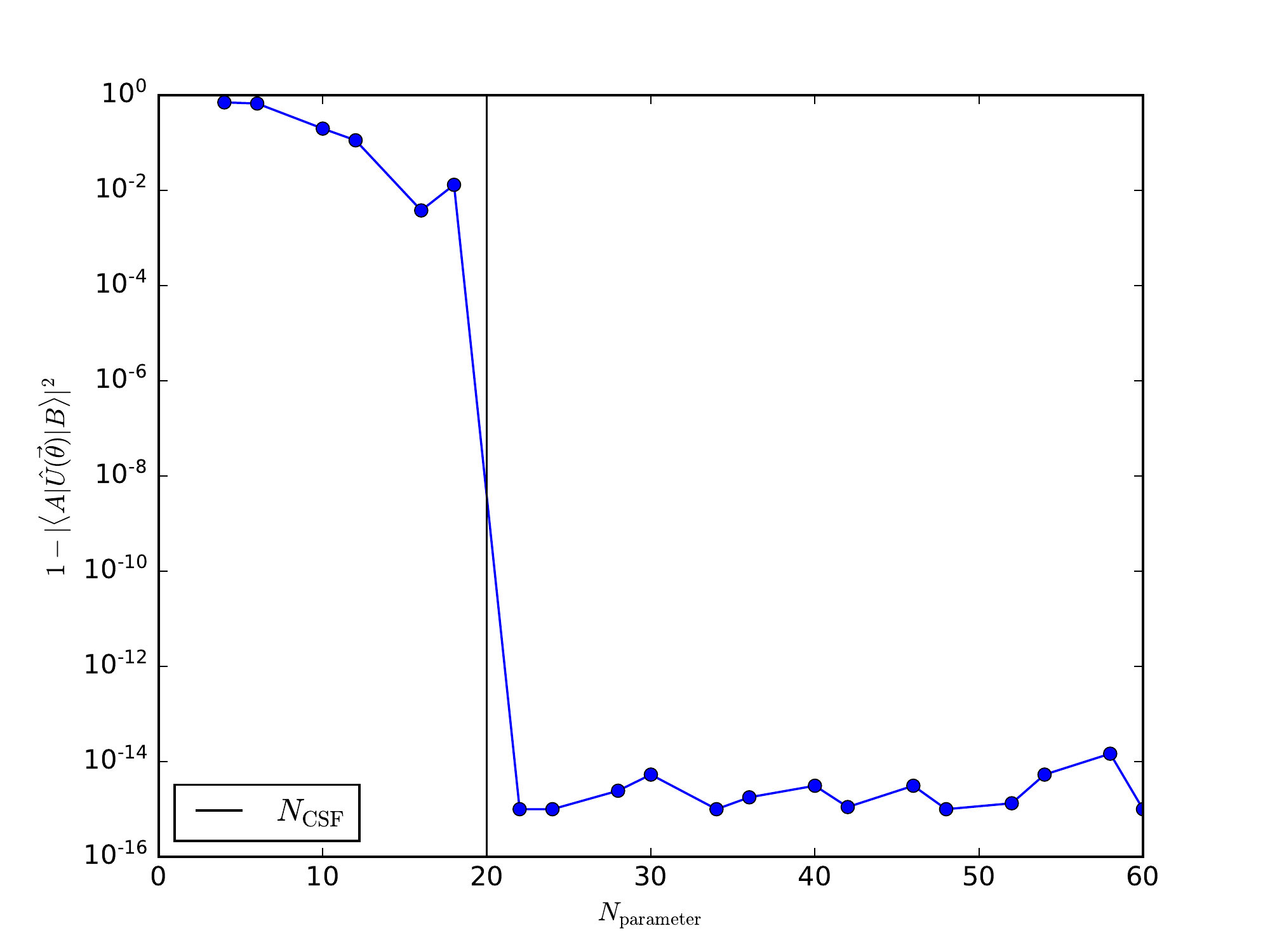}
\includegraphics[width=0.45\textwidth]{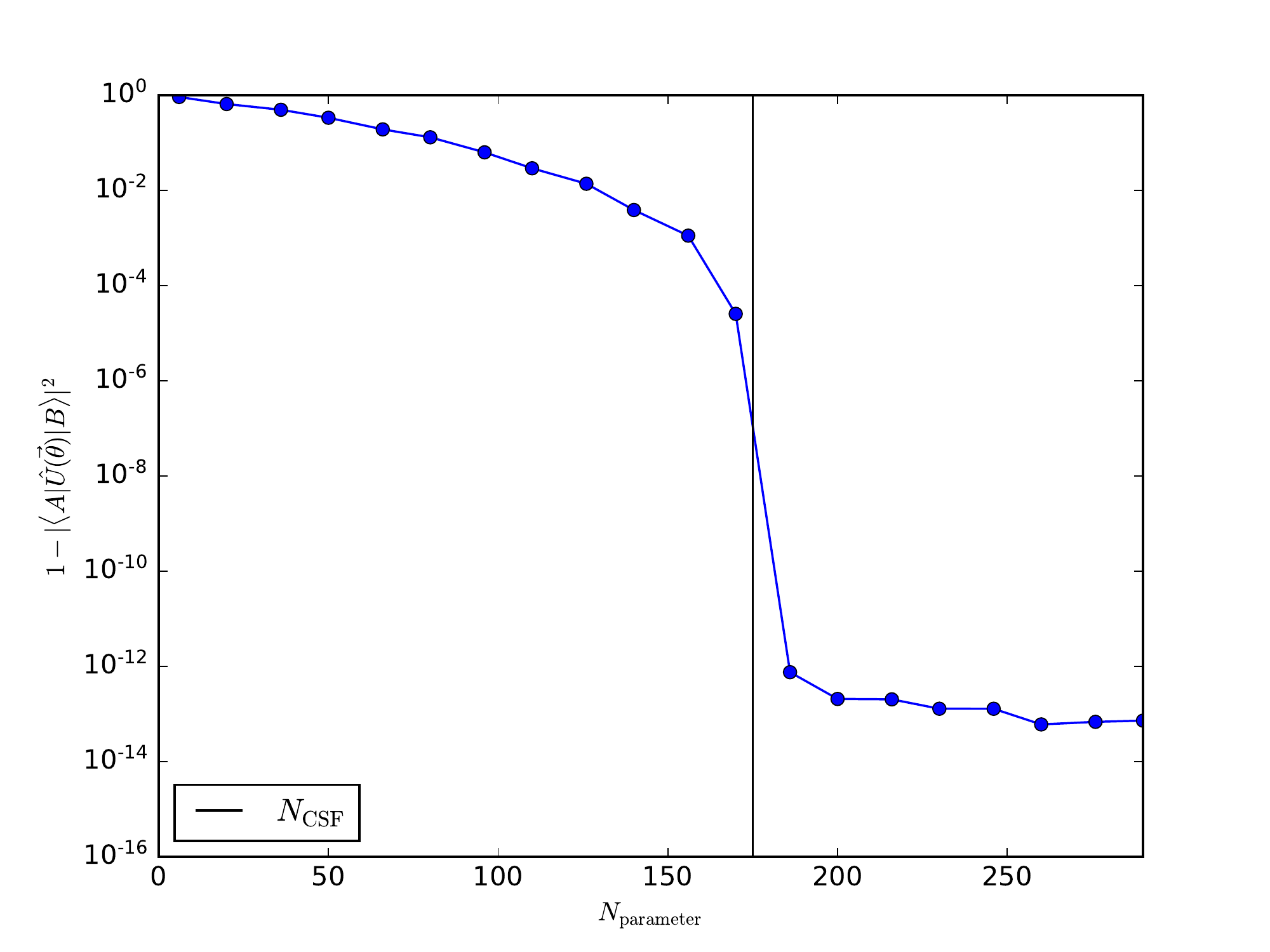}
\includegraphics[width=0.45\textwidth]{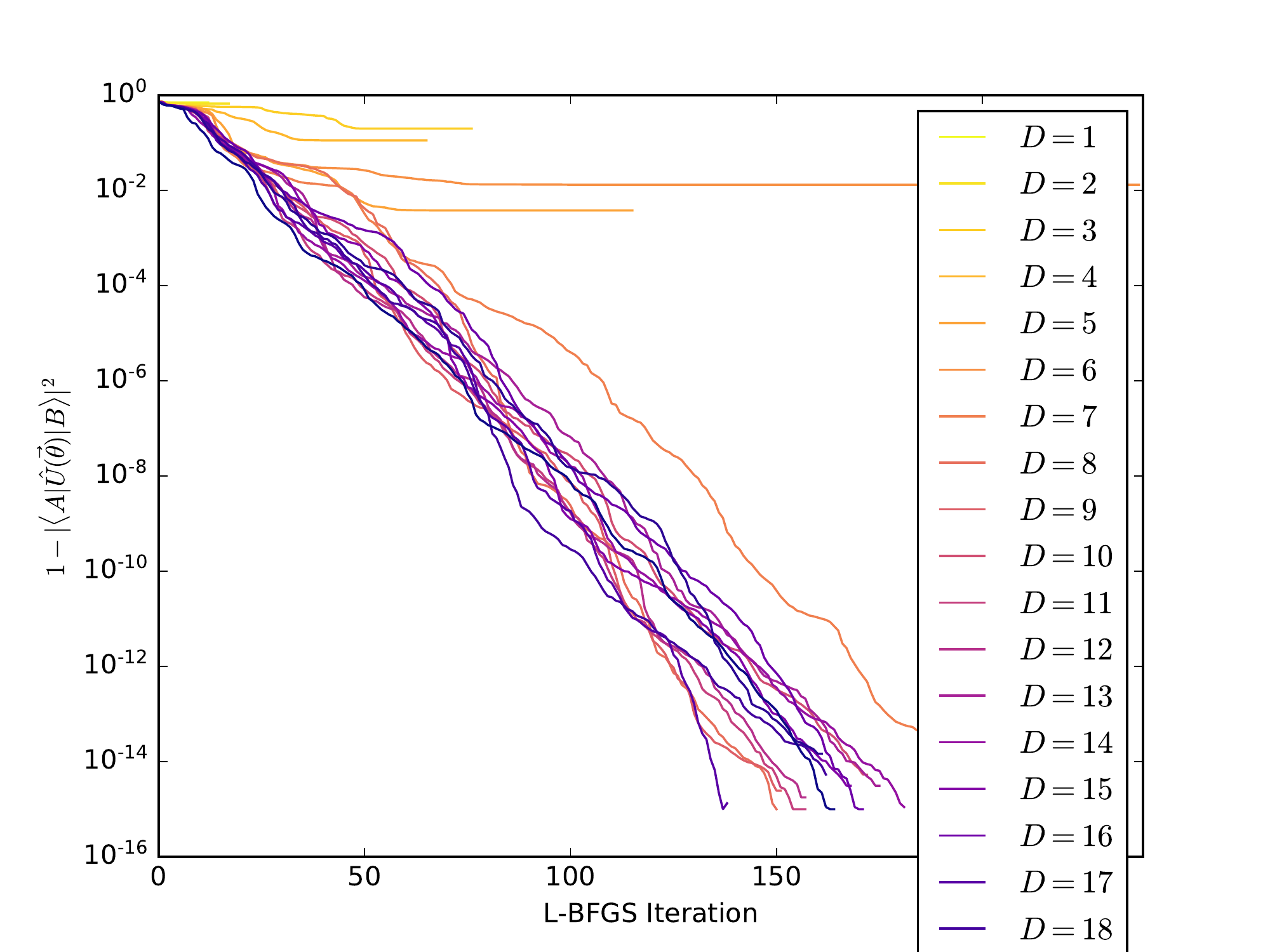}
\includegraphics[width=0.45\textwidth]{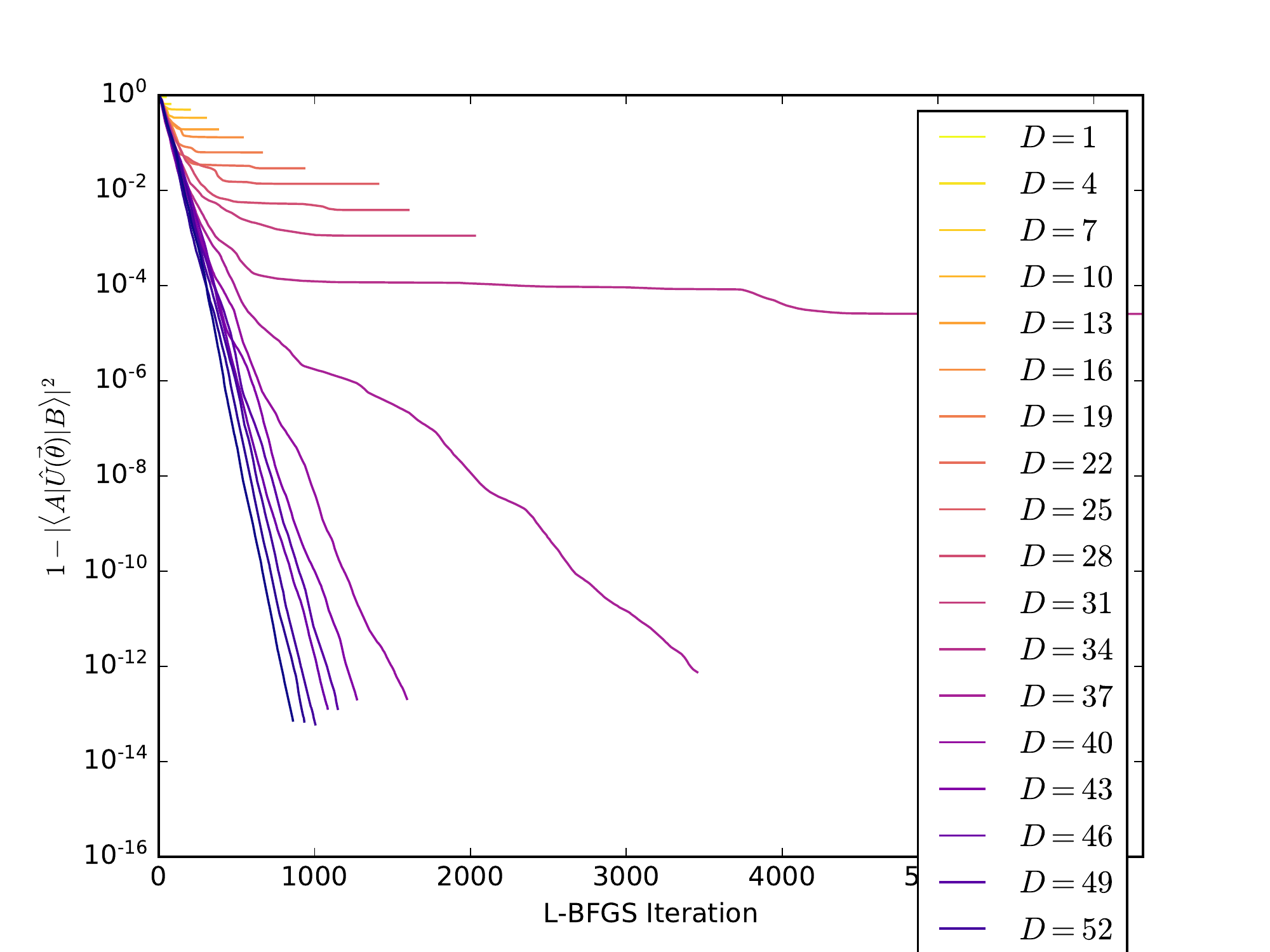}
\caption{
Numerical demonstration of universality of VQE gate fabric of the form of 
Figure \ref{fig:F} for Haar random states $|A\rangle$ and $|B\rangle$ within
$\mathcal{F}(2^{2M})$. The VQE gate fabric parameters were optimized via
L-BFGS with noise-free analytical gradients to maximize $|\langle A | \hat U | B
\rangle|^2$, where $\hat U$ is the VQE entangler circuit operator. 
Top row: Convergence of overlap $|\langle A|\hat U|B\rangle|$ with respect to
respect to gate fabric depth $D$ / number of parameters
$N_{\mathrm{parameter}}$ (roughly linearly proportional).
Bottom row: Convergence behavior of L-BFGS optimization procedure for each gate
fabric depth $D$ / number of parameters
$N_{\mathrm{parameter}}$ (roughly linearly proportional).
Left column: Results for $(M = 4, N_{\alpha} = 2, N_{\beta} = 2, S = 0)$ 8-qubit
example.
Right column: Results for $(M = 6, N_{\alpha} = 3, N_{\beta} = 3, S = 0)$ 12-qubit
example.
Each colormapped line in the lower panel corresponds to the L-BFGS numerical
convergence behavior of a single point in the upper panel.
Results are single random instances within each test case, and are wholly
representative of generic random instances and test cases in other quantum
number irreps.
}
\label{fig:haar-random}
\end{figure*}

\subsection{Non-universal edge cases}

It is important to note that while the $\hat Q$-type QNP gate fabrics of main
text are numerically universal for the vast majority of quantum number irreps in
the ``bulk'' of the Hilbert space, there are a limited number of edge cases for
which these gate fabrics are not universal. These cases constitute systems
where, after high-spin constraints are accounted for, there are only holes or
particles left in the remaining orbitals. In these cases, the
$\mathrm{QNP_{PX}}$ gates have trivial action in the wholly hole or particle
space, and are unable to explore new configurations within the space. More
tangibly, for an irrep with dimensions $(M, N_{\alpha}, N_{\beta}, S)$, we first
compute the ``unconstrained'' irrep $(M - S, N_{\alpha}', N_{\beta}', 0)$ where
$N_{\alpha}' + N_{\beta}' + S = N_{\alpha} + N_{\beta}$ and the larger of
$N_{\alpha}' \coloneqq N_{\alpha}$ or $N_{\beta}' \coloneqq N_{\beta}$ is
decremented first until $N_{\alpha}' = N_{\beta}'$, and then both $N_{\alpha}'$
and $N_{\beta}'$ are decremented together (in this line, $\coloneqq$ is read as
``initialized to''). The resulting unconstrained irrep will always have
$N_{\alpha}' = N_{\beta}'$. If the unconstrained irrep is all holes
$(N_{\alpha}' = N_{\beta}' = 0)$ or all particles $(N_{\alpha}' = N_{\beta}' = M
- S)$, then the $\hat Q$-type QNP gate fabric is not universal. A trivial
  exception is if only a single orbital with all holes or all particles remains
in the unconstrained irrep, in which case universality is still preserved.

Note that the number of irreps in the Hilbert space grows roughly as
$\mathcal{O}(M^3)$, while the required constraints $N_{\alpha}' =
N_{\beta}' = 0 \ \mathrm{or} \ M - S$ seem to indicate that the number of
non-universal irreps indicate that the number of irreps which are not universal
with $\hat Q$-type QNP gate fabrics will grow as roughly $\mathcal{O}(M)$.
Moreover, the non-universal irreps appear at the ``edge'' of the Hilbert space,
and consist of cases with severe high-spin constraints which are likely to be
either polynomially tractable classically, physically uninteresting, or both.
Interesting cases with roughly half-and-half filling of holes and particles
and moderately low total spin number will almost surely fall into irreps which
are universal with $\hat Q$-type QNP gate fabrics. Finally, it is worth noting
that any issues with these edge cases can be completely obviated by instead
working with the 5-parameter $\hat F$ gate fabrics discussed in Section
\ref{appendix:other_qnp_gate_fabrics} - these do not appear to exhibit any edge
case non-universalities, and are numerically universal for all cases we have
tested.

Tables \ref{tab:M4exceptions} and \ref{tab:M6exceptions} show explicitly the
irreps for $M=4$ and $M=6$ that were found numerically to be non-universal with
$\hat Q$-type QNP gate fabrics via numerical studies of the same type as the
previous section. The non-universality behavior was immediately apparent as
discrepancies of overlap of $1 - |\langle A | \hat U | B \rangle|^2$ of order of
$10^{-2}$, while the universal irreps exhibited maximum discrepancies of
overlap of order of $<10^{-13}$.

\begin{table*}[ht!]
\caption{Quantum number irreps for $M=4$ for which the $\hat Q$-type QNP gates
of the main text are not universal. Overall there are 35 unique irreps for $M=4$
with total dimension $D \equiv 2^{2M} = 256$. The 6 irreps with total dimension
$36$ listed below are not universal due to high-spin constraints. All other
irreps are numerically found to be universal to the essentially machine precision.}
\label{tab:M4exceptions}
\begin{tabular}{rrrrrr}
\hline \hline
$N_{\alpha}$ & $N_{\beta}$ & $S$ & Dimension \\
\hline
0 & 2 & 2 & 6 \\
1 & 1 & 2 & 6 \\
2 & 0 & 2 & 6 \\
2 & 4 & 2 & 6 \\
3 & 3 & 2 & 6 \\
4 & 2 & 2 & 6 \\
\hline \hline
\end{tabular}
\end{table*}

\begin{table*}[ht!]
\caption{Quantum number irreps for $M=6$ for which the $\hat Q$-type QNP gates
of the main text are not universal. Overall there are 84 unique irreps for $M=6$
with total dimension $D \equiv 2^{2M} = 4096$. The 24 irreps with total dimension
$400$ listed below are not universal due to high-spin constraints. All other
irreps are numerically found to be universal to the essentially machine precision.}
\label{tab:M6exceptions}
\begin{tabular}{rrrrrr}
\hline \hline
$N_{\alpha}$ & $N_{\beta}$ & $S$ & Dimension \\
\hline
0 & 2 & 2 & 15 \\
0 & 3 & 3 & 20 \\
0 & 4 & 4 & 15 \\
1 & 1 & 2 & 15 \\
1 & 2 & 3 & 20 \\
1 & 3 & 4 & 15 \\
2 & 0 & 2 & 15 \\
2 & 1 & 3 & 20 \\
2 & 2 & 4 & 15 \\
2 & 6 & 4 & 15 \\
3 & 0 & 3 & 20 \\
3 & 1 & 4 & 15 \\
3 & 5 & 4 & 15 \\
3 & 6 & 3 & 20 \\
4 & 0 & 4 & 15 \\
4 & 4 & 4 & 15 \\
4 & 5 & 3 & 20 \\
4 & 6 & 2 & 15 \\
5 & 3 & 4 & 15 \\
5 & 4 & 3 & 20 \\
5 & 5 & 2 & 15 \\
6 & 2 & 4 & 15 \\
6 & 3 & 3 & 20 \\
6 & 4 & 2 & 15 \\
\hline \hline
\end{tabular}
\end{table*}

\end{widetext}

\end{document}